\documentclass[%
 reprint,
superscriptaddress,
showpacs,preprintnumbers,
nofootinbib,
 amsmath,amssymb,
 bm,
 aps,
]{revtex4-1}

\usepackage{graphicx}
\usepackage{dcolumn}
\usepackage{bm}
\usepackage{hyperref}
\usepackage{natbib}
\usepackage{color}

\usepackage{array}

\usepackage{enumitem}

\usepackage{multirow}

\usepackage{tabularx}

\usepackage[normalem]{ulem}

\usepackage{aas_macros}

\begin{document}

\preprint{NORDITA-2019-084, UTTG-09-2019, LCTP-19-23}

\title{CMB $B$-mode non-Gaussianity: \\optimal bispectrum estimator and Fisher forecasts }





\def\beq{\begin{equation}}
\def\eeq{\end{equation}}
\newcommand{\stockholm}{The Oskar Klein Centre for Cosmoparticle Physics,
Department of Physics, Stockholm University, SE-106 91 Stockholm, Sweden}
\newcommand{\princeton}{Department of Physics:\ Joseph Henry Laboratories, 
Jadwin Hall, \\ Princeton University, 
Princeton, New Jersey 08542, USA}
\newcommand{\VSI}{Van Swinderen Institute for Particle Physics and Gravity,\\ University of Groningen,
Nijenborgh 4, 9747 AG Groningen, The Netherlands}
\newcommand{\kavli}{Kavli Institute for Cosmology, Madingley Road, Cambridge, UK, CB3 0HA}
\newcommand{\damtp}{DAMTP, Centre for Mathematical Sciences, Wilberforce Road, Cambridge, UK, CB3 0WA}
\newcommand{\michigan}{Department of Physics, University of Michigan, Ann Arbor, MI 48109, USA}
\newcommand{\austin}{Department of Physics, University of Texas, Austin, TX 78712, USA}

\author{Adriaan J.\ Duivenvoorden}
\affiliation{\princeton}

\author{P.\ Daniel Meerburg}
\affiliation{\VSI}
\affiliation{\kavli}
\affiliation{\damtp}
\author{Katherine Freese}
\affiliation{\stockholm}
\affiliation{\austin}
\affiliation{\michigan}

\date{\today}

\begin{abstract}
Upcoming cosmic microwave background (CMB) data can be used to explore harmonic $3$-point functions that involve the $B$-mode component of the CMB polarization signal. We focus on bispectra describing the non-Gaussian correlation of the $B$-mode field and the CMB temperature anisotropies ($T$) and/or $E$-mode polarization, i.e.\ $\langle TTB \rangle$, $\langle EEB \rangle$, and $\langle TEB \rangle$. Such bispectra probe violations of the tensor consistency relation: the model-independent behavior of cosmological correlation functions that involve a large-wavelength tensor mode (gravitational wave). An observed violation of the tensor consistency relation would exclude a large number of inflation models.
We describe a generalization of the Komatsu-Spergel-Wandelt (KSW) bispectrum estimator that allows statistical inference on this type of  primordial non-Gaussianity with data of the CMB temperature and polarization anisotropies. The generalized estimator shares its statistical properties with the existing KSW estimator and retains the favorable numerical scaling with angular resolution. 
In this paper we derive the estimator and present a set of Fisher forecasts. 
We show how the forecasts scale with various experimental parameters such as lower and upper angular band-limit, relevant for e.g. the upcoming ground-based Simons Observatory experiment and proposed \emph{LiteBIRD} satellite experiment. We comment on possible contaminants due to secondary cosmological and astrophysical sources. 
\end{abstract}

\maketitle


\section{\label{sec:intro}Introduction}

Inflationary cosmology was proposed \cite{Guth:1980zm,linde_1982,Albrecht:1982wi} to solve several
cosmological puzzles: an early period of accelerated expansion explains
the homogeneity, isotropy, and flatness of the Universe, as well as the
lack of relic monopoles.  One of the great successes of the inflationary paradigm is the production of small density inhomogeneities that
 grow to create the large-scale structure of the Universe today~\cite{Mukhanov:1981xt, Hawking:1982cz, starobinsky_1982, kamionkowski_1997, zaldarriaga_1997}.  In addition, tensor modes produced during inflation lead to primordial gravitational waves that are potentially detectable in the 
polarization of the Cosmic Microwave Background (CMB)~\cite{starobinsky_1979, grishchuk_1975, Rubakov:1982df, Mukhanov:1990me}. 
Observations of the CMB provide tests of these predictions of inflation and can serve to distinguish between specific inflationary models.

As yet, CMB observations are consistent with a single slowly rolling scalar field as the inflaton, the field responsible for inflation~\cite{planck_2018_x}.  
For these single-field slow roll (SFSR) models, the fluctuations are described by primordial density fluctuations, which are nearly Gaussian, adiabatic, and nearly scale invariant~\cite{linde_1982, Albrecht:1982wi}.
Gaussianity implies that the $2$-point correlation function of the density fluctuations
uniquely determines all higher even $n$-point functions while all odd $n$-point functions vanish.  
In principle, inflation could be described by variants other than SFSR that introduce significant non-Gaussianity, such as multi-field inflation; models with non-canonical kinetic terms or non Bunch-Davies vacua~\cite{Chen:2010xka}.  As yet no evidence for primordial non-Gaussianity has been found in the \emph{Planck} data~\cite{Akrami:2019izv}; hence many of these models have been ruled out. Conversely, evidence for  non-Gaussian statistics in upcoming data would imply deviations from SFSR inflation and would provide an informative probe of the inflationary dynamics and the associated high-energy physics~\cite{Meerburg:2019qqi}.

While the usual searches for primordial non-Gaussianity focus on the $n$-point statistics of scalar fluctuations, in this paper we concentrate on the relatively unexplored observational signatures of non-Gaussian correlations involving tensor fluctuations, as previously discussed by~\cite{Meerburg:2016ecv}. 
We propose to extend the search for primordial non-Gaussianity from one that only looks for the `scalar-scalar-scalar'  correlation to one that also searches for the `scalar-scalar-tensor' correlation~\cite{Meerburg:2016ecv, Maldacena:2002vr, Akhshik:2014bla, Dimastrogiovanni:2014ina, Bordin:2016ruc, Domenech:2017kno, Endlich:2013jia}:  
the non-Gaussian correlation between two modes of the primordial scalar perturbation and a mode of the tensor perturbation produced during inflation.  To enable this goal, we generalize the statistical inference framework used for primordial non-Gaussianity.

The scalar-scalar-tensor correlation is parameterized in terms of the Fourier coefficients of the curvature (scalar) perturbation $\zeta$~\cite{Bardeen:1983qw, Wands:2000dp} and the two helicity modes ${}^{\pm2}h_{\bm{k}}$ that describe the tensor perturbation~\cite{Maldacena:2002vr}:
\begin{align}
\langle \zeta_{\bm{k}_1} \zeta_{\bm{k}_2} \, {}^{\pm2}h_{\bm{k}_3} \rangle  =  (2\pi)^3 \, \delta^{(3)} (\bm{q}) \, {}^{\pm2}F(\bm{k}_1, \bm{k}_2, \bm{k}_3) \, ,
\end{align}
with $\bm{q} = \bm{k}_1 + \bm{k}_2 + \bm{k}_3$. The ${}^{\pm2}F$ functions depend on the inflationary dynamics and can differ between models.  
Both $\zeta_{\bm{k}}$ and ${}^{\pm2}h_{\bm{k}}$ are described in the early radiation-dominated Universe at a time when their comoving wavelength $2\pi / k$ (with $k \equiv |\bm{k}|$) is larger than the Universe's `comoving Hubble radius' $(aH)^{-1}$ in natural units.  $H(t)$ and $a(t)$ are the Hubble parameter and the Robinson-Walker scale factor as function of cosmic time. Both types of perturbations are assumed to be `adiabatic', implying that they do not evolve on these `super-horizon'  scales~\cite{Weinberg:2003sw}.

Evidence for a nonzero $\zeta \zeta h$ $3$-point function would not only point towards a deviation from SFSR inflation~\cite{Maldacena:2002vr}, but would potentially rule out the majority of currently formulated models of inflation~\cite{Bordin:2016ruc}.  The reason for this is a robust consistency relation for the `squeezed limit': $|\bm{k}_3| \ll |\bm{k}_1| \approx |\bm{k}_2|$, of the $\zeta \zeta h$ correlation. In the squeezed limit, $\zeta \zeta h$ is completely determined by $P_{\zeta}(k)$ and $P_{h}(k)$, the power spectra of $\zeta_{\bm{k}}$ and ${}^{\pm2}h_{\bm{k}}$~\cite{Maldacena:2002vr, Baumann:2017jvh}:
\begin{align}\label{eq:tensor_cr}
\begin{split}
\frac{{}^{\pm2}F(\bm{k}_1, \bm{k}_2 , \bm{k}_3 )}{ P_{\zeta}(k_1)\, P_{h}(k_3)} =  \left(\frac{4-n_s}{2}\right) (\hat{k}_1)^a (\hat{k}_2)^b \, e_{ab}^{\pm 2}(\bm{\hat{k}}_3) 
\end{split} \, .
\end{align}
The relation is independent from the dynamics of scalar fields present during inflation and holds as long as modes of the tensor perturbation become adiabatic directly after reaching a super-horizon scale during inflation~\cite{Bordin:2016ruc}.
The `polarization tensors' $e^{\pm 2}_{ab}$ with $a,b \in \{1,2,3 \}$ are two traceless, transverse tensor fields that will be precisely defined later. $n_s - 1$ parameterizes how much $P_{\zeta}(k)$ deviates from the scale-invariant form, see Appendix~\ref{sec:prim_temp_intro}. 

The tensor consistency relation in Eq.~\eqref{eq:tensor_cr} is powerful because its predictions are falsified if a significant $\zeta \zeta h$ correlation is detected in the squeezed limit~\cite{Creminelli:2004yq,Pajer:2013ana}. 
An observed violation of the tensor consistency relation would indicate that inflation is described by a non-standard variant. For example, the relation is violated by inflationary models with light, nonzero spin fields that do not decay quickly after leaving the horizon~\cite{Bordin:2016ruc}. 
As a consequence, falsification of the tensor consistency relation allows ruling out models that approximately respect the de Sitter isometries~\cite{Deser:2001xr}, except for  isometry-respecting models with  so-called partially massless spin fields~\cite{Lee:2016vti, Baumann:2017jvh}. 
Other inflationary models that cannot be ruled out by falsification are those that weakly break some of the de Sitter isometries and couple the extra spin fields to the resulting preferred spatial slicing~\cite{Cheung:2007st, Senatore:2010wk, Bordin:2018pca}. 
Furthermore, in models where a subset of the de Sitter isometries is strongly broken, there is no reason for the consistency relation to hold~\cite{Domenech:2017kno, Endlich:2013jia}. 
This last class includes models in which the tensor perturbations are produced by additional fields~\cite{Senatore:2011sp,  Adshead:2013qp}. These models generally also make predictions for large tensor non-Gaussianity in different forms than just the squeezed $\zeta \zeta h$ type~\cite{Agrawal:2017awz}. 

It should be noted that tests for the consistency relation of the squeezed scalar-scalar-scalar ($\zeta \zeta \zeta$) correlation~\cite{Hinterbichler:2013dpa, Mirbabayi:2014zpa}, that are similar to the tests for the tensor consistency relation, are already underway~\cite{Akrami:2019izv}. The consistency relation for the squeezed  $\zeta \zeta \zeta$ correlation holds for single-field inflation models~\cite{Creminelli:2004yq}.\footnote{The exception are single-field models that relax the standard assumption of a Bunch-Davies vacuum state~\cite{Agarwal:2012mq, Berezhiani:2014kga}. 
Single-field non-attractor models~\cite{Namjoo:2012aa, Martin:2012pe} also do not conform to  the consistency relation, but still do not produce an observable $\zeta \zeta \zeta $ correlation in the squeezed limit~\cite{Bravo:2017gct}.} 
A detection of a significant $\zeta \zeta \zeta$ correlation in the squeezed limit would experimentally rule out the validity of the consistency relation and would provide evidence for the presence of more than one time-evolving scalar field during inflation. 
The tensor consistency relation in Eq.~\eqref{eq:tensor_cr} is arguably more general than the $\zeta \zeta \zeta$ counterpart as it will, in principle, still hold for models with multiple scalar fields~\cite{Bordin:2016ruc, Creminelli:2014wna}.

The CMB contains cosmological information both in its temperature anisotropies ($T$) and linear polarization. The polarization field can be divided in two components: 
the parity-even $E$-mode and parity-odd $B$-mode field~\cite{kamionkowski_1997, zaldarriaga_1997}.
Primordial scalar perturbations source $T$ and $E$-mode polarization, while primordial tensor perturbations source the $T$, $E$-, and $B$-mode fields. 
Observational searches using $T$ and $E$ constrain both scalar and tensor perturbations. However, the contributions to $T$ and $E$ from scalars are much larger than those of tensors, and so cosmic variance, due to the limited number of measurable modes, prohibits strong constraints on tensor perturbations with $T$ and $E$ data. The inclusion of $B$-mode data allows for much tighter constraints on tensor perturbations~\cite{planck_2018_cosmo}.
Furthermore, unlike $T$, current $B$-mode observations are not cosmic-variance limited; hence sensitivity to primordial tensor perturbations can significantly increase with $B$-mode polarization data~\cite{Abazajian:2016yjj}.  

For these reasons, this paper focuses on bispectra, the harmonic equivalent of $3$-point correlation functions, that describe how a single $B$-mode perturbation is correlated to perturbations in the $E$-mode field or the CMB temperature, i.e\ the $\langle TTB \rangle$, $\langle EEB \rangle$ and $\langle TEB \rangle$ bispectra. 
These correlation are currently  unconstrained, but will be within reach of observations by currently operating~\cite{fraisse_2013, bicep_2018, 2010SPIE.7741E..1SN, 2010SPIE.7741E..1EA}, 
 upcoming~\cite{Ade:2018sbj, Henderson:2015nzj, Benson:2014qhw, 2014SPIE.9153E..1IE, 2018SPIE10708E..07H, 2014SPIE.9153E..1NA} and proposed~\cite{Abazajian:2016yjj, 2018SPIE10698E..1YS} experiments.  
Since $B$-modes are sourced by primordial tensor modes,
these bispectra directly probe the $\zeta \zeta h$ correlation. 
The use of the $\langle TTB \rangle$, $\langle EEB \rangle$, and $\langle TEB \rangle$ bispectra avoids much of the scalar-induced cosmic variance that plagues current constraints on the $\zeta \zeta h$ correlation.\footnote{The only relevant dedicated searches have been for a parity-violating 3-point tensor-tensor-tensor  correlation using the \emph{Planck} data in~\cite{Akrami:2019izv} and a search in the \emph{WMAP} data in~\cite{Shiraishi:2017yrq} for a $\zeta \zeta h$ correlation  that violates the tensor consistency relation.} 
These constraints are expected to improve by an order of magnitude with the inclusion of current $B$-mode data~\cite{Meerburg:2016ecv, Domenech:2017kno, Abazajian:2016yjj}. CMB constraints on $\zeta \zeta \zeta$, already close to the cosmic-variance limit, will not see such improvements.\footnote{
While inference on certain standardized types of $\zeta \zeta \zeta$ non-Gaussianity will only improve by a factor of approximately two with upcoming CMB data~\cite{Abazajian:2016yjj}, it is possible that more complicated non-Gaussian features would  still be hidden in the data. 
This is especially true for models with oscillating or non-smooth inflationary potentials (see e.g. \cite{Chen:2006xjb,Holman:2007na,Meerburg:2009ys,Flauger:2010ja,Achucarro:2012sm}) or models that predict non-Gaussian $n$-point correlation functions with $n>3$~\cite{Shiraishi:2013vja, Flauger:2016idt, Baumann:2017jvh}. In terms of improving constraints on (especially squeezed) $\zeta \zeta \zeta$ correlation functions, 
 observables such as galaxy clustering~\cite{Dalal:2007cu, Baldauf:2010vn}, $21\ \mathrm{cm}$ tomography~\cite{Cooray:2006km}, the cross-correlation between CMB lensing and galaxy clustering~\cite{Schmittfull:2017ffw} or the cross-correlation between the primary CMB anisotropies and small-scale spectral distortions of the CMB~\cite{Pajer:2012vz} have the potential to significantly improve constraints in the (far) future.
} 
Future constraints on $\zeta \zeta h$ will benefit from the ongoing, unified experimental effort to collect $B$-mode data in order to constrain the ratio of the primordial tensor-to-scalar ratio $r$: 
\begin{align}
r_{k_0} \equiv \frac{P_{h}(k_0)}{P_{\zeta}(k_0)} \, . 
\end{align}
Besides the fact that a detection of a roughly scale-invariant tensor power spectrum $P_{h}(k)$ would provide a strong argument against a range of alternatives to  inflation~\cite{Khoury:2001wf, Gasperini:2002bn, Brandenberger:2006xi, Ijjas:2018qbo}, constraints on $r$ are used to differentiate between models of inflation~\cite{planck_2018_x}. For slow-roll models, $r$  also provides the energy scale of inflation $V^{1/4}$ : $V^{1/4} \sim r^{1/4} \times 10^{16}\ \mathrm{GeV}$~\cite{Kamionkowski_2016}. 
The upper limit on $r$ is determined by the \textsc{Bicep}2/\emph{Keck Array} and \emph{Planck} CMB data to be $r_{0.002} < 0.064$ (at $95\%$ confidence level)~\cite{planck_2018_x}. 
In the case of a non-detection, upcoming $B$-mode observations have the potential to improve over the current $95\%$ CL upper limit by factors of approximately $ 10$~\cite{Ade:2018sbj, 2018SPIE10708E..07H} and $ 30$~\cite{Abazajian:2016yjj, 2018SPIE10698E..1YS}.

Statistical inference on primordial non-Gaussianity is generally done using statistical `estimators'. Loosely speaking, an estimator is a rule to  transform observed data into a statistical estimate of a parameter of interest. Here we concentrate on a CMB bispectrum estimator that transforms CMB data into an estimate of the amplitude of a given bispectrum and, simultaneously, the amplitude of the primordial $3$-point function responsible for this bispectrum. 
There is a complication associated with the $\zeta \zeta h$ $3$-point function that prohibits a straightforward implementation of the standard bispectrum estimator, see Eq.~\eqref{eq:ksw_scalar}~\cite{KSW,Creminelli:2005hu,Yadav:2005tf,Yadav:2007rk,Yadav:2007ny}. 
Existing bispectrum estimators rely on a summary statistic of the CMB bispectrum: the so-called reduced bispectrum $b_{\ell_1 \ell_2 \ell_3}$, defined in Sec.~\ref{sec:estimator.estimator} of this paper~\cite{komatsu2001acoustic}. 
Data are usually compared to a version of the reduced bispectrum that is separable (factorizable) in $\ell_1$, $\ell_2$, and $\ell_3$. For data with a large harmonic band-limit $\ell_{\mathrm{max}}$ this separable form reduces the computational scaling of the estimator from $\mathcal{O}(\ell_{\rm max}^5)$ to $\mathcal{O}(\ell_{\rm max}^3)$~\cite{KSW}.
The problem is that the $(\hat{k}_1)^a (\hat{k}_2)^b \, e_{ab}^{\pm 2}(\bm{\hat{k}}_3)$ term that is present in the $\zeta \zeta h $ $3$-point correlation function  results in reduced bispectra that are not separable into $\ell_1$, $\ell_2$, and $\ell_3$~\cite{shiraishi2011cmb}.   
Without a separable form of the reduced bispectrum, inference on $\zeta \zeta h$ likely becomes an enormous computational challenge.\footnote{Approximate methods that retain some computational efficiency without relying on a separable form do exist (the binned bispectrum estimator~\cite{bucher_2016} and the modal estimator~\cite{FergussonShellard2009,ShellardBispectrum2006,ShellardModeExpansion2009}) 
and have been successfully applied in the \emph{Planck} analysis \cite{Ade:2013ydc,ade2016planck, Akrami:2019izv}. The first constraint on the amplitude of the $\zeta \zeta h$ $3$-point function in~\cite{Shiraishi:2017yrq} was made with a modified~\cite{Shiraishi:2014roa} version of the modal estimator. 
Despite the fact that the binned and modal estimators are broadly applicable, 
they are relatively involved, not strictly statistically optimal and have an unnecessary computational overhead in the case of reduced bispectra that are already in separable form. For inference on such bispectra the dedicated estimator developed in Ref.~\cite{KSW,Creminelli:2005hu,Yadav:2005tf,Yadav:2007rk,Yadav:2007ny} provides a simpler and more efficient solution.}

We demonstrate that a numerically efficient estimation of the amplitude of the  $\zeta \zeta h$ $3$-point correlation is still possible by making use of the \emph{full} bispectrum instead of the reduced bispectrum, and propose a generalization of the standard bispectrum estimator (Eq.~\eqref{estimator_cubic_2}). This generalization, which can be seen as the main result of this paper,  allows for computationally efficient (and statistically optimal) estimation for all $\zeta \zeta h$ $3$-point functions that include the $(\hat{k}_1)^a (\hat{k}_2)^b \, e_{ab}^{\pm 2}(\bm{\hat{k}}_3)$ term in the following way:
\begin{align}
{}^{\pm2}F(\bm{k}_1, \bm{k}_2, \bm{k}_3) =  f(k_1, k_2, k_3)\, (\hat{k}_1)^a (\hat{k}_2)^b \, e_{ab}^{\pm 2}(\bm{\hat{k}}_3) \, .
\end{align}
Here it is assumed that $f$ can be expressed as (a sum of terms) separable in the three wave numbers $k_1$, $k_2$, and $k_3$. It is argued how numerical evaluation still scales as  $\mathcal{O}(\ell_{\mathrm{max}}^3)$ and how  
the proposed estimator is exact: it does not rely on lossy data compression or on the flat-sky approximation~\cite{Coulton:2017crj, Meerburg:2016ecv}. 
 
In Appendix~\ref{appendix:angular_templates} it is shown how the estimator can be adapted to other non-standard $3$-point correlation functions. We derive an estimator for scalar $3$-point functions that are sensitive to the presence of higher-spin fields during inflation~\cite{Shiraishi:2013vja, Franciolini:2018eno, Shiraishi:2013oqa} and provide estimators for 3-point functions that involve two or three tensor components. 3-point functions with multiple tensor components are relevant for inflation models with peusoscalar-gauge
field interactions~\cite{Agrawal:2017awz, Shiraishi:2013kxa, Cook:2013xea, Dimastrogiovanni:2018xnn}, models with higher-derivative terms in the inflationary gravitational sector~\cite{Bartolo:2018elp} and bimetric gravity models~\cite{Dimastrogiovanni:2018uqy}. 
 
To illustrate the potential of the generalized estimator for testing the tensor consistency relation we provide a number of Fisher forecasts that represent idealized experimental outcomes. These forecasts demonstrate the $\ell_{\rm min}$ and $\ell_{\rm max}$ dependence of constraints on the amplitude of the squeezed $\zeta \zeta h$ correlation. The forecasts also show the influence of the lensing $B$-mode power spectrum, the effects of reionization and the advantage of using both temperature and $E$-mode data in addition to the $B$-mode data. 
We comment on the expected contamination that is associated with $B$-mode data and the high-resolution data needed for squeezed $3$-point functions. In future work the generalized estimator  will be applied to simulated microwave sky data to evaluate the Fisher forecasts. 

The current paper is organized as follows. We first review the CMB anisotropies, bispectrum and the primordial $3$-point correlation function in Sec.~\ref{sec:preliminaries}. We then introduce the generalized bispectrum estimator in Sec.~\ref{sec:estimator.estimator} and present Fisher forecasts for the tensor-scalar-scalar bispectrum in Sec.~\ref{sec:Fisher}. We discuss  future work in Sec.~\ref{sec:discussion} and conclude in Sec.~\ref{sec:conclusion}.

\section{\label{sec:preliminaries} Preliminaries}

\subsection{CMB anisotropies}\label{sec:cmb_ani}

The data we consider are spherical harmonic modes of the CMB temperature and linear polarization anisotropies on the celestial sphere. After a brief review of the general properties of the harmonic modes, we will demonstrate the linear relation between the CMB anisotropies and the primordial scalar and tensor perturbations.

The temperature harmonic modes are related to the CMB temperature $T$ measured at position  $\bm{\hat{n}} \in S^2$ on the celestial sphere by:
\begin{align}\label{eq:alm_temp}
a_{T,\ell m} = \int_{S^2}  \mathrm{d}\Omega(\bm{\hat{n}}) \, T(\bm{\hat{n}}) \, Y^*_{\ell m}(\bm{\hat{n}}) \, ,
\end{align}
where $\mathrm{d}\Omega(\bm{\hat{n}})$ and $Y_{\ell m}^*$ are the differential solid angle and a complex-conjugated spherical harmonic function respectively. See appendix~\ref{sec:appendix_swsh} for a summary of our notation. 

The symmetric, traceless tensor field that describes the linearly polarized component of the microwave sky can be decomposed into two (real) fields: $Q(\bm{\hat{n}})$ and $U(\bm{\hat{n}})$. These fields are coordinate-dependent quantities that  transform among themselves when the local coordinate basis (the tangent space) on the sphere at $\bm{\hat{n}}$ is rotated. For that reason, it is convenient to combine these fields into a complex `spin-$2$' field on the sphere: ${}^{(\pm 2)}P$, which is  defined as follows: 
 \begin{align}
{}^{(\pm 2)}P(\bm{\hat{n}}) \equiv (Q\pm iU)(\bm{\hat{n}}) \, .
\end{align}
Under a right-handed rotation of the local coordinate system around the point $\bm{\hat{n}}$ we then have: 
\begin{align}\label{eq:p_spin_transform}
{}^{(\pm 2)}P(\bm{\hat{n}}) \mapsto {}^{(\pm 2)}P(\bm{\hat{n}}) e^{\mp 2 i \psi} \, ,
\end{align} 
where $\psi$ is the angle of rotation. The sign of the exponent is a convention.

Instead of directly using ${}^{(\pm 2)}P$, we will describe  polarization in terms of the harmonic modes of two fields that are scalars under coordinate rotations around $\bm{\hat{n}}$: the parity-even $E$ field and the parity-odd $B$ field. The harmonic modes of these two fields: the $E$- and $B$-modes, are related to the locally observable field as follows:
\begin{align} \label{eq_alm_EB}
\begin{split}
a_{E,\ell m} &= - \frac{1}{2}\sum_{s \in \pm2}  \int_{S^2} \! \! \mathrm{d}\Omega(\bm{\hat{n}}) \, {}^{(s)}P(\bm{\hat{n}}) \, {}_{s}Y^*_{\ell m}(\bm{\hat{n}}) \, , \\
a_{B,\ell m} &= - \frac{1}{2i} \sum_{s\in \pm2} \! \! \mathrm{sgn}(s) \! \!  \int_{S^2} \! \! \mathrm{d}\Omega(\bm{\hat{n}})\,  {}^{(s)}P(\bm{\hat{n}}) \, {}_{s}Y^*_{\ell m}(\bm{\hat{n}}) \, .
\end{split}
\end{align}
The spin-weighted spherical harmonics ${}_{s}Y_{\ell m}$ form a complete and orthonormal basis for spin-$s$ functions on the sphere, analogous to the regular spherical harmonics. See Appendix~\ref{sec:appendix_swsh} for a brief overview.

The parity-even $E$ and parity-odd $B$ harmonic modes transform differently under the parity transformation of the underlying spherical coordinates. Under parity, the odd moments of the temperature anisotropies and the $E$-mode field gain  a minus sign. The opposite behavior holds for the $B$-mode field:
\begin{align}\label{eq:parity}
\begin{split} 
a_{T, \ell m}  &\mapsto (-1)^{\ell} a_{T, \ell m} \,, \\
a_{E, \ell m}  &\mapsto (-1)^{\ell} a_{E, \ell m} \,, \\
a_{B, \ell m}  &\mapsto (-1)^{\ell+1} a_{B, \ell m} \, .
\end{split}
\end{align}

To describe the primordial adiabatic scalar perturbations that source the CMB anisotropies, we use the gauge invariant curvature  perturbation $\zeta$~\cite{Bardeen:1983qw, Wands:2000dp}.\footnote{The invariance under choice of gauge (the choice of constant-time spacelike hypersurfaces and constant-position timelike worldlines) of $\zeta$ explains why it can simultaneously be interpreted as e.g.\ the spatial curvature on hypersurfaces with constant energy density or as the energy density perturbation on spatially flat hypersurfaces~\cite{Lyth:2009zz}.} As the initial adiabatic state is constant on super-horizon scales, we only need to consider the amplitude of $\zeta$ on some spacelike hypersurface in the early radiation-dominated era when all Fourier modes of interest were super horizon. 
The Fourier coefficients of this amplitude at early time $t_i$ are given by:
\begin{align}\label{eq:zeta_fourier}
\zeta_{\bm{k}} \equiv 
\int \! \mathrm{d}^3\bm{x} \, \zeta(\bm{x}, t)\big |_{t = \tilde{t}(t_i, \bm{x})} e^{-i \bm{k} \cdot \bm{x}}  \, ,
\end{align}
where $\tilde{t} = t + \delta t(\bm{x}, t) $  parameterizes weakly perturbed spacelike hypersurfaces relative to comoving coordinates $\{ \bm{x}, t\}$ of the flat FLRW background. 
Throughout this work $\bm{k}$ denotes a 3D comoving wave vector.

The primordial tensor perturbation $h$ is the traceless and divergenceless linear perturbation to the flat FLRW metric:
\begin{align}
\mathrm{d}s^2 = - \mathrm{d}t^2  + a^2(t) [\delta_{ab} + h_{ab}(\bm{x},t)]  \mathrm{d}x^a \mathrm{d}x^b  \, ,
\end{align}
with $h^a_{\phantom{a}a} = \partial_a h^{ab} = 0$. 
Instead of using the coordinate basis to describe the tensor perturbation, we use a basis that sits perpendicular to the unit wave vector $\hat{\bm{k}}$, spanned by the $\hat{e}_{(\pm)}$ unit vectors.\footnote{To relate the basis vectors of the comoving coordinates $\hat{e}_{(a)}$ to those of the noncoordiate basis, we introduce a set of `polarization' vectors: $\{e_{+}, e_{-}, e_{0} \}$, such that 
$\hat{e}_{(\lambda)} = e^{\phantom{a}a}_{\lambda} \hat{e}_{(a)}$ with $\lambda \in \{+,-,0 \}$. 
 Geometrically, the $\hat{e}_{(\pm)}$ basis vectors span the plane perpendicular to the wave vector, while $\hat{e}_{(0)}$ points along the wave vector. The three vectors form a complete orthonormal basis. We let $\hat{e}_{(\pm)}$ describe states of circular polarization, i.e.\ the  polarization vectors obey $(e^{\phantom{a}a}_{\pm })^* = e^{\phantom{a}a}_{\mp }$.} 
On this new basis, the tensor perturbation conveniently reduce to two helicity states with Fourier coefficients given by: 
\begin{align} \label{eq:h_ij}
{}_{(\pm 2)}h_{\bm{k}} \equiv \frac{e^{ab}_{\pm  2}(\bm{\hat{k}})}{2} \int \! \mathrm{d}^3\bm{x}  \, h_{ab} ( \bm{x}, t)\big|_{t = \tilde{t}(\bm{x}, t_i)} e^{-i \bm{k} \cdot \bm{x}}  \, .
\end{align}
 The polarization tensors $e_{\pm2}$ are two symmetric, traceless and transverse tensor fields that transform $h$ from the comoving coordinate basis to the $\hat{e}_{(\pm)}$ basis. The polarization tensors have the following properties:
 \begin{align}
 (e^{ab}_{\pm 2})^*(\bm{\hat{k}}) &= e^{ab}_{\mp 2}(\bm{\hat{k}}) \label{eq:pol_tens_cc} \, , \\
 e_{ab}^{\lambda}(\bm{\hat{k}}) e_{\lambda'}^{ab}(\bm{\hat{k}}) &= 2 \delta_{-\lambda'}^{\lambda} \quad (\lambda \in \pm 2) \label{eq:pol_tens_ortho} \, .
 \end{align}
 
The tensor perturbation $h$ is gauge invariant (in the same sense as $\zeta$ is)~\cite{Bardeen:1980kt}. The helicity components ${}_{(\pm2)}h$ are scalars under coordinate transformations up to a phase factor depending on the orientation of the basis spanned by $\hat{e}_{(\pm)}$.\footnote{
The polarization tensors are defined in terms of the $\pm$ polarization vectors as $e_{\pm 2}^{ab} \equiv \sqrt{2} e^{\phantom{a}a}_{\pm} e^{\phantom{b}b}_{\pm}$. In the Cartesian basis, we may define the polarization vector as $e_{\pm} = \{1, \pm i, 0 \}/\sqrt{2} $ for a wave vector aligned with the $\bm{\hat{z}}$ direction. 
The addition of a complex phase $\exp(-i \psi)$ to this definition amounts to an equally suitable basis that is simply rotated around the wave vector. The polarization tensors and helicity components are thus defined up to $\exp(-2 i \psi)$.}

Let us categorize the stochastic primordial (super-horizon) amplitudes in terms of their helicity $\lambda$: 
\begin{align} \label{eq:init_amps}
{}^{(\lambda)}\xi_{\bm{k}} = 
\begin{cases}
\zeta_{\bm{k}} \quad &\text{for } \lambda=0\\ 
{}^{(\lambda)}h_{\bm{k}} &\text{for } \lambda=\pm2 
\end{cases}\,.
\end{align}
Following the notation set by \cite{shiraishi2011cmb}, we then write down a compact expression for the observed CMB modes in terms of  these helicity-dependent super-horizon amplitudes and a set of rotationally invariant transfer functions~$\mathcal{T}_{\ell}(k)$: 
\begin{align}
\begin{split}
&a_{X,\ell m}^{(Z)} = \ 4 \pi (-i)^{\ell} \sum_{\lambda} \mathrm{sgn}(\lambda)^{\lambda + x}  \\
&\quad \quad \times \int \frac{\mathrm{d}^3 \bm{k}}{(2 \pi)^3} \, {}^{(-\lambda)}\xi_{\bm{k}} \mathcal{T}^{(Z)}_{X,\ell}(k) \,  {}_{-\lambda}Y^*_{\ell m } ({\bm{\hat{k}}}) \, ,
\label{eq:gen_alm}
\end{split}
\end{align}
with $Z \in \{\zeta\, (\mathrm{scalar}), h \, (\mathrm{tensor}) \}$, $\mathrm{sgn}(0) \equiv 0$, $0^0\equiv 1$, $X \in \{T, E, B\}$, and helicity and parity determined by:
\begin{align*}\lambda =
\begin{cases}
0 \quad &\text{for } Z=\zeta\\ 
\pm2 \quad &\text{for } Z=h
\end{cases} \,, \; \;
x =\begin{cases}
0 \quad &\text{for } X=T, E\\ 
1 \quad &\text{for } X=B
\end{cases} \,.
\end{align*}
Note that by defining ${}_{\mp2}Y^*_{\ell m}$ in Eq.~\eqref{eq:gen_alm} on the transverse basis spanned by $\hat{e}_{(\pm)}$, we ensure that the $a_{X, \ell m}$ for $Z=h$ are independent of the orientation of this basis. This approach is fully analogous to the decomposition of the spin-$2$ polarization field in~Eq.~\eqref{eq_alm_EB}.

The transfer functions $\mathcal{T}_{\ell}(k)$  transform the super-horizon amplitudes $\zeta_{\bm{k}}$ and ${}^{(\lambda)}h_{\bm{k}}$ to the CMB radiation and its polarization seen today~\cite{Seljak:1996is, zaldarriaga_1997}. In short, once the comoving Hubble radius (growing after inflation has ended) becomes larger than the comoving wavelengths of $\zeta_{\bm{k}}$ and ${}^{(\lambda)}h_{\bm{k}}$, they `enter the horizon' and start to evolve with time. The scalar perturbations sourced by $\zeta$ begin to oscillate under the effects of gravity and photon pressure, resulting in the acoustic oscillations seen in the CMB angular power spectra. The helicity components ${}^{\pm 2}h$ start to propagate through space as the two polarization states of a gravitational wave, virtually decoupled from the other components of the Universe, and decay away with the expansion of space~\cite{polnarev_1985, Weinberg:2003ur, Kamionkowski_2016}. As a result, the most prominent difference between the scalar and tensor transfer functions is that the latter result in small values for CMB fluctuations on small ($\ell > 100$) angular scales. Small-scale tensor perturbations that entered the horizon before recombination decay significantly before leaving their imprint on the CMB. 
The transfer functions only depend on the unperturbed background cosmology and are readily available through numerical Einstein-Boltzmann solvers such as CAMB~\cite{Lewis:1999bs,Howlett:2012mh} or CLASS~\cite{Blas:2011rf}.\footnote{See \url{https://camb.info} and \url{http://class-code.net}.} The projection onto the celestial sphere is also handled by the transfer functions. See Appendix~\ref{sec:prim_temp_intro} for more details on the transfer functions used in this work. 

With Eq.~\eqref{eq:gen_alm} we have quantified the relation between the CMB anisotropies and the primordial scalar and tensor fields. 
The relation reiterates an important point:  
the primordial scalar fluctuations do not source the parity-odd $B$-mode field at linear order~\citep{zaldarriaga_1997}. Higher-order cosmological effects, such as weak lensing by matter along the line of sight~\cite{Zaldarriaga:1998ar} or second order time-evolution of the scalar perturbations~\cite{Beneke_2011, Mollerach:2003nq, Fidler:2014oda}, create a $B$-mode signal even in the absence of a primordial tensor contribution. Such effects are not included in the linear transfer functions so their influence has to be described separately. The same is true for signal from astrophysical foregrounds. We will briefly discuss these contributions in Sec.~\ref{sec:discussion} but will consider them in more detail in a future paper.

\subsection{Bispectrum and the primordial $3$-point function}

In Sec.~\ref{sec:cmb_bisp_intro}, we summarize general properties of the observable of interest: the CMB bispectrum. As we are interested in bispectra that include a $B$-mode component, we explicitly discuss the inclusion of $B$-mode polarization. In Sec.~\ref{sec:cmb_lin_prop_bisp} we then introduce the concept of a linearly propagated, or primary, bispectrum: a primordial $3$-point correlation function that is evolved to the CMB bispectrum  today by the linear transfer functions introduced in Sec.~\ref{sec:cmb_ani}. In addition, we describe the primordial $\zeta \zeta h$ $3$-point correlation function in more detail.

\subsubsection{General properties of the bispectrum}\label{sec:cmb_bisp_intro}
The bispectrum is defined as the isotropic $3$-point correlation function represented in terms of spherical harmonic coefficients. The bispectrum is proportional to the multivariate generalization of the skewness of a probability distribution and thus vanishes for purely Gaussian coefficients. 

We can formulate a bispectrum for every combination of the temperature and polarization components $X_1, X_2, X_3 \in \{T, E, B\}$:
\begin{align}
B_{m_1 m_2 m_3, X_1 X_2 X_3}^{\ell_1 \ell_2 \ell_3} \equiv \big\langle a_{X_1, \ell_1 m_1} a_{X_2, \ell_2 m_2} a_{X_3, \ell_3 m_3}\big\rangle \, .
\end{align}
The $a_{X, \ell, m}$ are defined in Eq.~\eqref{eq:alm_temp} and Eq.~\eqref{eq_alm_EB}. 
Statistical isotropy constrains the azimuthal dependence such that the bispectrum may always be factored into a Wigner 3-$j$ symbol and a factor independent of $m_1$, $m_2$, and $m_3$ \cite{Spergel:1999xn, Hu_2001}: 
\begin{align}\label{eq:isotropy_b}
B_{m_1 m_2 m_3, X_1 X_2 X_3}^{\ell_1 \ell_2 \ell_3} = 
\begin{pmatrix}\ell_1 & \ell_2 & \ell_3 \\ m_1 & m_2 & m_3 \end{pmatrix} B_{\ell_1 \ell_2 \ell_3}^{X_1 X_2 X_3} \, .
\end{align}
We will refer to the l.h.s.\ as the bispectrum, while $B$ on the r.h.s.\ is the  angle-averaged bispectrum. See Appendix~\ref{sec:appendix_fourier} for an overview of the Wigner $3$-$j$ symbols.

It is possible to construct a parity-invariant bispectrum from three fields regardless of the parity behavior of the individual fields. This means that we can form a parity-invariant bispectrum for all combinations of $T$, $E$, and $B$. This is not the case for the angular power spectrum.\footnote{Given the parity transformation rules in Eq.~\eqref{eq:parity}, we see that the $2$-point cross correlation function between a $B$-mode coefficient and a $T$ coefficient transforms under parity as: \begin{align*}\langle a_{B, \ell_1 m_1} a_{T, \ell_2 m_2}^* \rangle \mapsto \langle a_{B, \ell_1 m_1} a_{T, \ell_2 m_2}^* \rangle (-1)^{\ell_1 + \ell_2 + 1} \, .\end{align*} Taken together with isotropy, which demands that the cross-correlation is proportional to $\delta_{\ell_1 \ell_2} \delta_{m_1 m_2}$, we see that there is no parity-invariant configuration. The $BE$ power spectrum vanishes by extension. }  
From  Eq.~\eqref{eq:parity}, we see that invariance under parity  alone imposes that $\ell_1 + \ell_2 + \ell_3 = \mathrm{even}$ for bispectra with an even number of $B$-mode contributions and $\ell_1 + \ell_2 + \ell_3 = \mathrm{odd}$ otherwise; see Table~\ref{tab:parity}~ \cite{kamionkowski_2011, Meerburg:2016ecv}. 

\begingroup
\renewcommand*{\arraystretch}{1.5}
\begin{table}[h]
\caption{\label{tab:parity}Factor gained after a parity transformation $\mathsf{P}: \bm{\hat{n}} \mapsto - \bm{\hat{n}}$, for bispectra $B_{m_1 m_2 m_3, X_1 X_2 X_3}^{\ell_1 \ell_2 \ell_3}$ grouped by $X_1, X_2, X_3$ polarization indices.
}
\begin{tabularx}{\columnwidth}{@{\hspace{0em}} >{\raggedright}p{17em} l @{\hspace{0em}}}
\hline
\hline
& $\mathsf{P}: \bm{\hat{n}} \mapsto - \bm{\hat{n}}$ \\
\hline
$TTT$, $TT E$, $T E E$, $TBB$, $ EEE $, $EBB$   & $(-1)^{\ell_1 + \ell_2 + \ell_3}$ \\
$TT B$, $T EB$, $ EEB $, $BBB$   & $(-1)^{\ell_1 + \ell_2 + \ell_3 + 1}$  \\
\hline
\hline
\end{tabularx}
\end{table}
\endgroup

Isotropy forces the $\ell_1 + \ell_2 + \ell_3 = \mathrm{even}$ components of bispectra to be real while the $\ell_1 + \ell_2 + \ell_3 = \mathrm{odd}$ parts are purely imaginary. This constraint can be deduced from the condition for isotropy in Eq.~\eqref{eq:isotropy_b} and the reality condition of the harmonic coefficients: 
\begin{align}
a^*_{X,\ell m} = a_{X,\ell -m} (-1)^m \, ,
\end{align}
which holds because the underlying $X = \{T,E, B \}$ fields are real-valued. The combination of these two conditions together with the reality of the $3$-$j$ symbols then implies: 
\begingroup
\begin{align*}
(B^*)_{m_1 m_2 m_3}^{\ell_1 \ell_2 \ell_3}\!  = \begin{pmatrix}\ell_1 & \ell_2 & \ell_3 \\ -m_1 & -m_2 & -m_3 \end{pmatrix}  B_{\ell_1 \ell_2 \ell_3}  (-1)^{\sum_{n=1}^3 m_n} \, ,
\end{align*}
which, through the property of the $3$-$j$ symbol  in Eq.~\eqref{wigner3j:parity}, means that complex-conjugating the bispectrum results in the following behavior:
\begin{align}
(B^*)_{m_1 m_2 m_3}^{\ell_1 \ell_2 \ell_3} = B_{m_1 m_2 m_3}^{\ell_1 \ell_2 \ell_3}  (-1)^{\sum_{n=1}^3 (m_n + \ell_n)} \nonumber \, .
\end{align}
\endgroup
Note that the Wigner $3$-$j$ symbol vanishes for $m_1 + m_2 + m_3 \neq 0$.  
Clearly, the above relation also holds for the angle-averaged bispectrum $B_{\ell_1 \ell_2 \ell_3}$. We have suppressed the $X$ indices as the above  holds for all combinations of  polarization indices. See Table~\ref{tab:parity_2} for an overview of the geometric constraints on parity-invariant, isotropic bispectra. The fact that the bispectra of interest here: $\langle TTB \rangle$, $\langle TEB \rangle$, and $\langle EEB \rangle$, are purely imaginary is a consequence of the complex representation of the spherical harmonics that we use. Expressed in terms of the Stokes parameters $Q$ and $U$, the corresponding $3$-point correlations would be real-valued and thus observable. 

\begingroup
\renewcommand*{\arraystretch}{1.5}
\begin{table}[h]
\caption{\label{tab:parity_2} Parity conservation forces  the bispectrum to be purely real, purely imaginary or to vanish,  depending on its $\ell_1$, $\ell_2$, and $\ell_3$ multipole indices and its $X_1$, $X_2$, and $X_3$ polarization indices. }
\begin{tabularx}{\columnwidth}{@{\hspace{0em}} >{\raggedright}p{11em} c c @{\hspace{0em}}}
\hline
\hline
& $\sum_{n=1}^3 \! \ell_n = \mathrm{odd}$ & $\sum_{n=1}^3 \! \ell_n = \mathrm{even}$\\
\hline
$TTT$, $TT E$, $T E E$, $TBB$, $ EEE $, $EBB$   & $\mathrm{Vanish}$ & $\mathrm{Real}$ \\
$TT B$, $T EB$, $ EEB $, $BBB$   & $\mathrm{Imag}$ & $\mathrm{Vanish}$ \\
\hline
\hline
\end{tabularx}
\end{table}
\endgroup

\subsubsection{Linearly propagated bispectrum and primordial $3$-point correlation function}\label{sec:cmb_lin_prop_bisp}

We start by defining the linearly propagated, or primary, bispectrum in its most general form. As mentioned before, the linearly propagated bispectrum is formed by time-evolving a primordial $3$-point correlation function to the CMB bispectrum today using the linear transfer functions introduced in Sec.~\ref{sec:cmb_ani}. We then introduce the standard scalar-only ($\zeta \zeta \zeta$) primordial $3$-point correlation function as well as our main focus: the $\zeta \zeta h$ $3$-point correlation function. 

Let us parameterize the super-horizon $3$-point correlation function, the object we are ultimately interested in, as a helicity-dependent quantity using the amplitudes introduced in Eq.~\eqref{eq:init_amps}:
\begin{align} 
\begin{split}
{}^{(\lambda_1 \lambda_2  \lambda_3)}\! B(\bm{k}_1, \bm{k}_2, \bm{k}_3)  \equiv \big\langle {}^{(\lambda_1)}\xi_{\bm{k}_1} {}^{(\lambda_2)}\xi_{\bm{k}_2} {}^{(\lambda_3)}\xi_{\bm{k}_3} \big\rangle \, ,
\end{split}
\end{align}
where the helicity $\lambda$ is $0$ for scalar perturbations and $\pm2$ for tensor perturbations. 
We can then, using Eq.~\eqref{eq:gen_alm}, form the linearly propagated bispectrum~\cite{WangKamionkowski}:
\begin{align}
\begin{split}
&B_{m_1 m_2 m_3, X_1 X_2 X_3}^{ \ell_1 \ell_2 \ell_3 (Z_1 Z_2 Z_3)}\! \!  = \! \! \Bigg[ \! \prod_{n=1}^3 4 \pi (-i)^{\ell_n} \! \! \sum_{\lambda_n} \mathrm{sgn}(\lambda_n)^{\lambda_n + x_n} \\ &\quad \quad \times \int \frac{\mathrm{d}^3 \bm{k}_n}{(2 \pi)^3}  \, {}_{-\lambda_n}Y^*_{\ell_n m_n } ({\bm{\hat{k}}_n})  \mathcal{T}^{(Z_n)}_{X_n,\ell_n}(k_n) \Bigg] \\ 
&\quad \quad \times {}^{(-\lambda_1 -\lambda_2  -\lambda_3)} \!B(\bm{k}_1, \bm{k}_2, \bm{k}_3) \, .
\label{eq:theoretical_bispectrum}
\end{split}
\end{align}
 Note that the three $Z$ indices of the bispectrum in Eq.~\eqref{eq:theoretical_bispectrum} may each be either $\zeta$ or $h$. 

We now consider the symmetries of the primordial $3$-point function. 
The assumed translational invariance of the process generating the primordial fluctuations implies momentum conservation in Fourier space:
\begin{align}\label{eq:prim_3point_hom}
\begin{split}
{}^{(\lambda_1 \lambda_2 \lambda_3)} B(\bm{k}_1, &\bm{k}_2, \bm{k}_3) =  (2\pi)^3 \, \delta^{(3)}(\bm{k}_1 + \bm{k}_2 + \bm{k}_3)  \\ &\times {}^{(\lambda_1, \lambda_2 , \lambda_3)}F(\bm{k}_1, \bm{k}_2, \bm{k}_3)\, .
\end{split}
\end{align}
What remains now is to consider certain expressions for the helicity-dependent ${}^{(\lambda_1, \lambda_2 , \lambda_3)}F(\bm{k}_1, \bm{k}_2, \bm{k}_3)$ functions. In a regular analysis, these functions would be given by the model under consideration. Here we are more interested in classes of models, so we use general parameterizations.

For the scalar-only ($\zeta \zeta \zeta$) $3$-point function, isotropy demands that $F$ only depends on scalar products of the three wave vectors: the individual amplitudes and $\bm{k}_1 \cdot \bm{k}_2$, $\bm{k}_1 \cdot \bm{k}_3$, and $\bm{k}_2 \cdot \bm{k}_3$.  $F$ cannot depend on a pseudoscalar like $\bm{k}_1 \cdot (\bm{k}_2 \times \bm{k}_3 )$ in case of a parity-invariant $3$-point correlation function. For simplicity, we use the following template:
\begin{align}
{}^{(000)}F(\bm{k}_1, \bm{k}_2, \bm{k}_3) =   f^{(\zeta\zeta \zeta)}(k_1, k_2, k_3)\, ,
\label{eq:sss_general}
\end{align}
where $f$ is generally referred to as the shape of the bispectrum. We will make use of this standard $\zeta \zeta \zeta$ template to introduce the reader to existing estimation techniques later in this paper.

For the $\zeta \zeta h$ case, we use the following parameterization:
\begin{align}
\begin{split}\label{eq:b_template_sst}
{}^{(00\pm 2)}F(\bm{k}_1, &\bm{k}_2, \bm{k}_3) =    f^{(\zeta\zeta h)}(k_1, k_2, k_3) \\ &\times (\hat{k}_1)^a (\hat{k}_2)^b \, e_{ab}^{\pm 2}(\bm{\hat{k}}_3) \, .
\end{split}
\end{align} 
Recall that roman indices denote three-dimensional spatial comoving coordinates; they are summed over when repeated.   Note that ${}^{(00+2)}F$ and ${}^{(00-2)}F$ correspond to two independent $3$-point functions; by denoting the shape function $f^{(\zeta\zeta h)}$ independent of helicity, we however implicitly assume parity invariance.

The class of $\zeta \zeta h$ $3$-point functions  described by Eq.~\eqref{eq:b_template_sst} include those predicted by SFSR inflation~\cite{Maldacena:2002vr}. The amplitude of the $\zeta \zeta h$ $3$-point function will be too small to be observable with CMB data in the SFSR case. More importantly, the template in Eq.~\eqref{eq:b_template_sst} also applies to the majority of mentioned models that violate the tensor consistency relation in Eq.~\eqref{eq:tensor_cr} and thus potentially produce an observable signal~\cite{Akhshik:2014bla, Dimastrogiovanni:2014ina, Bordin:2016ruc, Domenech:2017kno, Endlich:2013jia}. We may therefore use Eq.~\eqref{eq:b_template_sst} as the basis for inference on such models.

To gain intuition for the characteristics of the $\zeta \zeta h$ template, it is useful to realize that the delta function in Eq.~\eqref{eq:prim_3point_hom} imposes that $\bm{k_1} + \bm{k_2} + \bm{k_3} = \bm{0}$, i.e.\ the $3$-point function is defined on triangular configurations of the three wave vectors.  The $f^{(\zeta \zeta h)}(k_1, k_2, k_3)$ part of the $\zeta \zeta h$ template thus assigns a weight to each triangle based on the lengths of the three sides. While these weights completely determine the $3$-point function in the $\zeta \zeta \zeta$ case, the $\zeta \zeta h$ case requires that two more aspects are taken into account. First, the $\zeta \zeta h$ $3$-point function is always suppressed in triangular configurations wherein the wave vector of the $h_{\bm{k}}$ Fourier mode is roughly (anti)parallel to the wave vector(s) of one or both of the scalar modes. This suppression is not due to the  $f^{(\zeta\zeta h)}$ weight function but is a consequence of the nature of the polarization tensors. Their transverse property demands that $(\hat{k})^a e^{\pm2}_{ab}(\bm{\hat{k}'})$ vanishes as $\bm{\hat{k}}$ becomes equal to $\bm{\hat{k}'}$. We thus see a suppression when $\bm{\hat{k}}_3$ aligns with $\bm{\hat{k}}_2$ and/or $\bm{\hat{k}}_1$ in Eq.~\eqref{eq:b_template_sst}.  Secondly, the transverse traceless behavior of the $\zeta \zeta h$ $3$-point function is also reflected in its helicity dependence. 

We have demonstrated how the CMB is affected by a nonzero primordial $3$-point correlation function through the bispectrum. We have also introduced the $\zeta \zeta h$ $3$-point correlation function in  Eq.~\eqref{eq:b_template_sst}.
The bulk of this work will focus on this $3$-point function. 
Note that the estimation technique that will be presented in the following sections is, in principle, also applicable to other types of $3$-point functions. 
For conciseness, the discussion of some other templates (including the SFSR scalar-tensor-tensor and tensor-tensor-tensor $3$-point functions~\cite{Maldacena:2002vr}) is placed in Appendix~\ref{appendix:angular_templates}.

\section{\label{sec:estimator.estimator}  Estimator}
This section is organized as follows. We first introduce the general form of the bispectrum estimator in~\ref{sec:estimator.cmb_general}. In~\ref{sec:estimator.templates}, we then summarize the existing numerically efficient implementation of the estimator and Sec.~\ref{sec:fast_estimator_sst} we present our new work: the generalization of the fast implementation to the $\zeta \zeta h$ case.

\subsection{\label{sec:estimator.cmb_general} General bispectrum estimation}

We summarize the properties of the now standard CMB bispectrum estimation method~\cite{komatsu2001acoustic, Creminelli:2005hu, babich2005optimal}: a parametric search for the amplitudes of theoretically motivated bispectrum templates using an estimator that consists of a cubic and a linear statistic. This method has been the basis for the \emph{Planck} non-Gaussianity analysis~\cite{Akrami:2019izv}. A derivation of the estimator and discussion of its properties can be found in Appendix~\ref{app:estimator_der}.

The estimator yields an estimate of the overall (dimensionless) amplitude $f_{\mathrm{NL}} \in \mathbb{R}$ of a bispectrum. We thus parameterize the bispectrum of interest as:
\begin{align}\label{eq:b_fnl_scal}
B(f_{\mathrm{NL}}) = f_{\mathrm{NL}} B_1 \, ,
\end{align}
where $B_1 \equiv B(f_\mathrm{NL} = 1)$ is a fixed theoretical template with suppressed $\ell$ and $m$ indices. 

In searches for primordial non-Gaussianity, the template $B_1$ is given by a normalized version of the linearly propagated bispectrum in  Eq.~\eqref{eq:theoretical_bispectrum}. The linear nature implies that the  $f_{\mathrm{NL}}$ parameter corresponds to the overall amplitude of the primordial $3$-point correlation function ${}^{(-\lambda_1 -\lambda_2-\lambda_3)}B$ in Eq.~\eqref{eq:theoretical_bispectrum}. In principle, the amplitudes of several templates can be jointly estimated (see Appendix~\ref{app:estimator_der}). Here we only need the single parameter variant.

The estimator for $f_{\mathrm{NL}}$ is given by:
\begin{align}
\begin{split}\label{eq:estimator_single_par}
\hat{f}_{\mathrm{NL}} &= \frac{1}{6 \, \mathcal{I}_0} \sum_{\mathrm{all}\,\ell, m  }   \sum_{\mathrm{all} X} (B_1)_{m_1 m_2 m_3, X_1 X_2 X_3}^{\ell_1 \ell_2 \ell_3} \\ 
&\times \bigg\{ 
\left[(C^{-1}a)^{X_1}_{\ell_1 m_1} (C^{-1}a)^{X_2}_{\ell_2 m_2} (C^{-1}a)^{X_3}_{\ell_3 m_3} \right] \\
&\ \ \ - \left[(C^{-1})^{X_1 X_2}_{\ell_1 m_1 \ell_2 m_2} (C^{-1}a)^{X_3}_{\ell_3 m_3} + \mathrm{cyclic} \right] \bigg\} \, ,
 \end{split}
\end{align}
where $X \in \{T, E, B \}$. The data: $a_{X,\ell m}$, only enter in inverse-covariance-weighted form: 
\begin{align}\label{eq:c_inv_a_body}
(C^{-1}a)^{X}_{\ell m} =  \sum_{X'}\sum_{\ell' , m'}(C^{-1})^{X X'}_{\ell m  \ell'm'}a_{X',\ell' m'} \, .
\end{align}
Here $C^{-1}$ is the inverse of the block matrix:
\begin{align}\label{tot_cov}
C_{\ell m \ell' m'} \equiv \begin{pmatrix} 
	C_{TT} & C_{TE} & C_{TB} \\ 
    C_{ET} & C_{EE} & C_{EB} \\
    C_{BT} & C_{BE} & C_{BB} 
    \end{pmatrix}_{\ell m \ell' m'} \, .
\end{align}
Each element is defined as:
\begin{align}
C_{X X', \ell m \ell' m'} = \langle a_{X,\ell m} a^*_{X',\ell' m'} \rangle \, ,
\end{align}
with $X, X' \in \{T, E, B \}$. This covariance matrix includes both the signal and noise covariance and is therefore generally not diagonal.  
The estimating procedure considers the  covariances as fixed and known a priori. 

Intuitively, the first and second line, the `cubic term', in Eq.~\eqref{eq:estimator_single_par}  serve as a matched filter that correlates the observed bispectrum with the theoretical template $B_1$. The terms linear in the data (first times third line) are usually jointly referred to as the `linear term' and effectively serve to counter the estimator variance induced by the anisotropic parts of the covariance matrix \cite{Creminelli:2005hu, Yadav:2007ny}. Only the cubic part of the estimator is needed in cases where the covariance matrix in Eq.~\eqref{tot_cov} is rotationally invariant.\footnote{One can check that the rotational invariance of the bispectrum forces the linear term to be proportional to the (unobservable) CMB monopole perturbation when the covariance matrix is rotationally invariant~\cite{babich2005optimal}.}
With weakly anisotropic covariance, the linear term can be neglected for non-squeezed bispectrum templates and/or analyses without large-scale ($\ell \alt 100 $) data~\cite{Coulton:2017crj}.

The normalization of the estimator is given by the following (dimensionless) number:
\begin{align}\label{eq:fisher_full_scal}
\begin{split}
\mathcal{I}_0 &=  \frac{1}{6} \sum_{\mathrm{all}\,\ell, m  }   \sum_{\mathrm{all} X} \left(B_1\right)_{m_1 m_2 m_3, X_1 X_2 X_3}^{ \ell_1 \ell_2 \ell_3} \\
&\times \! \! \left[ (C^{-1})^{X_1 X_4}_{\ell_1 m_1 \ell_4 m_4} (C^{-1})^{X_2 X_5}_{\ell_2 m_2 \ell_5 m_5} (C^{-1})^{X_3 X_6}_{\ell_3 m_3 \ell_6 m_6} \! \right] \\
&\times   (B_1^{*} )_{m_4 m_5 m_6, X_4 X_5 X_6}^{\ell_4 \ell_5 \ell_6} \, .
\end{split}
\end{align}
Note that $\mathcal{I}_0$ is completely independent from the observed data.

The estimator is often referred to as `optimal'. The word `optimal' refers to the fact that, in the appropriate limit, the estimator yields an unbiased point estimate of $f_{\mathrm{NL}}$ with variance given by the inverse of the model's Fisher information on $f_{\mathrm{NL}}$. It should be noted that this behavior is strictly true only in the limit where all non-Gaussian signal vanishes, this includes $f_{\mathrm{NL}} \rightarrow 0$. The expression in Eq.~\eqref{eq:fisher_full_scal} becomes equal to the Fisher information on $f_{\mathrm{NL}}$ in this limit. In Appendix~\ref{app:estimator_der} we specify the likelihood function of the data to make the above statements more precise. 

The estimator in Eq.~\eqref{eq:estimator_single_par} is well-suited to estimate upper limits on $f_{\mathrm{NL}}$. When a weak non-Gaussian signal is present, the estimator is still usable, but one has to be wary of biases and non-optimal variance~\cite{Creminelli:2006gc, Liguori:2007sj}. This is especially relevant for $B$-mode data contaminated by Galactic signal or  high-resolution data with relatively strong non-Gaussian contributions from e.g.\ weak lensing. See the discussion in Sec.~\ref{sec:discussion} for more details.

We end this summary with a practical note on the inverse covariance matrix $C^{-1}$ that appears in the linear term and normalization of the estimator. This matrix is typically approximated by a Monte Carlo average over inverse-covariance-weighted Gaussian $a_{X, \ell m}$ with the same signal covariance,  noise covariance, masking etc.\ as the data, i.e.\ as if drawn from the distribution specified by Eq.~\eqref{tot_cov}:
\begin{align}\label{eq:mc_cov}
(C^{-1})^{X X'}_{\ell m  \ell'm'} \approx \left \langle  (C^{-1}a)^{X}_{\ell m} (C^{-1}a^{\dagger})^{X'}_{\ell' m'} \right \rangle_{\mathrm{MC}} \, ,
\end{align}
which converges because:
\begin{align}
\left\langle (C^{-1}a)^{X}_{\ell m} (C^{-1}a^{\dagger})^{X'}_{\ell' m'}\right\rangle = (C^{-1})^{X X'}_{\ell m  \ell'm'} \, ,
\end{align}
where ($\langle \dots \rangle$) denotes the ensemble average over the multivariate $\mathcal{N}(0, C)$ distribution. 
The need for this complication arises because the high-dimensional covariance matrix is usually too dense for regular matrix operations.\footnote{The signal and noise covariance matrices are typically sparse in the spherical harmonic and (pixel) coordinate basis respectively. The combined matrix is then sparse in neither of the two bases.} On the other hand, it is generally possible to evaluate Eq.~\eqref{eq:c_inv_a_body} with iterative methods.

\subsection{\label{sec:estimator.templates} Fast bispectrum estimation}

In this section we motivate the need for an efficient way to evaluate the estimator in Eq.~\eqref{eq:estimator_single_par} and review the standard method to do so: the KSW estimator. When used to estimate the amplitude of primordial $3$-point functions, the KSW estimator applies to the $\zeta \zeta \zeta$ correlation but not to our main interest: the $\zeta \zeta h$ correlation. We will introduce the generalized version of the KSW estimator that can be used for $\zeta \zeta h$ in the next section.

The number of numerical operations needed to evaluate the estimator in  Eq.~\eqref{eq:estimator_single_par} quickly grows to enormous  sizes as the resolution of the data, i.e.\ $\ell_{\mathrm{max}}$, increases. Even when the costs of computing $C^{-1}a$ are ignored, direct evaluation of the estimator in Eq.~\eqref{eq:estimator_single_par} will asymptotically scale as $\mathcal{O}(\ell_{\mathrm{max}}^6)$. The isotropy of the bispectrum may be used to reduce this scaling to $\mathcal{O}(\ell_{\mathrm{max}}^5)$ by, for instance, fixing $m_3 = -(m_1 + m_2)$, but this scaling is still unmanageable.

To avoid the $\mathcal{O}(\ell_{\mathrm{max}}^5)$ scaling, bispectrum estimation generally focuses on separable bispectrum templates to reduce the scaling to $\mathcal{O}(\ell_{\mathrm{max}}^3)$ (albeit possibly with a relatively large prefactor). The most straightforward implementation of this idea is formulated by Komatsu, Spergel and Wandelt~\cite{KSW}, in what we will refer to as the KSW estimator. See Ref.~\cite{smith2011algorithms} for technical details and Ref.~\cite{Yadav:2007rk, Yadav:2007ny} for a generalization that uses $E$-mode data in addition to $T$ data.

Simply put, the KSW estimator exploits the idea that for a hypothetical bispectrum template: 
\begin{align}\label{eq:fact_b_naive}
B_{m_1 m_2 m_3}^{\ell_1 \ell_2 \ell_3} = F_{\ell_1, m_1} G_{\ell_2, m_2} H_{\ell_3, m_3} \, ,
\end{align}
the sum in Eq.~\eqref{eq:estimator_single_par} can be factored into three independent parts, thereby reducing the scaling to $\mathcal{O}(\ell_{\mathrm{max}}^2)$. Of course, this hypothetical bispectrum template is not suitable, as it is not rotationally invariant. The decomposition in Eq.~\eqref{eq:isotropy_b} forbids isotropic templates that are explicitly factored like this. In reality, the KSW approach therefore uses a slightly modified version of the above decomposition. The numerical advantage is largely maintained with the modified version.

The modification comes in the form of the Gaunt integral expression. It allows the rotationally invariant part of the product of three (spin-weighted) spherical harmonics to be expressed in terms of Wigner $3$-$j$ symbols. The general expression can be found in  Eq.~\eqref{ext_gaunt_fin}. Here we only need the following version:
\begin{align}
\begin{pmatrix}\ell_1 & \ell_2 & \ell_3 \\ m_1 & m_2 & m_3 \end{pmatrix} J^{0 0 0}_{\ell_1 \ell_2 \ell_3} = \int_{S^2} \! \mathrm{d} \Omega(\bm{\hat{n}}) \prod_{i=1}^3 Y_{\ell_i m_i} (\bm{\hat{n}}) \, ,
\end{align}
with $J_{\ell_1 \ell_2 \ell_3}^{000}$ given by:
\begin{align}
\begin{split}
J_{\ell_1 \ell_2 \ell_3}^{000}  =& \ \sqrt{ \frac{ (2\ell_1 +1) (2\ell_2 +1) (2\ell_3 +1) }{4 \pi} } \\ &  \times \begin{pmatrix}\ell_1 & \ell_2 & \ell_3 \\ 0 & 0 & 0 \end{pmatrix} \!  .
\end{split}
\end{align}
Now consider the reduced bispectrum~\cite{komatsu2001acoustic}: 
\begin{align}\label{eq:reduced_bispectrum}
b_{\ell_1 \ell_2 \ell_3} \equiv \frac{B_{\ell_1 \ell_2 \ell_3}}{J_{\ell_1 \ell_2 \ell_3}^{000}} \, ,
\end{align}
where $B_{\ell_1 \ell_2 \ell_3}$ is the angle-averaged bispectrum. We have suppressed the polarization indices for simplicity. Note that the reduced bispectrum is only defined for $\ell_1 + \ell_2 + \ell_3 = \mathrm{even}$.\footnote{Restricting to $\ell_1 + \ell_2 + \ell_3 = \mathrm{even}$ does not introduce a loss of generality for the parity-invariant $\langle TTT \rangle$, $\langle TTE \rangle$, $\langle TEE \rangle$, and $\langle EEE \rangle$ angle-averaged bispectra that are usually considered (see Table~\ref{tab:parity_2}), but, as was shown in the previous section, angle-averaged bispectra can in general be nonzero for $\ell_1+\ell_2 + \ell_3 = \mathrm{odd}$.} By expressing the bispectrum in Eq.~\eqref{eq:isotropy_b} in terms of the reduced bispectrum, we may insert the Gaunt integral as follows:
\begin{align}
B_{m_1 m_2 m_3}^{\ell_1 \ell_2 \ell_3} =& 
\begin{pmatrix}\ell_1 & \ell_2 & \ell_3 \\ m_1 & m_2 & m_3 \end{pmatrix} J^{000}_{\ell_1 \ell_2 \ell_3} b_{\ell_1 \ell_2 \ell_3}\, , \nonumber \\
\begin{split}
=& \int_{S^2} \! \mathrm{d} \Omega(\bm{\hat{n}}) \Big[ \prod_{i=1}^3  Y_{\ell_i m_i} (\bm{\hat{n}}) \Big] b_{\ell_1 \ell_2 \ell_3} \, . \label{eq:b_local_fact}
\end{split}
\end{align}

The crucial insight is that isotropy does not constrain the reduced bispectrum in any way. Given a reduced bispectrum that is separable in $N_{\mathrm{fact}}$ sets of functions as:
\begin{align}\label{eq:fact_red_b}
b_{\ell_1 \ell_2 \ell_3} = \frac{1}{6} \sum_{i=1}^{N_{\mathrm{fact}}} f_{\ell_1}^{(i)} g_{\ell_2}^{(i)} h_{\ell_3}^{(i)} \ + (5 \ \mathrm{perm.}) \, ,
\end{align}
we may thus express the bispectrum as: 
\begin{align}\label{eq:b_loc_fact}
\begin{split}
&B_{m_1 m_2 m_3}^{\ell_1 \ell_2 \ell_3} = \int_{S^2} \! \mathrm{d} \Omega(\bm{\hat{n}}) \Bigg[ \frac{1}{6} \sum_{i=1}^{N_{\mathrm{fact}}}  f_{\ell_1}^{(i)}Y_{\ell_1 m_1}  \\ &\ \ \ \   \quad \quad \  \times g_{\ell_2}^{(i)}Y_{\ell_2 m_2} \, h_{\ell_3}^{(i)}Y_{\ell_3 m_3} \!  +  (5 \ \mathrm{perm.}) \Bigg]  (\bm{\hat{n}})  \, .
\end{split}
\end{align}
We will refer to bispectra that can be written as the above expression as `locally separable'. This name refers to the fact that the integrand of the angular integral is separable in $(\ell_1, m_1)$, $(\ell_2, m_2$), and $(\ell_3, m_3)$. 

We conclude that, while factored bispectra as in Eq.~\eqref{eq:fact_b_naive} are forbidden, isotropy allows a locally separable template like Eq.~\eqref{eq:b_loc_fact}.
As we will see, the alluded computational benefits are largely retained for such templates. It is important to note that up to now we have not assumed that the bispectra are sourced by primordial $3$-point functions. 

With regards to primordial non-Gaussianity, the above construction is only useful when actual theoretical bispectrum templates can be reduced to the form of Eq.~\eqref{eq:fact_red_b}. Fortunately, this is  the case for a large class of linearly propagated bispectra sourced by the $\zeta \zeta \zeta$ correlation. 
For such bispectra, the condition in Eq.~\eqref{eq:fact_red_b} is met when the shape of the $3$-point function in Eq.~\eqref{eq:sss_general} is separable in~$k$:
\begin{align}\label{eq:fact_cond}
\begin{split}
f^{(\zeta\zeta \zeta)}(k_1, k_2, k_3) = \frac{1}{6} \sum_{i=1}^{N_{\mathrm{prim}}} & \, f^{(i)}(k_1) g^{(i)}(k_2) h^{(i)}(k_3) \\ & + (5 \ \mathrm{perm.}) \, .
\end{split}
\end{align}
Note that the $N_{\mathrm{fact}}$ parameter in Eq.~\eqref{eq:fact_red_b} is linearly related to $N_{\mathrm{prim}}$; the constant of proportionality between $N_{\mathrm{prim}}$ and $N_{\mathrm{fact}}$ will be detailed later in Sec.~\ref{sec:estimator_cubic}. 
The local shape in Eq.~\eqref{eq:local_template} is an example of a separable shape template. The equilateral and orthogonal shape templates used in the \emph{Planck} analysis~\cite{Akrami:2019izv} have been specifically derived to be separable~\cite{Creminelli:2005hu, Senatore:2009gt}.

Going back to the general, not-necessarily primordially sourced case, it remains to be demonstrated how separable reduced bispectra  actually lead to a reduction in computational cost. To see this, we insert Eq.~\eqref{eq:b_loc_fact} into  Eq.~\eqref{eq:estimator_single_par} and write down the cubic part of the resulting expression:
\begin{align}\label{eq:ksw_scalar}
\begin{split}
&\hat{f}_{\mathrm{NL}, \mathrm{cubic}} = \frac{1}{6\, \mathcal{I}_0}  \int_{S^2} \! \mathrm{d}\Omega (\bm{\hat{n}}) \\ &\quad \times \Bigg[     \sum_{i=1}^{N_{\mathrm{fact}}} 
\mathcal{A}{[f_{\ell}^{(i)} ]}  \mathcal{A} {[g_{\ell}^{(i)} ]} \, \mathcal{A}{[h_{\ell}^{(i)} ]}   \Bigg]  (\bm{\hat{n}}) \, .
\end{split}
\end{align}
The $\mathcal{A}$ functionals yield spin-$0$ fields on the sphere given by the inverse covariance-weighted data, weighted by the factors of the reduced bispectrum ($f_{\ell}$, $g_{\ell}$, and $h_{\ell}$, see Eq.~\eqref{eq:fact_red_b}). For example:
\begin{align}\label{eq:a_func_scal}
\mathcal{A}[f_{X, \ell}](\bm{\hat{n}}) = \sum_{\ell , m} \sum_{X} f_{X, \ell} (C^{-1}a)^{X}_{\ell m}  Y_{\ell m} (\bm{\hat{n}}) \, .
\end{align}
Note that we have reintroduced the polarization indices and assume they only run over $X \in \{ T, E\}$ here. 
Evaluating Eq.~\eqref{eq:ksw_scalar} does not quite scale like $\mathcal{O}(\ell_{\mathrm{max}}^2)$ as one might expect but as $\mathcal{O}(N_{\mathrm{fact}}\ell_{\mathrm{max}}^3)$.  Simply put, the scaling is determined by the $\mathcal{O}(\ell_{\mathrm{max}}^3)$ scaling of the recursive algorithms needed to compute the spherical harmonics which have to be recomputed $N_{\mathrm{fact}}$ times.\footnote{It should be noted that Ref.~\cite{smith2011algorithms} describes an alternative, significantly more efficient $\mathcal{O}(N_{\mathrm{fact}} \ell_{\mathrm{max}}^3)$ algorithm for  Eq.~\eqref{eq:ksw_scalar} that only runs the expensive $Y_{\ell m}$ recursion once. }
This is still an significant improvement over the general   $\mathcal{O}(\ell_{\mathrm{max}}^5)$ scaling.

The Monte Carlo expression for the linear term in Eq.~\eqref{eq:estimator_single_par} becomes equal to:
\begin{align}\label{eq:ksw_scalar_lin}
\begin{split}
\hat{f}_{\mathrm{NL, lin}} &= \ \frac{1}{6\, \mathcal{I}_0} \int_{S^2} \mathrm{d}\Omega (\bm{\hat{n}})  \Bigg[ \sum_{i=1}^{N_{\mathrm{fact}}}   \mathcal{A}{[f_{\ell}^{(i)} ]}  \\ &\quad \quad  \times \big \langle  \mathcal{A} {[g_{\ell}^{(i)} ]} \, \mathcal{A}{[h_{\ell}^{(i)} ]} \big\rangle_{\mathrm{MC}} \!  + \mathrm{cyclic} \Bigg] (\bm{\hat{n}}) \, .
\end{split}
\end{align}
The two additional terms denoted by `$\mathrm{cyclic}$' are obtained by cyclic permutations of $f_{\ell}^{(i)}$, $g_{\ell}^{(i)}$, and $h_{\ell}^{(i)}$.
Evaluation of this linear term scales as $\mathcal{O}( N_{\mathrm{sim}} N_{\mathrm{fact}} \ell_{\mathrm{max}}^3)$, where $100 \alt N_{\mathrm{sim}} \alt 1000$ iterations are typically needed for a sufficiently accurate estimate~\cite{smith2011algorithms}.

The estimator normalization $I_0$ in Eq.~\eqref{eq:fisher_full_scal} is evaluated by a Monte Carlo estimate. We omit the details of this aspect of the estimation procedure, and only mention the two methods that are used in practical applications. The most straightforward estimate of $\mathcal{I}_0$ is given by the variance of the unnormalized estimator applied to an ensemble of simulated Gaussian data effectively drawn from the distribution specified by Eq.~\eqref{tot_cov}. A similar but slightly more involved Monte Carlo procedure is described in~\cite{smith2011algorithms}. This second method is shown to converge for smaller ensembles than the first method.

In summary, a primordial $\zeta \zeta \zeta$ $3$-point correlation function described by a separable shape function will source a separable reduced bispectrum. We have established that the separability of the reduced bispectrum allows the use of the KSW estimator, see Eq.~\eqref{eq:ksw_scalar}. Finally, the KSW estimator is a prescription that alleviates the scaling of the estimator in Eq.~\eqref{eq:estimator_single_par} from $\mathcal{O}(\ell_{\mathrm{max}}^5)$ to a more manageable $\mathcal{O}(N_{\mathrm{fact}} \ell_{\mathrm{max}}^3)$.

\subsection{Fast scalar-scalar-tensor bispectrum estimation}\label{sec:fast_estimator_sst}

\subsubsection{\label{sec:estimator.fast.intro} Overview}

We now turn to the situation for the $\zeta \zeta h$ $3$-point correlation function. We explain why the standard KSW estimator, derived in Sec.~\ref{sec:estimator.templates}, does not apply to this type of correlation. We then come to the main new result of this paper: we introduce an alternative approach that allows the construction of an efficient estimator for the $\zeta \zeta h$ correlation.

Recall that for the $\zeta \zeta \zeta$ correlation the necessary condition for a separable reduced bispectrum is given by Eq.~\eqref{eq:fact_cond}: a separable shape function. Unlike the $\zeta \zeta \zeta$ $3$-point correlation function, the $\zeta \zeta h$ correlation is not uniquely specified by a shape function. 
It turns out that  when the reduced bispectrum for the $\zeta \zeta h$ template in Eq.~\eqref{eq:b_template_sst} is computed, the result is non-separable in $\ell_1$, $\ell_2$, and $\ell_3$~\cite{Shiraishi:2017yrq}. This holds true even when the  $f^{(\zeta \zeta h)}$ shape function in Eq.~\eqref{eq:b_template_sst} is separable in $k_1$, $k_2$, and $k_3$, which means that the responsible piece is the angular term:
\begin{align}\label{eq:ang_coupling_sst}
\big\langle \zeta_{\bm{k}_1} \zeta_{\bm{k}_2} {}^{(\pm 2)}h_{\bm{k}_3} \big\rangle \propto (\hat{k}_1)^a (\hat{k}_2)^b \, e_{ab}^{\pm 2}(\bm{\hat{k}}_3) \, . 
\end{align}
Despite the angular dependence, this term is a scalar under spatial coordinate transformations. The term provides a weight and complex phase to each $\{ \hat{\bm{k}}_1, \hat{\bm{k}}_2\}$ configuration relative to the wave vector of the tensor perturbation but has no preference for a global orientation of the three wave vectors. The associated CMB bispectrum is therefore isotropic and has a trivial dependence on its $m_1$, $m_2$, and $m_3$ azimuthal numbers, given by Eq.~\eqref{eq:isotropy_b}. With the azimuthal numbers constrained by isotropy, the geometrical coupling between the wave vectors in Eq.~\eqref{eq:ang_coupling_sst} can then only manifest itself in an explicit coupling between the $\ell_1$, $\ell_2$, and $\ell_3$ multipole orders, which in turn prevents the reduced bispectrum to be separable.

Without a separable reduced bispectrum we cannot construct the KSW estimator for the $\zeta \zeta h$ template by simply inserting the factors of the reduced bispectrum into Eq.~\eqref{eq:ksw_scalar}.
To derive a generalized KSW estimator for this template, let us observe that each term in the sum over spatial indices in Eq.~\eqref{eq:ang_coupling_sst} is factored in the three wave vectors. 
Of course, unlike the summed expression, the individual terms are not 3-scalars; the decomposition is coordinate dependent. By itself, each term can be interpreted as a homogeneous but anisotropic $3$-point function. Homogeneity is still preserved by the overall delta function in Eq.~\eqref{eq:prim_3point_hom}.
$3$-point functions of this form result in anisotropic bispectra\footnote{The bispectrum is isotropic by definition so an anisotropic bispectrum should be understood as a shorthand for a harmonic $3$-point function that does not obey Eq.~\eqref{eq:isotropy_b}.} that are locally separable in the sense of Eq.~\eqref{eq:b_loc_fact}. The anisotropic expressions differ from the isotropic one in Eq.~\eqref{eq:b_loc_fact} by the  $f_{\ell}$, $g_{\ell}$, and $h_{\ell}$ factors; they gain a dependence on $m$ in addition to~$\ell$. 

Roughly speaking, we thus exchange isotropy for separability. The estimates of the amplitudes of the anisotropic terms combine into an estimate of the amplitude of the original isotropic template. The trade-off is that several anisotropic templates have to be considered for one isotropic template. Constructing analogues of the cubic and linear estimator terms in Eq.~\eqref{eq:ksw_scalar} and Eq.~\eqref{eq:ksw_scalar_lin} for an anisotropic template will turn out to be rather straightforward. The generalizations of the $\mathcal{A}$ functionals in Eq.~\eqref{eq:a_func_scal} will transform the data in an anisotropic manner, but note that this operation does not scale differently than the regular isotropic transformation. The overall scaling of the estimator with $\ell_{\mathrm{max}}$ will thus be unchanged. The number of anisotropic terms needed for $3$-point functions of the type in Eq.~\eqref{eq:ang_coupling_sst} turns out to be only five. The amount of extra computations compared to the $\zeta \zeta \zeta$ estimator is thus rather insignificant.

Guided by the rough arguments provided in this section, we now turn to the actual derivation of the proposed estimator. We will first derive the expression for the linearly propagated bispectrum for the $\zeta \zeta h$ $3$-point function and demonstrate how it is indeed given by a sum of anisotropic bispectra. We will then construct the actual estimator.

\subsubsection{Full bispectrum for the scalar-scalar-tensor template}\label{sec:full_bispec}

In this section we derive the linearly propagated bispectrum for the $\zeta \zeta h$ $3$-point correlation function. As mentioned in the previous section, we require an expression for the full bispectrum instead of the angle-averaged or reduced bispectrum.

The general expression for the linearly propagated bispectrum in Eq.~\eqref{eq:theoretical_bispectrum} is most easily evaluated by separating the integrals over the three wave vectors in angular and radial integrals. In order to do so we need to rewrite the delta function that imposes  momentum conservation in  Eq.~\eqref{eq:prim_3point_hom}. Additionally, we express all angular terms of the $3$-point function as spin-weighed spherical harmonics in order to simplify the angular integrals.

We start with the delta function. We make use of the plane wave expansion in terms of spherical harmonics and spherical Bessel functions:
\begin{align}\label{eq:plane_wave_decomp}
e^{i \bm{k} \cdot \bm{x}} = 4\pi  \sum_{L, M} i^{L} j_{L}(kr) Y^*_{L M} (\bm{\hat{k}}) Y_{L M} (\bm{\hat{n}}) \, ,
\end{align}
with $\bm{k}=k\bm{\hat{k}}$ and $\bm{x}=r\bm{\hat{n}}$. The unit vector $\bm{\hat{n}}$ represents the direction of the line of sight from the origin of the comoving coordinate system (our location). Using this expansion we decompose the delta function into radial and angular parts: 
\begin{align}
\begin{split}
 \delta^{(3)}&(\bm{k}_1 + \bm{k}_2 + \bm{k}_3) = \\ 
 & 8 \sum_{L_1, M_1} \sum_{L_2, M_2} \sum_{L_3, M_3}  \int_{S^2} \! \mathrm{d}\Omega(\bm{\hat{n}}) \! \left[\prod_{i=1}^3 Y_{L_i, M_i}(\bm{\hat{n}}) \right] \\
 & \times \int_0^{\infty} r^2 \mathrm{d}r \left[ \prod_{i=1}^3  i^{L_i} 
 j_{L_i} (k_i r) Y^*_{L_i M_i}(\bm{\hat{k}}_i)  \right] \, . \label{eq:delta_ball} 
 \end{split}
\end{align}
See Appendix~\ref{sec:app_delta} for details. Although the integral over $\bm{\hat{n}}$ is given by the Gaunt integral expression in Eq.~\eqref{ext_gaunt_fin}, it will turn out to be  important to leave the expression factorizable in $L_1$, $L_2$, and $L_3$ so we do not solve the angular integral.

We then move on to the angular part of the $\zeta \zeta h$ template in Eq.~\eqref{eq:b_template_sst}. As discussed, this part is already expressed as a sum of factorized terms, so we leave it in its uncontracted form. However, we express the unit vectors and polarization tensor in terms of spherical harmonics. In a general coordinate system, not necessary aligned with $\bm{k}_1$, $\bm{k}_2$, or $\bm{k}_3$, the two unit vectors in Eq.~\eqref{eq:b_template_sst} are decomposed into dipole ($\ell=1$) moments with a longitudinal ($m=0$) and two solenoidal ($m=\pm1$) modes, while the polarization tensor is decomposed into quadrupole ($\ell=2$) moments with longitudinal ($m=0$), solenoidal ($m=\pm1$) and transverse ($m=\pm2$) modes. To retain the correct transformation properties, the quadrupole moment is expressed in terms of spin-$\pm2$ spherical harmonics on the plane perpendicular to $\bm{\hat{k}}_3$. As the $45$ resulting combinations have to sum to a $3$-scalar, each combination has to be weighted by the appropriate Wigner $3$-$j$ symbol. The resulting expression is given by~\cite{shiraishi2011cmb}:
\newpage 
\begin{align}\label{eq;pol_tens_decomp}
\begin{split}
(\hat{k}_1)^a (\hat{k}_2)^b \, e_{ab}^{\pm 2}(&\hat{\bm{k}}_3) = \frac{(8\pi)^{3/2}}{6} \! \!  \sum_{\substack{m_a, m_b, \\ M}} \!
\begin{pmatrix}1 & 1 & 2 \\ m_a & m_b & M \end{pmatrix} \\ &\ \, \times 
Y^*_{1 m_a} (\bm{\hat{k}}_1) Y^*_{1 m_b}(\bm{\hat{k}}_2) \, {}_{\mp 2}Y^*_{2 M}(\bm{\hat{k}}_3) \, .
\end{split} 
\end{align}
The selection rules of the $3$-$j$ symbol limit the azimuthal modes to only nine combinations: those that obey $ m_a + m_b + M = 0$.

We may now use Eq.~\eqref{eq:delta_ball} and  Eq.~\eqref{eq;pol_tens_decomp} to decompose the $\zeta \zeta h$ $3$-point function in radial and angular parts, resulting in the following expression:
\begin{widetext}
\begin{align}
\begin{split}
&{}^{(0 0 \pm 2)} B(\bm{k}_1, \bm{k}_2, \bm{k}_3)  = (2\pi)^3 \frac{(8\pi)^{3/2}}{6} 8 \sum_{m_a, m_b, M}  
\begin{pmatrix}1 & 1 & 2 \\ m_a & m_b & M \end{pmatrix}
Y^*_{1 m_a}(\bm{\hat{k}}_1) Y^*_{1 m_b}(\bm{\hat{k}}_2) \, {}_{\mp 2}Y^*_{2 M}(\bm{\hat{k}}_3)  \\
&\quad \quad \quad \times \int_{0}^{\infty} r^2 \mathrm{d}r  \int_{S^2} \mathrm{d}\Omega(\bm{\hat{n}}) \,  \left[ \prod_{i=1}^3 \sum_{L_i, M_i} (-1)^{L_1/2} 
 j_{L_i} (k_i r) Y^*_{L_i, M_i}(\bm{\hat{k}}_i) Y_{L_i M_i}(\bm{\hat{n}}) \right] f^{(\zeta\zeta h)}(k_1, k_2, k_3)  \, .
 \label{maldacena_general_2}
 \end{split}
\end{align}
\end{widetext}
As is required for the KSW estimator, we assume that the shape function $f^{(\zeta \zeta h)}$ is separable, i.e.\ it obeys: 
\begin{align}\label{factor_f}
\begin{split}
f^{(\zeta\zeta h)}(k_1, k_2, k_3) = \frac{1}{6} \sum_{i=1}^{N_{\mathrm{prim}}} & \, f^{(i)}(k_1) g^{(i)}(k_2) h^{(i)}(k_3) \\ & + (5 \ \mathrm{perm.}) \, .
\end{split}
\end{align}
The $N_{\mathrm{prim}}$ sets of $f$, $g$, and $h$ functions depend on the model under investigation so we leave them unspecified.

We have now gathered all ingredients to form the linearly propagated CMB bispectrum for the $\zeta \zeta h$ $3$-point correlation function. We  do so by combining Eq.~\eqref{maldacena_general_2} and Eq.~\eqref{factor_f} and inserting the result into  Eq.~\eqref{eq:theoretical_bispectrum}. Because we have separated the $3$-point function in radial and angular parts, the expression neatly factors into six independent integrals. We evaluate the angular integrals using the generalized Gaunt integral relation in Eq.~\eqref{ext_gaunt_fin}. The resulting contribution to the CMB bispectrum is then as follows:
\begin{widetext}
\begin{align}\label{eq:factored_sst_cmb}
\begin{split}
&B_{m_1 m_2 m_3 X_1 X_2 X_3}^{ \ell_1 \ell_2 \ell_3 (\zeta \zeta h)} =  \frac{(8 \pi)^{3/2}}{36}
\sum_{m_a, m_b, M}  \begin{pmatrix}1 & 1 & 2 \\ m_a & m_b & M \end{pmatrix} 
 \int_{S^2} \mathrm{d}\Omega (\bm{\hat{n}})  \sum_{i=1}^{N_{\mathrm{prim}}} \int_{0}^{\infty} \! r^2 \mathrm{d}r \\
& \quad \quad \quad \quad \times  \sum_{L_1 , M_1} \left[  i^{\ell_1+L_1}  J_{1 L_1 \ell_1}^{000}  
\begin{pmatrix}1 & L_1 & \ell_1 \\ m_a & M_1 & m_1 \end{pmatrix} \big( \mathcal{K}^{(\zeta)}_{(X_1)}[ f^{(i)} ]\big)_{\ell_1, L_1} (r) \right] Y_{L_1 M_1}(\bm{\hat{n}}) \\
& \quad \quad \quad \quad \times  \sum_{L_2 , M_2} \left[  i^{\ell_2+L_2}  J_{1 L_2 \ell_2}^{000}  
\begin{pmatrix}1 & L_2 & \ell_2 \\ m_b & M_2 & m_2 \end{pmatrix} \big( \mathcal{K}^{(\zeta)}_{(X_2)}[ g^{(i)} ]\big)_{\ell_2, L_2} (r)   \right] Y_{L_2 M_2}(\bm{\hat{n}}) \\
& \quad \quad \quad \quad \times  \sum_{L_3 , M_3} \left[ i^{\ell_3+L_3}  
J_{2 L_3 \ell_3}^{- 2 0 2} \left[1 + (-1)^{x_3 + L_3 + \ell_3} \right]
\begin{pmatrix}2 & L_3 & \ell_3 \\ M & M_3 & m_3 \end{pmatrix} \big( \mathcal{K}^{(h)}_{(X_3)}[ h^{(i)} ]\big)_{\ell_3, L_3} (r) \right]  Y_{L_3 M_3}(\bm{\hat{n}}) \, \\ & \quad \quad \quad \quad 
\quad \quad \quad \quad \quad \quad \quad \quad \quad \quad \quad \quad \quad \quad \quad \quad \quad \quad \quad \quad \quad \quad \quad \quad \quad + (5 \ \mathrm{perm.}) \, .
\end{split}
\end{align}
\end{widetext}
Here we have, as a shorthand, defined the following set of functionals for all $Z \in \{\zeta, h\}$, $X \in \{T, E, B\}$:
\begin{align}\label{eq:radial_functional}
\big(\mathcal{K}^{(Z)}_{(X)} [ f ]\big)_{\ell, L}\equiv \frac{2}{\pi} \int_0^{\infty} \! k^2 \mathrm{d}k \, f(k)  \mathcal{T}^{(Z)}_{X,\ell}(k) j_{L}(kr) \,  .
\end{align}
The $\mathcal{T}_{\ell}(k)$ transfer functions were introduced in Eq.~\eqref{eq:gen_alm}.
To evaluate the sum over the tensor helicities we have made use of the following relation:
\begin{align*}
\sum_{\lambda_3 \in \pm 2} \mathrm{sgn}(\lambda_3)^{\lambda_3 + x_3} J_{2 L_3 \ell_3}^{- \lambda_3 0 \lambda_3} =  
J_{2 L_3 \ell_3}^{- 2 0 2} \left[1 + (-1)^{x_3 + L_3 + \ell_3} \right] \, ,
\end{align*}
which reflects that the $f^{(\zeta \zeta h)}$ shape function in  Eq.~\eqref{maldacena_general_2} is helicity-independent.
Recall that $x_3 \in \{0, 1\}$ indicates whether the $X_3$ CMB field is parity-even or parity-odd. The $J$ symbols are defined in Eq.~\eqref{eq:prefactor_gaunt}.

The expression for the bispectrum in Eq.~\eqref{eq:factored_sst_cmb} is a bit verbose, but this expanded form will make it easier to construct the estimator in the next section. The expression shows how the bispectrum can be separated into factors that only depend on $\ell_1$, $\ell_2$, or $\ell_3$. Of course, the expression, taken as a whole, ought to be isotropic. This may be checked by summing over all azimuthal dummy indices ($m_a$, $m_b$, $M$, $M_1$, $M_2$, $M_3$).\footnote{First express the angular integral over $Y_{L_1 M_1}$,  $Y_{L_2 M_2}$, and  $Y_{L_3 M_3}$ in terms of the Gaunt integral and then sum over the five $3$-$j$ symbols that depend on azimuthal numbers using Eq.~\eqref{eq:9j_in_3j}~\cite{shiraishi2011cmb}.}
As expected, the resulting expression reduces to the isotropic form in Eq.~\eqref{eq:isotropy_b} but yields a non-separable angle-averaged/reduced bispectrum. 

Each term in the sum over $m_a$, $m_b$, and $M$ in  Eq.~\eqref{eq:factored_sst_cmb} describes an anisotropic bispectrum. 
Each of these bispectra is `locally' separable in the sense of Eq.~\eqref{eq:b_loc_fact}. The integral over the comoving radial coordinate $r$ in Eq.~\eqref{eq:factored_sst_cmb} may be replaced with a weighted sum over $N_{\mathrm{quad}}$ integration points. Combined with the  $N_{\mathrm{prim}}$ terms in the primordial shape function there will then be $N_{\mathrm{fact}} = N_{\mathrm{prim}} N_{\mathrm{quad}}$ locally separable terms.

The allowed combinations of $L_1$, $L_2$, and $L_3$  per $(\ell_1, \ell_2, \ell_3)$ triplet in  Eq.~\eqref{eq:factored_sst_cmb} are quite limited; depending on the polarization indices of the bispectrum only eight or twelve combinations are allowed~\cite{shiraishi2011cmb}. Recall that the capital $L$'s arise from the expansion of the delta function in Eq.~\eqref{eq:delta_ball}. The specific values can be found by systematically going over the $3$-$j$ symbols, including the ones hidden in the $J$ symbols (see Eq.~\eqref{eq:prefactor_gaunt}). First, note that $J^{000}_{1L_1 \ell_1}$ and $J^{000}_{1 L_2 \ell_2}$ require $L_1 + \ell_1$ and $L_2 + \ell_2$ to be odd. The triangle conditions of the $3$-$j$ symbols in the second and third line then enforce $L_1 = |\ell_1 \pm 1|$ and  $L_2 = |\ell_2 \pm 1|$. The term in square brackets in the fourth line forces $L_3 + \ell_3$ to be even when $x_3=0$ or odd when $x_3=1$. The triangle condition of the $3$-$j$ symbol in the fourth line then requires  $L_3 = \{\ell_3, |\ell_3 \pm 2| \}$ for $x_3=0$ and $L_3 = \{|\ell_3 \pm 1| \}$ for $x_3=1$. Finally, when the angular integral over $Y_{L_1 M_1}$,  $Y_{L_2 M_2}$, and  $Y_{L_3 M_3}$  is performed using  Eq.~\eqref{ext_gaunt_fin} it becomes clear how $ L_1 + L_2 + L_3 = \mathrm{even}$ is (again) imposed as well as $|L_1 - L_2| \leq L_3 \leq L_1 + L_2$.

We have derived the linearly propagated bispectrum for the $\zeta \zeta h$ $3$-point correlation function: a crucial ingredient for the derivation of the estimator. The resulting bispectrum is given in Eq.~\eqref{eq:factored_sst_cmb}. We have showed that the bispectrum can be viewed as a sum of anisotropic bispectra. As a sanity check of the derivation one may verify that the bispectrum holds up to the general constraints due to parity invariance that were formulated in Sec.~\ref{sec:cmb_bisp_intro}. For polarization triplets $X_1$, $X_2$, $X_3$ with even parity, i.e.\ $X_3 \neq B$, the bispectrum is real and nonzero when  $ \ell_1 + \ell_2 +\ell_3 = \mathrm{even}$. On the other hand, when $X_3 = B$ (so $x_3 =1$), the bispectrum becomes purely imaginary and nonzero only for $ \ell_1 + \ell_2 +\ell_3 = \mathrm{odd}$. 

\begin{figure}[htbp]
   \centering
   \includegraphics[width=0.5\textwidth]{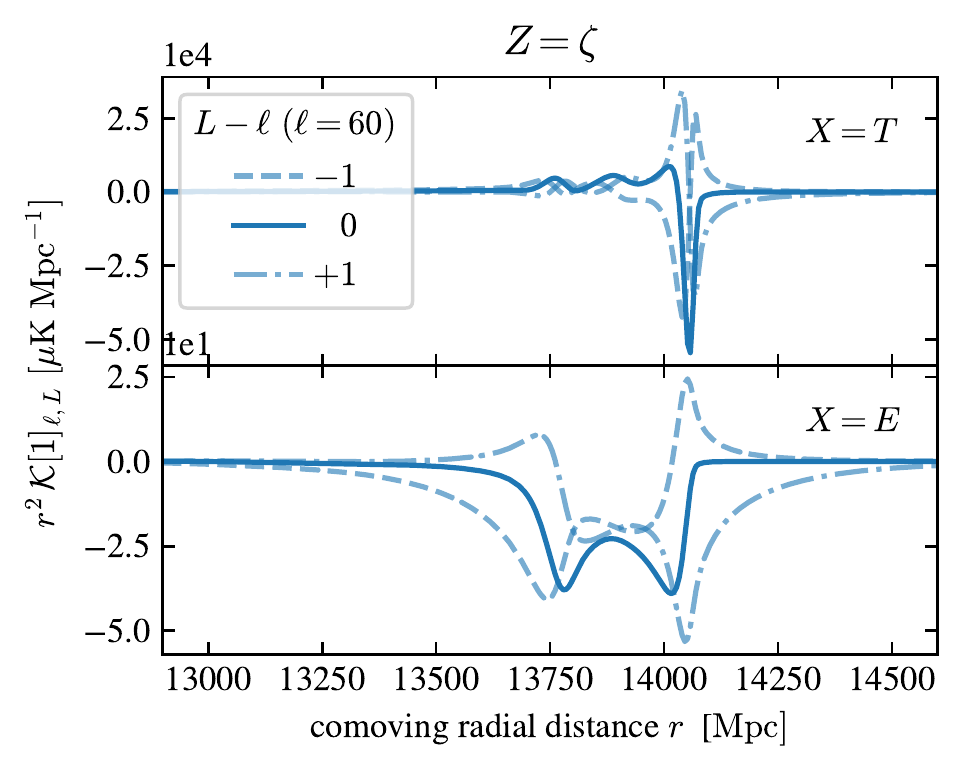}
   \caption{Radial transfer functions (solid lines) demonstrating the response of the $\ell=60$  temperature (top) and $E$-mode (bottom) CMB anisotropies  to the curvature perturbation at comoving radial distance $r$. The response shown here corresponds to the epoch of recombination. The dashed and dot-dashed lines show the radial parts of the functions used to project a dipole moment 
   constructed from the curvature perturbation to the ($\ell=60$) CMB harmonic modes. For low multipole orders, like the one depicted here, these functions are significantly less localized in $r$ than the transfer function and thus require a wider range of integration points.    }
   \label{fig:alpha_scal}
\end{figure}

\subsubsection{$\mathcal{K}$ functionals}\label{sec:k_functionals}
Before constructing the estimator it is instructive to take a more detailed look at the $\mathcal{K}_{\ell, L}$ functionals defined in Eq.~\eqref{eq:radial_functional}. They will become an important part of the estimator. We thus have a brief digression in which we illustrate the role of the functionals in Eq.~\eqref{eq:factored_sst_cmb}. Readers that are more interested in the actual estimator may skip this section.

The $\mathcal{K}$'s are a straightforward generalization of the $\alpha_{\ell} (r)$ and $\beta_{\ell}(r)$ functions introduced in the KSW description for the local model~\cite{KSW}.\footnote{The functions $\alpha_{\ell}(r)$ and $\beta_{\ell}(r)$  from Ref.~\cite{KSW} are given by $(\mathcal{K}^{(\zeta)}_{(T)}[ 1 ])_{\ell, \ell}$ and $(\mathcal{K}^{(\zeta)}_{(T)}[ P_{\Phi} ])_{\ell, \ell}$ respectively. $P_{\Phi}$ is the power spectrum of the gauge-invariant $\Phi_H$ Bardeen  potential~\cite{Bardeen:1980kt} instead of the curvature perturbation $\zeta$ we use; the two gauge-invariant quantities are related as $\zeta = -3 \Phi_H/2$ and $\zeta = -5 \Phi_H/3$ for super-horizon adiabatic perturbations in the radiation and matter dominated eras respectively~\cite{Lyth:2009zz}.} In the original KSW description the $\mathcal{K}$'s serve to transform the factors of the $3$-point function to the factors of the reduced bispectrum, i.e.\   $f(k) \mapsto \mathcal{K}[f ] = f_{\ell}$. For the $\zeta \zeta h$ estimator, the $\mathcal{K}$'s still serve to transform factors of the $3$-point function into factors of the bispectrum. The difference is that, as can be seen in Eq.~\eqref{eq:factored_sst_cmb}, the factors of the $3$-point function now each require multiple transformations to account for their non-scalar nature. 

Let us first focus on the $\mathcal{K}$ functionals that are relevant for regular $\zeta \zeta \zeta$ non-Gaussianity estimation: the $\mathcal{K}$'s with $L=\ell$ and transfer functions for $Z=\zeta$. It is convenient to consider a constant input function $f(k) = 1$, the resulting functions are equal to the $\alpha_{\ell}^X(r)$ functions defined in Ref.~\cite{Yadav:2007rk}:
\begin{align} \label{eq:alpha_ksw_pol}
\big(\mathcal{K}^{(\zeta)}_{(X)} [1 ]\big)_{\ell, \ell}(r) &= \alpha^X_{\ell}(r) \nonumber \\  &= \frac{2}{\pi} \int_{0}^{\infty} \! k^2 \mathrm{d} k  \,   \mathcal{T}^{(\zeta)}_{X,\ell}(k) j_{\ell}(kr) \, ,
\end{align}
where  $X \in \{T, E \}$ because of the $\zeta$ transfer function. 
The $\alpha^X_{\ell}(r)$ functions have a special interpretation: they serve as the  transfer functions in coordinate space instead of Fourier space.  Eq.~\eqref{eq:alpha_ksw_pol} is an inverse Fourier transform (i.e.\ inverse spherical Hankel transform) of the transfer function $\mathcal{T}_{\ell}(k)$ and it is true that the observable CMB harmonic modes sourced by $\zeta$ may be expressed as follows~\cite{Yadav:2005tf}:
\begin{align}\label{eq:alm_real_space}
a^{(\zeta)}_{X, \ell m} = \int_{0}^{\infty} \! r^2 \mathrm{d} r \, \zeta_{\ell m}(r) \alpha^X_{\ell} (r) \, ,
\end{align}
for $X\in \{T, E \}$. Here $\zeta_{\ell m}(r)$ are the spherical harmonic coefficients of the same initial amplitude of the curvature perturbation as in Eq.~\eqref{eq:zeta_fourier} but now decomposed on spherical shells around the origin of the comoving coordinate system:
\begin{align}\label{eq:radial_zeta_lm}
\zeta_{\ell m}(r) = \int_{S^2} \mathrm{d}\Omega (\bm{\hat{n}}) \, \zeta(\bm{x}, t)\big |_{t = \tilde{t}(\bm{x}, t_i)} Y^*_{\ell m}(\bm{\hat{n}})  \, .
\end{align}
Recall that $\tilde{t}(\bm{x}, t_i)$ denotes a spacelike hypersurface in the early radiation-dominated era.

The solid lines in  Fig.~\ref{fig:alpha_scal} show $\alpha^X_{\ell}(r)$ for $X=T$ and $X=E$ as function of comoving radius on the initial spatial hypersurface. The lines show how   $\zeta_{\ell m}(r)$ contributes to $a_{X, \ell m}$ for $\ell = 60$ over a range of comoving radii around $14000\ \mathrm{Mpc}$. In terms of the conformal time along the path of a radially traveling photon ($\Delta \tau =  r / c$), this range of $r$ is roughly centred around the epoch of recombination. 
Another response at $r\approx 9000 \ \mathrm{Mpc}$ corresponds to the rescattering of CMB photons at reionization. Finally, at $r \alt 3000\ \mathrm{Mpc}$ there is a slowly rising response as $r$ approaches zero for $X=T$ and $\ell \alt 150$  that corresponds to the late-time integrated Sachs-Wolfe (ISW) effect.

The fact that $\mathcal{K}^{(\zeta)}[1]$ yields the radial transfer functions provides a physical reason why the $\mathcal{K}$ functionals result in functions that are highly localized in $r$. 
During bispectrum estimation the integral over $r$ has to be evaluated as efficiently as possible; the localized nature of the radial functions is thus highly beneficial. We will now see how and why the radial functions used for the $\zeta \zeta h$ bispectrum differ from the ones used for regular scalar-sourced bispectra. These new functions will turn out to be slightly less localized in $r$, but the difference is minor.

Eq.~\eqref{eq:alm_real_space} must hold because the harmonic modes of the curvature perturbation on spherical shells $\zeta_{\ell m} (r)$ in  Eq.~\eqref{eq:radial_zeta_lm} are related to the harmonic modes of the Fourier representation of  $\zeta$ through the following simple relation:
\begin{align}\label{eq:radial_zeta}
\zeta_{\ell m}(k) = 4 \pi (-i)^{\ell} \int_0^{\infty} r^2 \mathrm{d}r \, \zeta_{\ell m} (r) j_{\ell}(kr) \, .
\end{align}
Here the $\zeta_{\ell m}(k)$ are the coefficients of the spherical harmonic decomposition of the angular part of $\zeta_{\bm{k}}$ in Eq.~\eqref{eq:zeta_fourier}:
\begin{align}\label{eq:four_zeta_lm}
 \zeta_{\bm{k}}  = \sum_{\ell , m} \zeta_{\ell m}(k) Y_{\ell m}(\hat{\bm{k}}) \, .
\end{align}
One can check that Eq.~\eqref{eq:alm_real_space} holds by inserting Eq.~\eqref{eq:four_zeta_lm} and  Eq.~\eqref{eq:radial_zeta} into Eq.~\eqref{eq:gen_alm} and making use of the orthonormality of the spherical harmonics.

In turn, Eq.~\eqref{eq:radial_zeta} is valid because  
$\zeta$ is a $3$-scalar, it has no intrinsic angular dependence. 
The projection from the Fourier basis to a basis of spherical shells at comoving radii $r$ is thus completely determined by the `orbital' angular momentum of the field, i.e.\ the projection is determined by the plane wave decomposition of the 3D Fourier basis functions in Eq.~\eqref{eq:plane_wave_decomp}. Simply put: projecting a Fourier mode of a $3$-scalar to an angular mode with multipole order $\ell$ and azimuthal mode $m$ sitting on a shell at radius $r$  only requires transformations involving $j_{\ell}$ and $Y_{\ell m}$. Inserting Eq.~\eqref{eq:radial_zeta} into Eq.~\eqref{eq:four_zeta_lm} demonstrates this behavior. 

For fields that are not $3$-scalars, a relation like Eq.~\eqref{eq:radial_zeta} will not hold. In these cases, the coupling between the intrinsic angular dependence of the field and that of the plane wave contributes to the projection. 
The exact expressions for these `total angular momentum' projection operators may be found in Ref.~\cite{Hu:1997hp, Dai:2012bc, Dai:2012ma}. We will use the general properties of these operators to gain a better understanding of the role of the second multipole index of the $\mathcal{K}_{\ell, L}$ functionals. 

In the above we argued that the projection of a single Fourier mode, i.e.\ a plane wave, to an angular mode with multipole order $\ell$ and azimuthal mode $m$ sitting on a shell at radius $r$ will only involve $j_{\ell}$ and $Y_{\ell m}$. The same projection for an intrinsically dipole-like ($\ell' = 1$) field that is modulated by a plane wave will involve operators constructed out of  $j_{\ell\pm1}$ and $Y_{\ell\pm1 m}$. Two distinct projections exist in this case: one for the longitudinal ($m'=0$) component of the dipole-like- field and one for the solenoidal ($m'=\pm1$) components~\cite{Hu:1997hp}. Similarly, the  projection of an intrinsically quadrupole-like ($\ell' = 2$) field  modulated by a plane wave will involve $j_{\ell}$, $Y_{\ell m}$; $j_{\ell\pm1}$, $Y_{\ell\pm1 m}$; and $j_{\ell\pm2}$, $Y_{\ell\pm2 m}$. Again, there are distinct projections for the longitudinal ($m'=0$), solenoidal ($m'=\pm1$), and transverse  ($m'=\pm2$) components of the field. This time, a projection using $\ell$ and $\ell\pm2$ only contributes to the parity-even component of the resulting field; the $\ell\pm1$ projections only contribute to the parity-odd component~\cite{Hu:1997hp}.

Having gained this intuition, it is now understood why only the terms with $L_1 = |\ell_1 \pm 1|$ and  $L_2 = |\ell_2 \pm 1|$ contribute in the second and third line of Eq.~\eqref{eq:factored_sst_cmb} respectively. Each of the two lines describes how a dipole moment constructed out of one of the two unit wave vectors in the $3$-point function template in Eq.~\eqref{eq:ang_coupling_sst} is projected to a set of angular modes on spherical shells at radius~$r$. The prefactor given to $\mathcal{K}_{\ell ,L} Y_{LM}$ in the second and third line of Eq.~\eqref{eq:factored_sst_cmb} will change depending on whether the longitudinal mode (e.g.\ $m_a = 0$) or the solenoidal modes (e.g.\ $m_a = \pm1$) are projected. 

The $\mathcal{K}$ functionals with $L = \ell \pm 1$ differ substantially from the $L = \ell $ variants used in the $\zeta \zeta \zeta$ KSW estimator. This is especially true for low ($\ell \alt 500$) multipole orders. We plot the $\mathcal{K}[1]_{\ell, \ell\pm 1}$ functions next to the regular radial transfer functions in Fig.~\ref{fig:alpha_scal} to illustrate this. Note that for $\ell \agt 500$ the functions with $L = \ell \pm 1$  converge to the shape of those with $L=\ell$ although there remains a small phase shift in~$r$ regardless of~$\ell$.

\begin{figure}[htbp]
   \centering
   \includegraphics[width=0.5\textwidth]{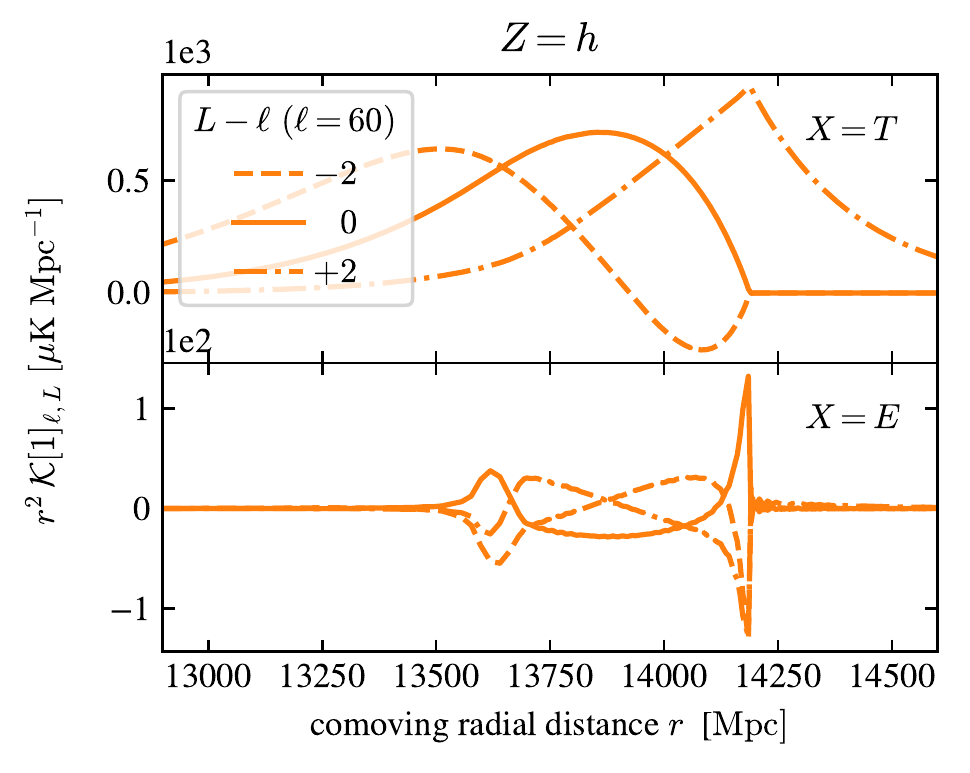}
   \caption{The three radial functions needed to compute the response of the $\ell=60$ temperature (top) and $E$-mode (bottom) CMB anisotropies to a quadrupole moment constructed from the $3$-tensor metric perturbation at comoving radial distance~$r$.}
   \label{fig:alpha_tens_ie}
\end{figure}

\begin{figure}[htbp]
   \centering
   \includegraphics[width=0.5\textwidth]{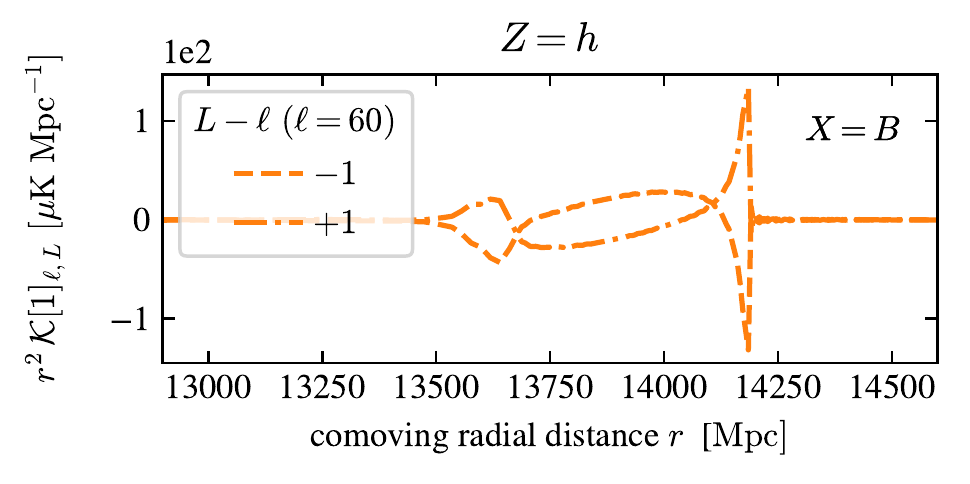}
   \caption{Similar to Fig.~\ref{fig:alpha_tens_ie} but instead showing the two radial functions needed to compute the response of the $\ell = 60$ $B$-mode CMB anisotropies to a quadrupole moment constructed from the $3$-tensor metric perturbation at comoving radial distance $r$.}
   \label{fig:alpha_tens_b}
\end{figure}

In a similar way, the fourth line of Eq.~\eqref{eq:factored_sst_cmb} describes the projection of the quadrupole moment constructed out of the polarization tensor in   Eq.~\eqref{eq:ang_coupling_sst}. As we established before, the $L = \ell \pm 1$ components are needed for the $B$-mode field while the $L = \ell, \ell \pm 2$ components are used for the parity-even $T$ and $E$ fields. The prefactor of $\mathcal{K}_{\ell_3, L_3} Y_{L_3M_3}$ is now dependent on $M$, which denotes whether the longitudinal ($M=0$), solenoidal ($M=\pm1$) or transverse ($M=\pm2$) components of the quadrupole are taken into account. To illustrate how the $\mathcal{K}_{\ell, L}$ functionals change when the $Z=h$ transfer functions are used instead of the $Z=\zeta$ transfer functions we considered before, we plot $\mathcal{K}_{\ell, L}[1]$ for $Z=h$ and $L = \ell, \ell\pm2 $ in Fig.~\ref{fig:alpha_tens_ie}. 
The plotted range again roughly corresponds to the recombination era. Not shown is another small response that corresponds to the reionization era. There is no equivalent for the late-time ISW effect. In Fig.~\ref{fig:alpha_tens_b} we plot the same functions but for $L = \{\ell\pm1 \}$. These functions are used to project the quadrupole moment of the $3$-point function to the CMB $B$-mode field.

The small aliasing effects seen in Fig.~\ref{fig:alpha_tens_ie} and Fig.~\ref{fig:alpha_tens_b} are purely numerical; both $j_{L}$ and $\mathcal{T}_{\ell}$ in Eq.~\eqref{eq:radial_functional} oscillate rapidly with $k$. The integral thus requires a large number of $k$ integration points to completely converge for each value of $r$. We have verified that the bispectrum and the results in Sec.~\ref{sec:Fisher} are not sensitive to these numerical artifacts. 

The point of this section was to explain the role of the  $\mathcal{K}_{\ell,L }$ functionals present in the $\zeta \zeta h$ bispectrum in Eq.~\eqref{eq:factored_sst_cmb}. As illustrated in the figures, the functionals with $\ell \neq L$, i.e.\ the ones needed for the $\zeta \zeta h$ bispectrum, differ substantially from the $\ell = L$ functionals that are used for the standard $\zeta \zeta \zeta$ bispectrum.

\subsubsection{Estimator} \label{sec:estimator_cubic}

Using Eq.~\eqref{eq:factored_sst_cmb}, the expression for the $\zeta \zeta h$ bispectrum, we now write down the estimator for the amplitude of this bispectrum template. For simplicity we start by neglecting the linear term in the estimator in Eq.~\eqref{eq:estimator_single_par} and focus on the cubic term. 

The expression for the bispectrum in Eq.~\eqref{eq:factored_sst_cmb} is sourced by the $ \zeta \zeta h$ template. The order matters, the observed CMB bispectrum is also sourced by the $3$-point functions with permuted $\zeta$ and $h$ indices. However, it will be convenient to keep ignoring the $\zeta h \zeta$ and $h \zeta \zeta$ contributions for now and start by constructing the estimator for the $\zeta \zeta h$ template only. We thus divide the (cubic part of) the estimator in three parts:
\begin{align}\label{eq:tot_cubic}
\hat{f}_{\rm NL, cubic}^{\mathrm{tot}} = \hat{f}_{\rm NL, cubic}^{\zeta \zeta h} + \hat{f}_{\rm NL, cubic}^{\zeta h \zeta} + \hat{f}_{\rm NL, cubic}^{h \zeta \zeta} \, ,
\end{align}
and start with the first term on the r.h.s.

We reap the benefits of our work in the previous sections; the cubic estimator is simply constructed by inserting the expression for the bispectrum, Eq.~\eqref{eq:factored_sst_cmb}, into the general expression for the estimator in Eq.~\eqref{eq:estimator_single_par} and keeping the terms cubic in the data. Let us again stress that this result is only achieved through the use of the full bispectrum as opposed to the angle-averaged or reduced bispectrum. The resulting expression for the cubic part of the estimator, and the main result of this paper, is given by: 
\begin{widetext}
\begin{align}
\begin{split}
\hat{f}_{\mathrm{NL, cubic}}^{\zeta \zeta h} 
= \ \frac{\sqrt{2}}{54\, \mathcal{I}_0}   \sum_{\substack{m_a, m_b, \\ M}} \! \! \begin{pmatrix}1 & 1 & 2 \\ m_a & m_b & M \end{pmatrix} 
\int_{S^2} \mathrm{d}\Omega(\bm{\hat{n}}) \sum_{i=1}^{N_{\mathrm{prim}}} \!
 \int_{0}^{\infty} \! r^2 \mathrm{d}r \, \Big(&
 \mathcal{A}^{(\zeta)}_{( 1, m_a )}[ f^{(i)}] \,
 \mathcal{A}^{(\zeta)}_{( 1, m_b )}[ g^{(i)}] \,
 \mathcal{A}^{(h)}_{( 2, M )}[ h^{(i)}] 
 \Big) (r, \bm{\hat{n}})  \\ 
 &+ \ (2 \ \mathrm{cyclic})\, .
\label{estimator_cubic_2}
\end{split}
\end{align}
\end{widetext}
We have again made use of the shape template in Eq.~\eqref{factor_f}. The two extra terms are cyclic  permutations of $f^{(i)}$, $g^{(i)}$, $h^{(i)}$. The six permutations of the shape function in  Eq.~\eqref{eq:factored_sst_cmb} thus reduce to three. This is possible due to the invariance under simultaneous interchange of $f^{(i)}$, $g^{(i)}$ and $m_a$, $m_b$ in Eq.~\eqref{estimator_cubic_2} or, more physically, the indistinguishability of the two scalar components of the $3$-point function.

The similarity of Eq.~\eqref{estimator_cubic_2} to the standard KSW estimator in Eq.~\eqref{eq:ksw_scalar} is evident. There are, however, two clear differences between the expressions. The most important difference is the anisotropy in the dependence on $m_a$, $m_b$, and $M$; as a reminder, in order to construct the equivalent of a KSW estimator, we need to construct pieces of the bispectrum separable in $l_1, l_2,$ and $l_3$, and this can only be done at the expense of introducing several anisotropic templates. 
As discussed previously, the estimates of the amplitudes of the anisotropic terms combine into an estimate of the amplitude of the original isotropic template. The anisotropy appears in the Wigner $3$-$j$ symbol and the $m_a$, $m_b$, and $M$ indices of the generalized $\mathcal{A}$ functionals in Eq.~\eqref{estimator_cubic_2}. We will discuss the meaning of the $m_a$, $m_b$, and $M$ indices and the $3$-$j$ symbol in more detail in the remainder of this section, but in short: each $(m_a, m_b,  M)$ triplet corresponds to a combination of the longitudinal, solenoidal and/or transverse angular modes of the contracted angular term, see Eq~\eqref{eq:ang_coupling_sst}, that is present in the $\zeta \zeta h$ $3$-point function. For a given $(m_a, m_b, M)$ triplet, Eq.~\eqref{estimator_cubic_2} estimates the contribution from the corresponding combination of angular modes to the data; the $3$-$j$ symbols then provides a relative weight to each contribution when all are summed into the final estimate $\hat{f}_{\mathrm{NL, cubic}}^{\zeta \zeta h}$. 
The second difference between Eq.~\eqref{estimator_cubic_2} and  Eq.~\eqref{eq:ksw_scalar} is the absence of the radial integral over comoving radius $r$ in Eq.~\eqref{eq:ksw_scalar}. This difference is mainly notational. As explained in the previous section (Sec.~\ref{sec:k_functionals}), the radial integral is also present for standard $\zeta \zeta \zeta$ estimation because it is part of the transformation between factors of the $3$-point function to the factors of the reduced bispectrum, e.g.\   $f(k) \mapsto f_{\ell}$. The radial integral could be implicitly included in Eq.~\eqref{eq:ksw_scalar} by letting the sum over $i$ run over $N_{\mathrm{fact}} = N_{r} N_{\mathrm{prim}}$ values, with $N_{r}$ denoting the number of quadrature point used to evaluate the radial integral and $N_{\mathrm{prim}}$ defined in Eq.~\eqref{eq:fact_cond}.

Before coming to the computational scaling of the estimator, let us focus our attention to the generalized $\mathcal{A}$ functionals in Eq.~\eqref{estimator_cubic_2}. For a given input function $f(k)$,  $\mathcal{A}^{(Z)}_{( S, n )}[ f]$ returns a scalar field on a spherical shell at comoving radial coordinate $r$. The $S$ index denotes whether the associated factor of the $3$-point function is a monopole ($S=0$), dipole ($S=1$) or quadrupole ($S=2$) source.  The $n$ index tells us whether we are considering the longitudinal ($n=0$), solenoidal ($n=\pm1$) or transverse ($n=\pm2$) part of the source. From Eq.~\eqref{estimator_cubic_2} we see that for the $\zeta \zeta h$ bispectrum we only need the $S=1$ functionals for the $Z=\zeta$ part and the $S=2$ functionals for the $Z=h$ part.

At each radial coordinate $r$ we may decompose the $\mathcal{A}$ functionals in terms of spherical harmonics:
\begin{align}\label{eq:a_map_def}
\mathcal{A}^{(Z)}_{( S, n )}[ f] (r, \bm{\hat{n}}) = \sum_{L,M}  \! \left(\mathcal{A}^{(Z)}_{( S, n )}[ f]   \right)_{LM} \! (r) \, Y_{L M} (\bm{\hat{n}}) \, .
\end{align}
The resulting harmonic modes are given by linear transformations of the inverse-covariance-weighted data. Based on the primordial index~$Z$, we identify two cases:
\begin{widetext}
\begin{align}\label{eq:a_harm_coeff}
\left(\mathcal{A}^{(Z)}_{( S, n )}[ f]   \right)_{LM} \! (r) \equiv  \! \!
\begin{cases}
\displaystyle   
 \! (4\pi)^{1/2} \sum\limits_{\ell, m} i^{\ell+L} \,  J_{S L \ell}^{000}  \
\begin{pmatrix}S & L & \ell \\ n & M & m \end{pmatrix} \! \! \sum\limits_{X}   \big( \mathcal{K}^{(\zeta)}[ f ]\big)^X_{\ell, L} (r) \, \, (C^{-1}a)^{X}_{\ell m}  \!  & Z\! =\! \zeta \, ,\\
\displaystyle  
\! (4\pi)^{1/2} \sum\limits_{\ell, m}  i^{\ell+L}  
J_{S L \ell}^{- 2 0 2}  \begin{pmatrix}S & L & \ell \\ n & M & m \end{pmatrix} \! \!
 \sum\limits_{X}  \left[1 + (-1)^{x + L + \ell} \right] \!
\big( \mathcal{K}^{(h)}[ f ]\big)^X_{\ell, L} (r) \, \,  (C^{-1}a)^{X}_{\ell m}     & Z\! =\! h \, .
\end{cases}  
\end{align}
\end{widetext}
Note that for the $Z=\zeta$ case, the sum over $X$ only runs over $\{T, E \}$, while for $Z=h$ it runs over $\{T, E, B\}$. The parity behavior associated with a given polarization index $X$ is denoted by $x$. The $\mathcal{K}$ functionals are defined in Eq.~\eqref{eq:radial_functional}. The data are filtered by the different $\mathcal{K}$ functionals in an anisotropic manner depending on the value of $n$. For example, the $(\mathcal{A}_{(S,2)})_{LM}$ modes are sourced by the $m=-(M+2)$ modes of the data.

The inverse spherical harmonic transformation needed to evaluate Eq.~\eqref{eq:a_map_def} scales as $\mathcal{O}(\ell_{\mathrm{max}}^3)$ and will in reality determine the overall scaling of the estimator evaluation. One might worry that the sums over $\ell$ and $m$ needed to construct the harmonic coefficients in Eq.~\eqref{eq:a_harm_coeff} will contribute significantly to the computational scaling. This is not the case, as the selection rules of the  Wigner $3$-$j$ symbols forbid most values of $\ell$ and $m$.  Only $\ell \in L\pm1$ and $m=-(M+n)$ are needed to compute $\mathcal{A}^{(\zeta)}_{L M}$ while for $\mathcal{A}^{( h)}_{L M}$ only $\ell \in \{L, L\pm 1, L\pm 2 \}$ and $m=-(M+n)$ are required.

To compute the angular integral in Eq.~\eqref{estimator_cubic_2}, the pixelization scheme used for the $\mathcal{A}[f]$ fields (or `maps') must support harmonic band-limits given by the sum of the  band-limits of the three individual maps (see Eq.~\eqref{eq:product_swshs}). In reality, the $\mathcal{A}^{( \zeta)}$ maps will be likely band-limited by the instrumental beam or noise covariance. On the other hand, the $\mathcal{A}^{(h)}$ maps only contain information on large ($\ell \alt 200$) scales; the tensor transfer functions suppress all information in the data on smaller scales. Small-scale tensor perturbations produced by an approximately scale-invariant process are inaccessible through the primary anisotropies. Unlike scalar perturbations,  small-scale tensor perturbations decay away with cosmic expansion before recombination.

It is instructive to take a closer look at how the symmetries of the spherical harmonics and the $3$-$j$ symbols relate the  harmonic coefficients of the $\mathcal{A}$ functionals with $(S,n)$ to those with $(S,-n)$.
This relation can be used to approximately half the number of  inverse harmonic transformations needed to evaluate Eq.~\eqref{estimator_cubic_2}. 
Assuming that the input function $f(k)$ is real-valued, the coefficients transform as follows under complex conjugation:
\begin{align}
\big(\mathcal{A}^{(Z)}_{(S, n)}[f]\big)^*_{LM} = 
\big(\mathcal{A}^{(Z)}_{(S, -n)}[f]\big)_{L-M} (-1)^{n+M+S} \, .
\end{align}
It follows that the functionals in Eq.~\eqref{eq:a_map_def} map input functions to  complex fields on the sphere that obey:
\begin{align}\label{eq:a_complex_conj}
\big(\mathcal{A}^{(Z)}_{(S, n)}[f]\big)^* (\bm{\hat{n}}) =  \big(\mathcal{A}^{(Z)}_{(S, -n)} [f]\big) (\bm{\hat{n}}) \, (-1)^{n+S}  \, .
\end{align}
Going back to the estimator in Eq.~\eqref{estimator_cubic_2}, we see that only five out of the nine allowed combinations of $m_a$, $m_b$, and $M$ need to be considered: the remaining terms may be found with the use of Eq.~\eqref{eq:a_complex_conj}. We may for example use the following five combinations:
\begin{align}\label{eq:dist_triplets}
\begin{split}
 (m_a,  m_b,  M) \ \in \  \big\{ &(1,1,-2),\, (1,0,-1),\, (0,1,-1), \\ &(1,-1,0), \, (0,0,0) \big\} \, .
\end{split}
\end{align}
The $m_a=m_b=M=0$ case is unique, the other four combinations in Eq.~\eqref{eq:dist_triplets} are related to the remaining four combinations by a factor $(-1)$. The $3$-$j$ symbol in Eq.~\eqref{estimator_cubic_2} does not change if this minus sign is added to its lower indices. For the $\mathcal{A}$ maps, Eq.~\eqref{eq:a_complex_conj} tells us that the addition of a minus sign to the $n$ index is equivalent to complex conjugation. For the products of $\mathcal{A}$ maps the following thus holds:
\begin{align}
\begin{split}\label{eq:comp_saving}
& \mathcal{A}^{(\zeta)}_{( 1, m_a )} 
 \mathcal{A}^{(\zeta)}_{( 1, m_b )} 
 \mathcal{A}^{(h)}_{( 2, M )} \! + \!  \mathcal{A}^{(\zeta)}_{( 1, -m_a )} 
 \mathcal{A}^{(\zeta)}_{( 1, -m_b )} 
 \mathcal{A}^{(h)}_{( 2, -M )} \\
 &=   2 \, \mathrm{Re} \left( \mathcal{A}^{(\zeta)}_{( 1, m_a )}
 \mathcal{A}^{(\zeta)}_{( 1, m_b )} 
 \mathcal{A}^{(h)}_{( 2, M )} \right) \, .
 \end{split}
\end{align}
Note that we have suppressed the $f^{(i)}$, $g^{(i)}$, and $h^{(i)}$ input functions to the $\mathcal{A}$'s. 
Eq.~\eqref{eq:comp_saving} implies that, instead of computing nine products, one can only calculate five products of complex $\mathcal{A}$ maps and discard the imaginary parts to evaluate Eq.~\eqref{estimator_cubic_2}.
The fact that only five out of nine terms are needed can be understood from the original expression for the angular term. Starting with the nine terms in the sum over $a$ and $b$ in Eq.~\eqref{eq:ang_coupling_sst}, the symmetry under simultaneous exchange of $a$, $b$ and $\zeta_{\bm{k}_1}$, $\zeta_{\bm{k}_2}$ removes three degrees of freedom. The vanishing trace of the polarization tensor removes the fourth.

It is easy to see that the two additional estimator terms with permuted indices in Eq.~\eqref{eq:tot_cubic} are constructed by permuting the columns of the $3$-$j$ symbol together with the $(m_a, \zeta)$, $(m_b, \zeta)$, and $(M, h)$ index pairs of the three $\mathcal{A}$ functionals in Eq.~\eqref{estimator_cubic_2}. The $3$-$j$ symbol is invariant under such permutations. The product of $\mathcal{A}$ maps is also invariant under such permutations because of the symmetrized form of the shape function in Eq.~\eqref{factor_f}. The total cubic term of the estimator is therefore simply given by:
\begin{eqnarray}
\hat{f}_{\rm NL, cubic}^{\mathrm{tot}} = 3 \hat{f}_{\rm NL, cubic}^{\zeta \zeta h} \, .
\end{eqnarray}

After deriving the cubic part of the estimator, the linear term is obtained in an analogous way. 
It can be found by inserting the bispectrum in Eq.~\eqref{eq:factored_sst_cmb} into Eq.~\eqref{eq:estimator_single_par} and keeping the terms linear in the data:
\begin{widetext}
\begin{align}
\begin{split}
\hat{f}_{\rm NL, lin}^{\zeta \zeta h} = &- \frac{\sqrt{2}}{54\, \mathcal{I}_0} \sum_{M, m_a, m_b}  \begin{pmatrix}1 & 1 & 2 \\ m_a & m_b & M \end{pmatrix}
\int_{S^2} \mathrm{d} \Omega(\bm{\hat{n}})  \sum_{i=1}^{N_{\mathrm{prim}}} \int_{0}^{\infty} r^2 \mathrm{d} r  \\ &\quad \quad \times \left(\Big\langle \mathcal{A}^{(\zeta)}_{( 1, m_a )}[ f^{(i)}] \,   \mathcal{A}^{(\zeta)}_{( 1, m_b )}[ g^{(i)}] \Big\rangle_{\mathrm{MC}} \, \mathcal{A}^{(h)}_{( 2, M )}[ h^{(i)}]  + 8 \ \mathrm{perm.} \right) (r, \bm{\hat{n}}) 
\, .
 \end{split}
 \label{estimator_lin_2}
\end{align}
\end{widetext}
We again assume an input shape function parameterized by Eq.~\eqref{factor_f}. 
The eight additional permutations in Eq.~\eqref{estimator_lin_2} are those constructed by cyclic  permutations of $f^{(i)}$, $g^{(i)}$, $h^{(i)}$ and  by varying which pair of $\mathcal{A}$'s sits in the $\langle \rangle_{\mathrm{MC}}$ brackets.

Similar to the total cubic term, it may be checked that including the two cyclic permutations of $\zeta \zeta h$ simply amounts to:
\begin{eqnarray}
\hat{f}_{\rm NL, lin}^{\mathrm{tot}} = 3 \hat{f}_{\rm NL, lin}^{\zeta \zeta h} \, .
\end{eqnarray}

Finally, the normalization of the estimator $\mathcal{I}_0$ may be estimated by simply applying the unnormalized estimator to an ensemble of simulated data. Given the expressions for the cubic and linear term presented here, the efficient algorithm from Ref.~\cite{smith2011algorithms} for the estimation of the normalization can also be used for this type of bispectrum. We omit the details of this implementation.

This concludes the derivation of the estimator for the $\zeta \zeta h$ $3$-point function. The resulting expression is given in Eq.~\eqref{estimator_cubic_2}. In Appendix~\ref{appendix:angular_templates} we show how one would repeat this effort for several more involved $3$-point functions.

\section{\label{sec:Fisher} Fisher Forecasts}

We forecast the expected uncertainty on an upper limit on the amplitude of a squeezed  $\zeta \zeta h$ $3$-point correlation function. We illustrate the constraining power of current and upcoming CMB experiments, and  demonstrate how the upper limit depends on certain instrumental effects. We expand on previous forecasts in Ref.~\cite{Meerburg:2016ecv, Domenech:2017kno} by taking into account the dependence on the lower harmonic band-limit of the data, the addition of $E$-mode data and the extra variance induced by weak lensing. In a future paper we apply the derived estimator to a set of map-based simulations to better judge the effects of foreground contamination, non-trivial noise covariances and secondary non-Gaussian contamination. In this light, the forecasts presented here should be considered as a baseline for more realistic forecasts.

\subsection{Procedure}

Before presenting the results from the Fisher forecasts, this section specifies the exact parameterization of the $\zeta \zeta h$ $3$-point function. We also explain the assumed experimental setup and the numerical implementation of  forecast calculation.

We parameterize the $k$-dependent part of the $\zeta \zeta h$ template in Eq.~\eqref{eq:b_template_sst} as follows:
\begin{align}\label{eq:fnl_def}
\begin{split}
f^{(\zeta\zeta h)}(k_1, k_2, k_3) =&\, 16 \pi^4 A_{s}^2 f^{\mathrm{tot}}_{\mathrm{NL}} \, f(k_1, k_2, k_3)   \, .
\end{split}
\end{align}
$A_s$ represents the amplitude of the curvature perturbation (see Appendix~\ref{sec:prim_temp_intro}). We imagine an analysis that looks for a deviation from the tensor consistency relation by placing an upper limit on the amplitude of the squeezed $3$-point function; we thus use the standard local shape as $f(k_1, k_2, k_3)$ template as a generic squeezed shape template. See Eq.~\eqref{eq:local_template} for the precise expression. The local shape differs slightly from the SFSR shape template \cite{Maldacena:2002vr} used in Ref.~\cite{Meerburg:2016ecv, Domenech:2017kno, Shiraishi:2017yrq}. However, the two templates give almost equal weight to squeezed configurations with a large-wavelength tensor perturbation. Given that the tensor perturbation only sources CMB anisotropies on large angular scales, we may, for all practical purposes, consider the shapes as equal here. This is reflected in the results we obtain: our forecasts agree with those in Ref.~\cite{Meerburg:2016ecv, Domenech:2017kno} when  parameters overlap.\footnote{The definition in Eq.~\eqref{eq:fnl_def} differs from the one used in Ref.~\cite{Meerburg:2016ecv, Lee:2016vti} by a factor $\sqrt{r}$: $f_{\mathrm{NL}}^{\mathrm{here}} = \sqrt{r} f_{\mathrm{NL}}^{\mathrm{there}}$, where $r$ is the tensor-to-scalar ratio. To compare our results to those in Ref.~\cite{Domenech:2017kno}, use  $f_{\mathrm{NL}}^{\mathrm{here}} = (\lambda_{sst} \epsilon)^{\mathrm{there}}$. }

For simplicity, we only consider the $\langle TTB \rangle$, $\langle EEB \rangle$, and $\langle TEB \rangle$ bispectra in the forecasts. We thus do not take into account the information contained in the $\langle TTT \rangle$, $\langle TTE \rangle$, $\langle TEE \rangle$, and $\langle EEE \rangle$ bispectra. The main justification for this choice is the associated extra cosmic variance due to the lack of a $B$-mode component. Additionally, it should be noted that the squeezed $\langle TTT \rangle$ bispectrum is expected to be relatively strongly contaminated by secondary non-Gaussian signal~\cite{Hill:2018ypf}. It is expected to be of limited use for our purpose; see the discussion in Sec.~\ref{sec:discussion}.

 We use the inverse Fisher information $\mathcal{I}_0$ as an estimate for the estimator variance. The $1\sigma$ upper limits that we will quote are simply given by $1 / \sqrt{\mathcal{I}_0}$. We calculate the Fisher information in the limit of no non-Gaussian signal contribution, i.e.\ we use Eq.~\eqref{eq:fisher_full_scal}. We further simplify the situation by assuming isotropic signal and noise covariances. The resulting diagonal covariance matrices, together with  the orthonormality relation of the Wigner $3$-$j$ symbols in Eq.~\eqref{eq:squared_3j} allow the Fisher information to be expressed in terms of angle-averaged bispectra. 
The effects from incomplete sky coverage are treated in a simplified manner by taking into account an increase in estimator variance proportional to the observed fraction of the sky ($f_{\mathrm{sky}}$). Given this trivial scaling, we assume $f_{\mathrm{sky}} = 1$ in all of the following. Finally, we use the lensed version of the CMB power spectra, but neglect the non-Gaussian aspects of CMB lensing. See the discussion in Sec.~\ref{sec:disc_var}. 

The resulting simplified expression for the Fisher information $\mathcal{I}_0$ is given by:
\begin{widetext}
\begin{align}\label{eq:fisher_iso}
\mathcal{I}_0 =   f_{\mathrm{sky}} \sum_{\ell_1 \leq \ell_2 \leq \ell_3 }   \sum_{\mathrm{all} X}  \frac{1}{\Delta_{\ell_1 \ell_2 \ell_3}} \left(B_1\right)_{\ell_1 \ell_2 \ell_3}^{X_1 X_2 X_3} 
\left[(C^{-1})^{X_1 X_4}_{\ell_1} (C^{-1})^{X_2 X_5}_{\ell_2 } (C^{-1})^{X_3 X_6}_{\ell_3 } \right] (B_1^{*})_{\ell_1 \ell_2 \ell_3}^{X_4 X_5 X_6} \, ,
\end{align}
with $(B_1^{*})_{\ell_1 \ell_2 \ell_3}^{X_1 X_2 X_3} = (B_1)_{\ell_1 \ell_2 \ell_3}^{X_1 X_2 X_3} (-1)^{\ell_1 + \ell_2 + \ell_3}$ and with total angle-averaged bispectrum given by:
\begin{align}\label{eq:tot_aa_bispec}
\left(B_1\right)_{\ell_1 \ell_2 \ell_3}^{X_1 X_2 X_3} = 
\left(B_1\right)_{\ell_1 \ell_2 \ell_3}^{X_1 X_2 X_3 (\zeta \zeta h)} + 
\left(B_1\right)_{\ell_1 \ell_2 \ell_3}^{X_1 X_2 X_3 (\zeta h \zeta)} +
\left(B_1\right)_{\ell_1 \ell_2 \ell_3}^{X_1 X_2 X_3 (h \zeta \zeta)} \, .
\end{align}
The factor $\Delta_{\ell_1 \ell_2 \ell_3}$ in Eq.~\eqref{eq:fisher_iso} simply results from using  $(1/6) \sum_{\ell_1, \ell_2, \ell_3} = \sum_{\ell_1 \leq \ell_2 \leq \ell_3} 1 / \Delta_{\ell_1 \ell_2 \ell_3}$ where $\Delta_{\ell_1 \ell_2 \ell_3}$ is defined to equal $6$ for identical $\ell$ indices, $1$ for unequal indices and $2$ otherwise. This simplification is possible because the bispectrum is invariant under all six permutations of its $(\ell, m)$ index pairs.\footnote{Note that the angle-averaged bispectrum used in Eq.~\eqref{eq:fisher_iso} is only invariant under cyclic permutations of $\ell_1$, $\ell_2$, and $\ell_3$. For odd permutations, it picks up a factor $(-1)^{\ell_1 + \ell_2 +\ell_3}$. Although we consider the $\ell_1+\ell_2+\ell_3 = \mathrm{odd}$ case here, the factors $(-1)$ cancel in the expression for the Fisher information, so we may still use the $1 / \Delta_{\ell_1 \ell_2 \ell_3}$ simplification.} Written as such, permutations of $\{X_1, X_2, X_3\}$, $\{X_4, X_5, X_6\}$, and $\{\zeta, h\}$ become distinct and have to be explicitly summed over.

As explained in Sec.~\ref{sec:full_bispec}, we may obtain the angle-averaged version of the $\zeta \zeta h$ bispectrum by summing over the $m_a$, $m_b$, $M$, $M_1$, $M_2$, and $M_3$ indices in Eq.~\eqref{eq:factored_sst_cmb} and inserting the resulting bispectrum into Eq.~\eqref{eq:isotropy_b}. 
This will yield the expression first derived in Ref.~\cite{shiraishi2011cmb}. The first term in Eq.~\eqref{eq:tot_aa_bispec} for the primordial shape in Eq.~\eqref{factor_f} is given by: 
\begin{align}\label{eq:ang_ave_w_beta}
\begin{split}
&\left(B_1\right)_{\ell_1 \ell_2 \ell_3}^{X_1 X_2 X_3 (\zeta \zeta h)} = \,  \frac{(8 \pi)^{3/2}}{3} \sum_{L_1, L_2, L_3} 
\left[ \prod_{i=1}^3   (-i)^{\ell_i-L_i}   \right]  J_{L_1 L_2 L_3 }^{000}  J_{\ell_1 L_1 1 }^{000}  J_{\ell_2 L_2 1 }^{000} J_{\ell_3 L_3 2 }^{20-2} 
 \begin{Bmatrix}\ell_1 & \ell_2 & \ell_3 \\ L_1 & L_2 & L_3 \\ 1 & 1 & 2 \end{Bmatrix} \\
 &\quad \quad \times \frac{1}{6} \sum_{i=1}^{N_{\mathrm{prim}}}   \int_{0}^{\infty} r^2 \mathrm{d}r \,  \left[ \big(\mathcal{K}^{(\zeta)}_{(X_1)}  [ f^{(i)} ]\big)_{\ell_1, L_1} \, \big(\mathcal{K}^{(\zeta)}_{(X_2)} [ g^{(i)} ]\big)_{\ell_2, L_2} \, \big(\mathcal{K}^{(h)}_{(X_3)} [ h^{(i)} ]\big)_{\ell_3, L_3}\right] (r) + (5 \ \mathrm{perm.})\, .
 \end{split}
\end{align}
\end{widetext}
The other two terms in Eq.~\eqref{eq:tot_aa_bispec} are obtained by permuting the $\zeta$ and $h$ indices. The five permuted terms in Eq.~\eqref{eq:ang_ave_w_beta} refer to permutations of the $f^{(i)}$, $g^{(i)}$ and $h^{(i)}$ functions. The $\mathcal{K}$ functionals were introduced in Eq.~\eqref{eq:radial_functional}.

The evaluation of Eq.~\eqref{eq:fisher_iso} has an overall $\mathcal{O}(\ell_{\mathrm{max}}^3)$ scaling. The computation is feasible because the $\mathcal{K}$ functionals in Eq.~\eqref{eq:ang_ave_w_beta} can be precomputed. However, for high band-limits (e.g.\ $\ell_{\mathrm{max}} = 5000$ used below) the procedure is unwieldy. This is especially true when multiple choices for the inverse signal+noise covariance matrix $C^{-1}$ are to be explored. The computation of multiple Wigner $9$-$j$ symbols at every valid $(\ell_1, \ell_2, \ell_3)$ triplet exacerbates the situation compared to the Fisher information for a $\zeta \zeta \zeta$ bispectrum.

To get around the computational complexity of Eq.~\eqref{eq:fisher_iso}, we split the problem in two parts: We first store a sparsely sampled representation of  Eq.~\eqref{eq:ang_ave_w_beta}. We then interpolate this representation over all multipole orders when the sums over $\ell_1$, $\ell_2$, and $\ell_3$ are performed. 
This approach results in an insignificant reduction in accuracy but reduces evaluation time significantly. Computing $\mathcal{I}_0$ with $\ell_{\mathrm{max}} = 5000$ takes roughly 30 CPU minutes. The method is effective because the smoothness of the primordial templates and transfer functions (in $k$ and $\ell$ respectively) translate into an angle-averaged bispectrum that is rather smooth with $\ell_1$, $\ell_2$, and~$\ell_3$.\footnote{This is only true when the factor $(-i)^{\ell_1 + \ell_2 + \ell_3}$ in Eq.~\eqref{eq:ang_ave_w_beta} is ignored. If required (for the cross-correlation of two different templates), this phase can be included after the interpolation step.}

The sparse sampling is determined by the following binning scheme: $\Delta \ell = 1$ for $\ell \leq 50$, $\Delta \ell = 4$ for $50 < \ell \leq 200$, $\Delta \ell = 12$ for $200 < \ell \leq 500$, $\Delta \ell = 24$ for $500 < \ell \leq 2000$, and finally $\Delta \ell = 40$ for $\ell > 2000$. This binning scheme is used for the $\ell_1$, $\ell_2$, and $\ell_3$ dimensions. In each resulting three-dimensional bin, a single valid sample (depending on the parity and triangle constraints) is selected. The angle-averaged bispectrum for each $(X_1, X_2, X_3)$ polarization tuple is then calculated over all selected samples. The integral over $r$ in Eq.~\eqref{eq:ang_ave_w_beta} is evaluated using the trapezoidal rule with $500$ integration points that span $0 \leq r \leq 18000$. Most points are placed around regions corresponding to the reionization and recombination eras. With some effort, we expect that the number of $r$ samples can be reduced by a factor of $10$. The resulting sparse, angle-averaged bispectra are compact enough to be saved to disk. Finally, to evaluate Eq.~\eqref{eq:fisher_iso} the sparse representations are interpolated over all valid multipole combinations using a three-dimensional linear interpolation scheme. The result is weighed by the (unbinned) inverse covariance matrices in Eq.~\eqref{eq:fisher_iso}.

The above algorithm is implemented in a  publicly available Python code library.\footnote{\href{https://github.com/adrijd/cmb_sst_ksw}{https://github.com/adrijd/cmb\_sst\_ksw}} The code makes heavy use of the scientific SciPy and NumPy libraries.\footnote{\href{https://www.scipy.org}{https://www.scipy.org}} performance-critical steps are compiled to optimized machine code at runtime by Numba: a just-in-time Python compiler~\cite{Lam:2015:NLP:2833157.2833162}. The Wigner symbols are evaluated using the WIGXJPF library~\cite{Johansson:2015cca}. The radiation transfer functions and CMB power spectra are computed using CAMB. Finally, every step of the code has been written with the Message Passing Interface (MPI) standard in mind;  computing in parallel on distributed memory systems is therefore possible. The code should be relatively easily adaptable to other (smooth) bispectrum templates. The repository also contains the necessary scripts to reproduce the results in the following section.

In summary, we use the Fisher information to forecast the expected upper-limits on the amplitude of the squeezed $\zeta \zeta h$ $3$-point function. The exact form of the $\zeta \zeta h$ correlation is specified in Eq.~\eqref{eq:b_template_sst} and Eq.~\eqref{eq:fnl_def} with the standard local shape template for $f(k_1, k_2, k_3)$.

\subsection{Results}

The results presented in this section fall into three categories. We first study how the expected upper-limits on the $\zeta \zeta h$ amplitude vary as function of upper and lower angular band limits. Second, we explore how advantageous it is to use both $T$ and $E$-mode data together with the $B$-mode data. Finally, we investigate the deterioration of the upper-limits due to gravitational lensing.

We start by exploring how the lower angular band-limit of the $B$-mode data affects the constraining power. The flat-sky forecasts in Ref.~\cite{Meerburg:2016ecv} did not probe this regime. The lowest achievable lower band-limit $\ell^B_{\mathrm{min}}$ is one of the main distinctions between ground-based and satellite CMB experiments. The atmosphere prohibits measurements over large angular scales. Current $B$-mode data from ground-based observatories reach $\ell^B_{\mathrm{min}} \approx 50$. Polarization modulation techniques, such as spinning half-wave plates, might allow  future efforts to reach an effective $\ell^B_{\mathrm{min}} \approx 30$ ~\cite{Ade:2018sbj}. Without atmospheric contamination satellite missions can in principle reach $\ell^B_{\mathrm{min}} =2$. In reality, it remains to be seen if uncertainty on systematic instrumental effects and Galactic foregrounds will allow such a challenging measurement to be made. A more conservative estimate for a satellite (or balloon-borne) experiment would be $\ell^B_{\mathrm{min}} \approx 20$. 

\begin{figure}[htbp]
   \centering
   \includegraphics[width=0.5\textwidth]{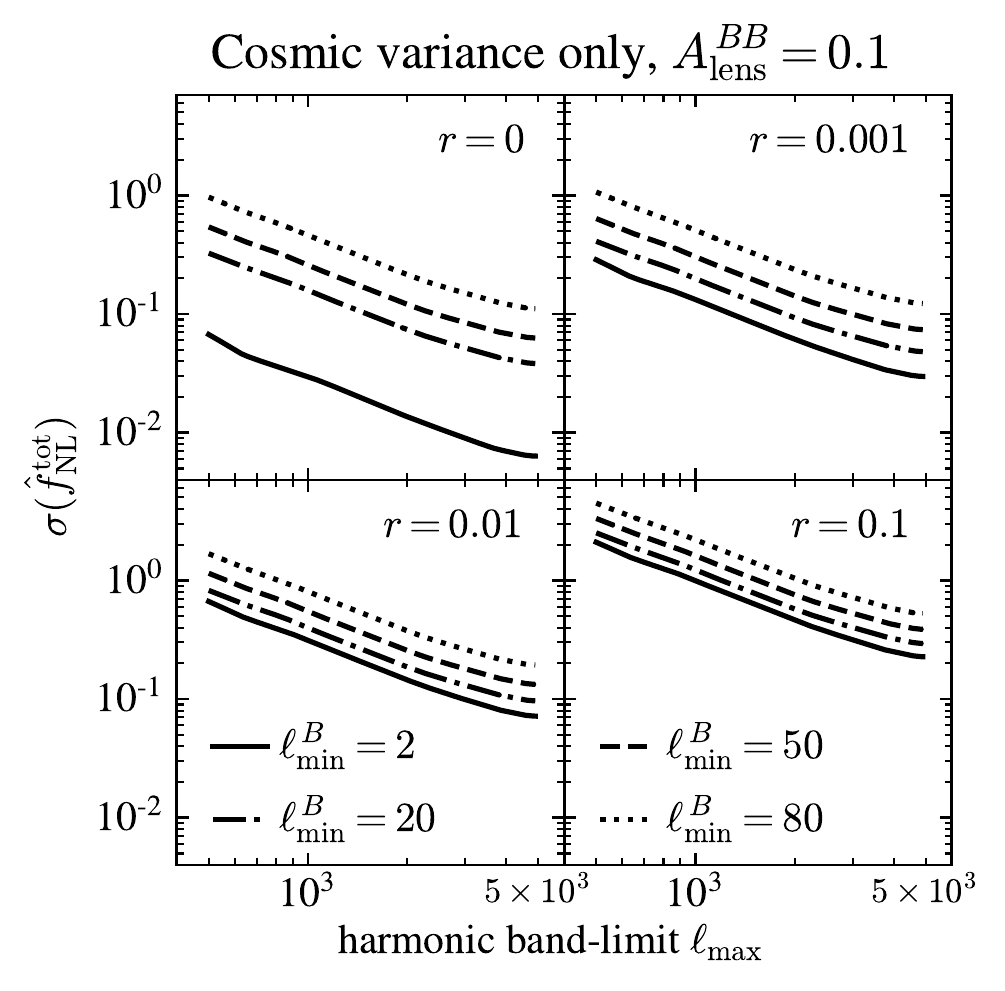}
   \caption{Achievable $1 \sigma$ upper limits on  $f_{\mathrm{NL}}^{\mathrm{tot}}$, i.e.\ the unavoidable errors solely caused by cosmic variance, as function of maximum harmonic band-limit $\ell_{\mathrm{max}}$. Here, the  $f_{\mathrm{NL}}^{\mathrm{tot}}$ parameter is the amplitude of the $\zeta \zeta h$ $3$-point function with a local shape function. The lines in each panel correspond to lower band-limits $\ell^{B}_{\mathrm{min}}$ of the $B$-mode data. The vast improvement due to low-multipole $B$-mode data seen in the upper-left panel is caused by the contribution from reionization to the bispectrum. When the tensor power is increased (the other three panels) the scaling with $\ell^{B}_{\mathrm{min}}$ becomes more regular: the contribution from reionization gets suppressed by the $B$-mode power spectrum. Still, the low multipole orders contain a significant amount of information on the $3$-point function. These results take into account the Fisher information in the $\langle TTB \rangle$, $\langle TEB \rangle$, and $\langle EEB \rangle$ CMB bispectra. The lensing contribution to the $B$-mode power spectrum is assumed to be `delensed' to only 10\% of the $\Lambda$CDM amplitude ($A_{\mathrm{lens}}^{BB} = 0.1$).}
   \label{fig:cv_lim}
\end{figure}

In Fig.~\ref{fig:cv_lim} we show the achievable $1\sigma$ upper limits on $f_{\mathrm{NL}}^{\mathrm{tot}}$ as function of overall band-limit $\ell_{\mathrm{max}}$ and lower band-limit $\ell^B_{\mathrm{min}}$. There is no contribution from instrumental noise, the only source of uncertainty is the cosmic variance induced by the Gaussian components of $\zeta$ and $h$. The lensing contribution to the $B$ power spectrum is assumed to be `delensed' to only 10\% of the $\Lambda$CDM amplitude ($A_{\mathrm{lens}}^{BB} = 0.1$). It is clear that as long as the Gaussian contribution to $h$ is neglected, i.e.\ $r=0$, the upper limits strongly benefit from a low $\ell_{\mathrm{min}}^{B}$. Scattering at reionization significantly contributes to the $\ell \alt 20$ $B$-mode components of the bispectrum for $r < 0.001$. The lensing contribution to $B$ is essentially negligible at such large angular scales, so the low-$\ell$ $B$-mode data become a highly sensitive probe of the squeezed bispectrum. When $r\neq 0 $, the additional cosmic variance induced by $h$ quickly closes this window, even though there still remains a significant dependence on $\ell_{\mathrm{min}}^{B}$ for $r \neq 0$. We find that for $r\geq 10^{-2}$, the $1\sigma$ upper limits conform rather well to the $\ell_{\mathrm{max}} (\log ( \ell^B_{\mathrm{max}} /  \ell^B_{\mathrm{min}}))^{1/2}$ scaling conjectured in Ref.~\cite{Bordin:2016ruc}. Here $\ell_{\mathrm{max}}$ refers to the band-limit of the $T$ and $E$-mode data, while $\ell^B_{\mathrm{max}}$ refers to the band-limit of the $B$-mode data. The scaling fits well when $\ell^B_{\mathrm{max}} \approx 150$: roughly the maximum multipole order that contains usable information on the primordial tensor perturbation for a 90\% delensed $B$-mode power spectrum. The curves in the two panels in Fig.~\ref{fig:cv_lim} that have $r< 10^{-2}$  do not fit the scaling: the relatively strong  contributions from reionization and lensing are not captured by the analytic relation.

\begin{figure}[htbp]
   \centering
   \includegraphics[width=0.5\textwidth]{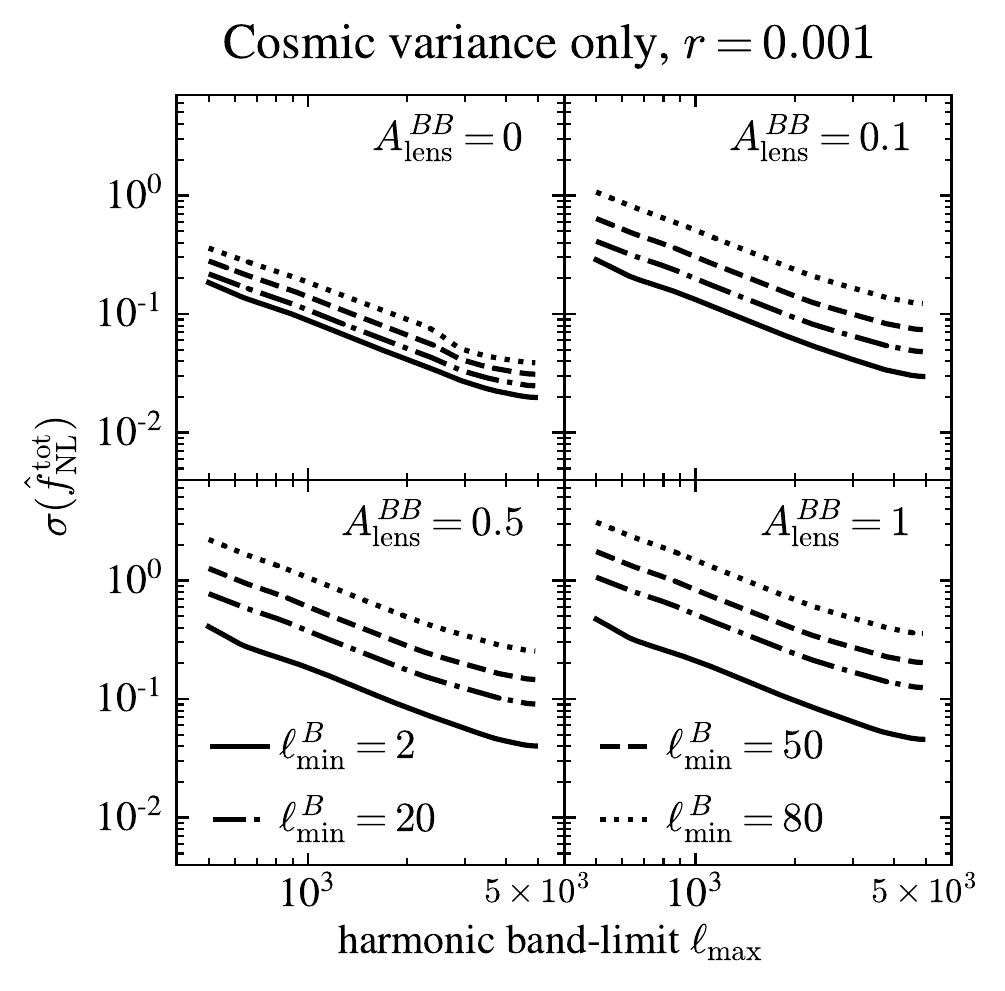}
   \caption{Cosmic variance limited $1 \sigma$ upper limits on  $f_{\mathrm{NL}}^{\mathrm{tot}}$ as function of maximum harmonic band-limit $\ell_{\mathrm{max}}$. Here, the  $f_{\mathrm{NL}}^{\mathrm{tot}}$ parameter is the amplitude of the $\zeta \zeta h$ $3$-point function with a local shape function. The lines in each panel correspond to lower band-limits $\ell^{B}_{\mathrm{min}}$ of the $B$-mode data.  As the lensing contribution to the $B$-mode power spectrum $A_{\mathrm{lens}}^{BB}$ is increased from the upper-left panel to the lower-right panel, upper limits worsen and become more dependent on the low-multipole $B$-mode data.
   These limits take into account the Fisher information in the $\langle TTB \rangle$, $\langle TEB \rangle$, and $\langle EEB \rangle$ CMB bispectra. The  tensor contribution to the CMB power spectra is sourced by an $r=0.001$ primordial tensor power spectrum.}
   \label{fig:cv_lim_alens}
\end{figure}

The relative importance of the low-$\ell$ $B$-mode data also grows when the lensing contribution to the $B$ power spectrum is increased. This behavior is depicted in Fig.~\ref{fig:cv_lim_alens}. As we move from no lensing $BB$ contribution to the full $\Lambda$CDM amplitude, the low-$\ell$ $B$-mode data becomes more relevant. This is a simple consequence of the shape of the lensing contribution relative to the bispectrum. The dominant lensing contribution to the estimator variance, i.e.\ the $B$ lensing power spectrum, is roughly constant with $\ell$ on large-scales while the $\langle TTB \rangle$, $\langle EEB \rangle$ and $\langle TEB \rangle$ bispectra peak at configurations with large-scale $B$-mode components.

Note that the lower band-limit used for the $T$ and $E$ data is set at $\ell=2$ for all results presented in this section. The rational behind this choice is that the $\emph{WMAP}$ and $\emph{Planck}$ data already provide cosmic-variance limited data for $T$ and $E$ on large angular scales. Note that this is not strictly true for the $E$-mode data. 
Current $\ell \alt 30$ $E$-mode data is systematic limited~\cite{Akrami:2018jnw, Aghanim:2018fcm}. We have checked that by conservatively removing the $\ell \leq 30$ $E$-mode data the curves do not visibly change.

\begin{figure}[htbp]
   \centering
   \includegraphics[width=0.5\textwidth]{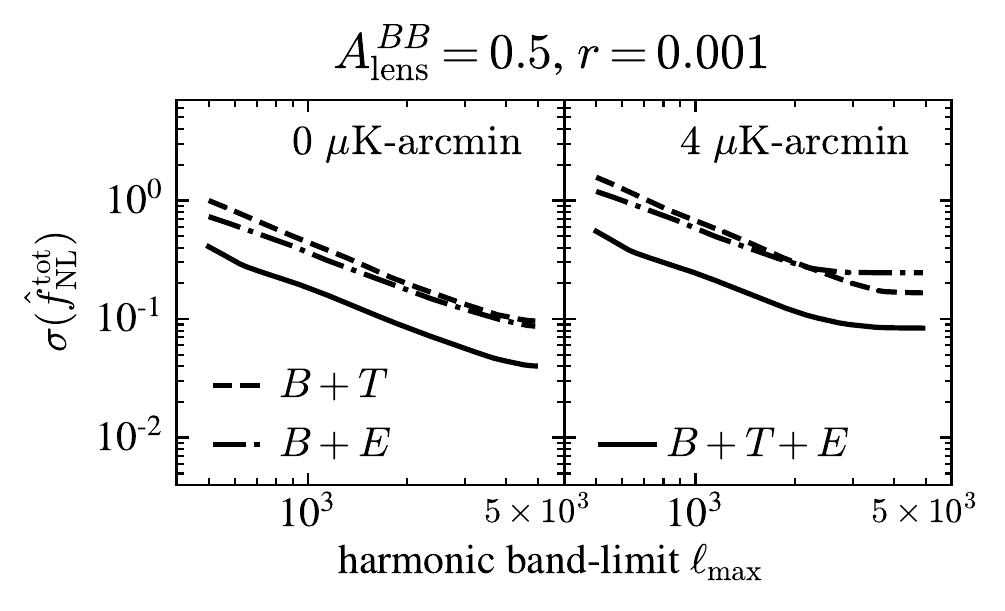}
   \caption{How the expected upper limits ($1\sigma$) on the amplitude $f_{\mathrm{NL}}^{\mathrm{tot}}$ of the $\zeta \zeta h$ $3$-point function change when $E$-mode data are excluded (dashed) or when $T$ data are excluded (dot-dashed). Combined constraints (i.e.\ from the Fisher information in the $\langle TTB \rangle$, $\langle TEB \rangle$, and $\langle EEB \rangle$ bispectra) (solid) are significantly stronger than those obtained from a naive addition of the $B+T$ and $B+E$ Fisher information. This effect holds when (white) noise is added to the data; the left panel shows the noiseless case, while in the right panel $4$ ($4\sqrt{2}$) $ \mu \mathrm{K}{\text -}\mathrm{arcmin}$ noise is added to the $T$ ($E$, $B$) harmonic modes. For these noise levels, the $T$ ($E$) data are cosmic-variance limited up to $\ell \approx 4000$ ($2500$). For data with higher band-limits ($\ell_{\mathrm{max}}$) the constraints saturate due to the noise. The addition of white noise to the $B$-mode data is responsible for the overall upward shift of the curves in the right panel. Note that the lower harmonic band-limit of the $B$-mode data is set to  $\ell=2$ for this figure.}
   \label{fig:pol}
\end{figure}

We now focus on the individual and combined contribution of the $\langle TTB \rangle$, $\langle TEB \rangle$ and $\langle EEB \rangle$ bispectra. In Ref.~\cite{Meerburg:2016ecv} only the $\langle TTB \rangle$ bispectrum was taken into account. Ref.~\cite{Domenech:2017kno} additionally calculated the Fisher information associated with the $\langle EEB \rangle$ bispectrum. In Fig.~\ref{fig:pol} we demonstrate how combining the information in $T$ and $E$ (in addition to $B$) yields much better results than the Fisher information of the individual cases would suggest. This effect is also seen in $\zeta \zeta \zeta$ non-Gausianity estimation and can be traced back to the fact that the $T$ and $E$ transfer fuctions for $\zeta$ are out of phase~\cite{Yadav:2005tf}. The same is true for the radial transfer functions we use, see Fig.~\ref{fig:alpha_scal}. This effect holds up under slightly more realistic circumstances: by adding $4\ \mu \mathrm{K}{\text -}\mathrm{arcmin}$ white noise to the $T$ harmonic modes and $4\sqrt{2} \ \mu \mathrm{K}{\text -}\mathrm{arcmin}$ to the $E$ and $B$ harmonic modes, we see the same behavior.

\begin{figure}[htbp]
   \centering
   \includegraphics[width=0.5\textwidth]{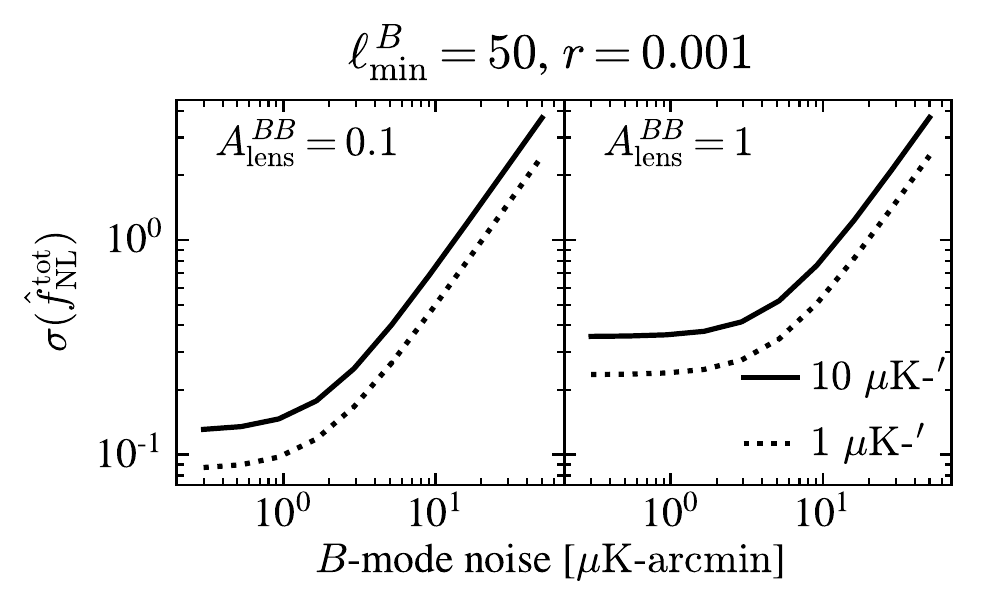}
   \caption{Expected upper limits ($1\sigma$) on the amplitude $f_{\mathrm{NL}}^{\mathrm{tot}}$ of the $\zeta \zeta h$ $3$-point function as function of the (white) noise amplitude of the $B$ harmonic modes. It can be seen how decreasing the $B$-mode noise is useful only up to a certain limit given by the amplitude of the lensing $BB$ contribution. For $B$-mode data that is `delensed' to only 10\% of the $\Lambda$CDM amplitude ($A_{\mathrm{lens}}^{BB} = 0.1$) (left panel), the constraining power saturates roughly below $1 \ \mu\mathrm{K}\text{-}\mathrm{arcmin}$. Without delensing, the constraints already start to saturate below $5\ \mu \mathrm{K}\text{-}\mathrm{arcmin}$. This behavior is essentially independent of the noise level of the $T$ and $E$ data: the same curve is seen regardless of whether $1$ ($\sqrt{2}$)  (solid) or $10$ ($10 \sqrt{2}$) (dotted) $ \mu \mathrm{K}{\text -}\mathrm{arcmin}$ noise is added to the $T$ ($E$) data. Note that the harmonic band-limit of the data is set to $\ell_{\mathrm{max}} = 5000$. }
   \label{fig:noise}
\end{figure}

Finally, we investigate the relation between lensing amplitude and instrumental noise level. As mentioned before, the lensing signal serves as a cosmic variance contribution to the estimator variance. The lensing contribution to the $T$ and $E$-mode power spectra provides a relatively minor contribution, while the contribution from the lensed $B$ power spectrum is significant. Fortunately, the lensing contribution to the $B$-mode field is not entirely irreducible: with knowledge of the lensing potential, the lensing contribution can be reduced, or `delensed'~\cite{Seljak:2003pn}. In Fig.~\ref{fig:noise} we show upper limits as function of instrumental $B$-mode noise for the case of only 10\% lensing contribution to the $B$-mode power spectrum ($A_{\mathrm{lens}}^{BB} = 0.1$) and for the full lensing contribution. The instrumental $B$-mode noise ranges from $50$ to $0.3$ $\mu\mathrm{K}\text{-}\mathrm{arcmin}$. To put this in context: the upper value roughly corresponds to the noise level in the \emph{Planck} data. The Simons Observatory~\cite{Ade:2018sbj} and \emph{LiteBIRD}~\cite{2018SPIE10698E..1YS} experiments aim to achieve a $B$-mode noise level of approximately $3$ $\mu\mathrm{K}\text{-}\mathrm{arcmin}$, while the CMB-S4 proposal~\cite{Abazajian:2016yjj} aims for approximately $1$ $\mu\mathrm{K}\text{-}\mathrm{arcmin}$. From the figure it becomes clear that the lensing $BB$ contribution starts to dominate over the instrumental noise for noise amplitudes below $5\ \mu\mathrm{K}\text{-}\mathrm{arcmin}$. This is unsurprising given that the large-scale $B$-mode lensing contribution is well approximated by $5\ \mu\mathrm{K}\text{-}\mathrm{arcmin}$ white noise~\cite{lewis2006weak}. We can thus infer that for the Simons Observatory or \emph{LiteBIRD} experiments the gain from $B$-mode delensing would be noticeable but relatively minor, while an experiment like CMB-S4 would need at least a factor of $10$ of delensing in the $B$-mode power to make use of the potential of the instrumental sensitivity.

In summary, the forecasts demonstrate that the statistical improvement with angular band-limit roughly follows the expected behavior for a squeezed $3$-point function, with the exception that a low $\ell_{\mathrm{min}}$ for the $B$-mode data is more advantageous than one would naively expect. Constraints benefit significantly from the simultaneous use of $T$ and $E$-mode data. Lastly, future experiments will need to delens their $B$-mode data significantly to keep improving upper-limits. It should be noted that these conclusions will likely  differ for shapes that are not squeezed.

\section{\label{sec:discussion}Discussion}

Generally, we expect two effects that will influence our ability to measure primordial non-Gaussianity. The first effect is a bias in the estimated amplitude of the primordial signal, i.e.\ a mismatch between the true amplitude and the expectation value of the estimate, due to other non-Gaussian signal that mimics the primordial signal.  Non-primordial, non-Gaussian signal is for example caused by secondary extragalactic sources and Galactic foregrounds. In some cases, these biases may be subtracted from the estimate or captured by a joint estimate, see e.g.\ the lensing-ISW bias in the \emph{Planck} analysis~\cite{Akrami:2019izv}. A second, more irreducible effect comes from the fact that non-Gaussian signal, primordial or secondary, will contribute to the estimator variance. When this contribution exceeds the contribution from (cosmic) variance from the Gaussian CMB component and detector noise, simulations of the responsible non-Gaussian signal are needed order to accurately characterize the estimator variance.  
While we will leave a detailed discussion of both effects to a future publication, here we provide a brief discussion. We focus on contaminants for squeezed bispectra with one large-scale $B$-mode component, as such bispectra will provide the largest constraining power for the primordial $\zeta \zeta h$ $3$-point function.

\subsection{Polarized Galactic foregrounds}
The large-scale polarization $B$- (and to lesser extent) $E$-mode fields are dominated by Galactic emission: at low frequencies by synchrotron radiation and at higher frequencies by polarized dust emission~\cite{Akrami:2018wkt}. Because the primordial $B$-mode signature is expected at large angular scales ($\ell \alt 100$), inference on the tensor-to-scalar ratio $r$ relies heavily on multi-frequency data to break the degeneracy between foreground and CMB power. Similarly, inference on the bispectra we are interested in would require uncontaminated large-scale $B$-mode data. 

One would naively expect that component-separated $B$-mode data suitable for constraints on $r$ is also suitable for constraints on bispectra with $B$-mode components. However, there is an extra complication for bispectrum inference: residual anisotropic or non-Gaussian correlations between foreground $B$ and foreground $T$ or $E$ signal. Residual correlations of this type might not be important for a power spectrum analysis but will bias a bispectrum analysis. Unfortunately, it is quite natural to expect  Galactic signal to source a squeezed bispectrum: small-wavelength foreground power in a given direction is likely not independent from the foreground signal on larger wavelengths in the same direction. The question is thus whether multi-frequency cleaning of the data will suppress such correlations enough.

Characterization of the non-Gaussian aspects of the polarized Galactic signal is relatively unexplored at this point. Early results obtained from the $\emph{Planck}$ data in Ref.~\cite{Coulton:2019bnz} suggest that there are indeed significant squeezed $\langle TTB \rangle$, $\langle TEB \rangle$, and $\langle EEB \rangle$ bispectra on large angular scales in the thermal dust component of the Galactic signal. No significant bispectrum is found in the synchrotron emission. Ref.~\cite{Coulton:2019bnz} does not find a  significant non-Gaussian correlation when foreground-cleaned \emph{Planck} $B$-mode data are correlated with the $T$ and/or $E$ components of the Galactic dust. Although this analysis omits the very large angular scales ($\ell \leq 40$), it does suggests that the standard component separation methods sufficiently suppress Galactic foregrounds given the \emph{Planck} noise level. It should also be noted that in a related study no evidence was found for a dust bispectrum template in the foreground-cleaned $\emph{Planck}$ temperature data~\cite{Jung:2018rgf}. More investigation is clearly still needed; just like it seems to be the case for inference on $r$, one would expect foreground uncertainty to be the limiting factor for inference on the $\zeta \zeta h$ $3$-point correlation function.

\subsection{Secondaries sourced by $\zeta$}

We now consider non-Galactic secondary non-Gaussian signals that are sourced by the curvature perturbation $\zeta$ (as opposed to $h$). 
We again focus on squeezed bispectra with a large wavelength $B$-mode, as such bispectra may bias the inference on the primordial signal.

The most well-studied secondary signal is sourced by the correlation between the late-time ISW effect and the lensing potential~\cite{Goldberg:1999xm}. A similar correlation exists between the quadrupole perturbation that sources the polarized reionzation signal and the lensing potential~\cite{Lewis:2011fk}.  
The ISW effect and the polarization generated at reionization only affect the CMB over large angular scales. On the other hand, the lensing potential modulates small-scale power. The associated bispectra are thus of the squeezed type. The ISW effect only affects the temperature anisotropies, the polarized reionization signal is purely $E$. This means that although $\langle TTB \rangle$, $\langle TEB \rangle$, $\langle EEB \rangle$ bispectra are produced~\cite{Hu:2000ee,Lewis:2011fk}, the only significant configurations will have large-scale $T$ or $E$-mode components instead of a $B$-mode component. 

In general, the requirement of a squeezed bispectrum with a large-scale $B$-mode contribution is highly constraining. There are no obvious (non-Galactic) candidates that preferentially source a $B$-mode signal on large angular scales. Non-linear effects other than lensing that produce $B$-mode signal, such as patchy reionization~\cite{2009PhRvD..79j7302D} and the polarized Sunyaev Zel'dovich (pSZ) effect~\cite{Kamionkowski:1997na, Deutsch:2017cja}, do so only at relatively small angular scales. Unclustered, extragalactic point sources may be weakly polarized and have a reduced bispectrum that is approximately constant with multipole order~\cite{Curto:2013hi}. They thus contaminate all bispectra, regardless of shape. However, especially for squeezed models, the point-source bias is found to be negligible: the two types of bispectra can be estimated independently~\cite{Curto:2013hi, Akrami:2019izv}.

\subsection{Secondaries sourced by $h$}

The $\zeta \zeta h$ $3$-point correlation function is contingent upon the existence of the primordial tensor perturbation $h$. For completeness, we thus  briefly discuss possible secondary non-Gaussian signal sourced by a purely Gaussian tensor perturbation $h$. 

In this case, the most obvious single-$B$-mode bispectrum candidate will be due to the interplay between two effects: 1) the standard correlation between the lensing potential $\phi$ and the ISW and polarized reionization signal, together with 2) the fact that lensing will now convert some of the (Gaussian) primordial $B$-mode signal to $E$-mode polarization~\cite{Okamoto:2003zw}. In the resulting $\langle B E X \rangle$ bispectrum, $B$ is the standard primordial $B$-mode signal, $E$ is the primordial $B$-mode signal lensed to an $E$-mode signal and $X$ is the standard scalar-induced $T$ or $E$ signal. To first order in the lensing potential, the bispectrum should be given by the triangular configurations of $C^{BB}_{\ell} C^{\phi X}_{\ell'}$. The suppression by $r$, due to the presence of the primordial $B$-mode power spectrum, makes this bispectrum lower in amplitude than the standard lensing-ISW bispectrum discussed in the previous section. More importantly however, the fact that the lensing-ISW and lensing-reionization correlation $C^{\phi X}_{\ell}$ is only nonzero for $\ell \alt 100$~\cite{Lewis:2011fk} means that there will be no significant bispectrum configurations with a large-scale $B$-mode component and two small-scale ($\ell > 100$) $T$ and/or $E$ components: the relevant configuration for a bias.

Analogous to the $E$-mode-lensing correlation in the previous section, the $B$-mode signal from reionization, present when $r \neq 0$, is also correlated to small-scale power through a correlation with the lensed signal. The difference is that isotropy and parity invariance forbid a correlation between $B$ and the regular gradient-type lensing potential. Instead the $B$-mode signal is correlated to the curl-type lensing potential sourced by the $h$ perturbation~\cite{Dodelson:2003bv, Cooray:2005hm}. 
Unlike the $\zeta \zeta \zeta$ case, there will now exist $\langle B X X'\rangle$ bispectra, where $B$ is the unlensed $B$-mode field and $X$ the curl-lensed $T$ or $E$-mode field (and $X' \in \{T,E \}$). To leading-order we expect such bispectra to be proportional to the triangular configurations of $ C^{B\omega}_{\ell} C^{XX'}_{\ell'}$, where $C^{B\omega}_{\ell}$ is the cross-correlation between the curl component of the lensing deflection angle and the reionization $B$ signal. 
The power spectrum of the tensor-induced $\omega$, i.e.\ $C^{\omega \omega}_{\ell}$, is strongly suppressed compared to scalar-induced lensing and decays rapidly for $\ell > 2$~\cite{Dodelson:2003bv, Cooray:2005hm}. One would expect similar behavior for the amplitude of $C^{B\omega}_{\ell}$ and thus expect that $C^{B\omega}_{\ell} C^{XX'}_{\ell'}$ is negligible. Still, the associated bispectra are maximized in the squeezed limit with a large-scale $B$-mode, so they should be considered as a potential bias to a primordial signal.

The tensor-induced temperature quadrupole on the last-scattering surface seen by galaxy clusters will source the pSZ effect~\cite{alizadeh_2012, Deutsch:2018umo}. The resulting small-scale power will be correlated with the primary $B$-mode field from reionization and will thus source a squeezed $\langle BEE \rangle$ bispectrum (among others). The $B$-mode component is on large angular scales, which means that the bispectrum has the right shape to be a potentially relevant contaminant of the primordial bispectrum.

\subsection{Contributions to the covariance}\label{sec:disc_var}

In the previous three sections we focused on possible biases to the estimator. All discussed effects will also contribute to the covariance of the estimate. Fortunately, in most cases these effects are subdominant to the Gaussian contribution to the covariance, given by the inverse of  Eq.~\eqref{eq:fisher_full_scal}. 
However, as we illustrate in Appendix~\ref{app:estimator_der}, the covariance of the estimator receives additional contributions from any connected $4$- and $6$-point correlation function present in the data. For example: for the $\zeta \zeta \zeta$, temperature-only bispectrum, the connected moments due to lensing will introduce significant additional covariance on small angular scales. The variance due to the connected $4$-point function alone is expected to dominate the cosmic-variance induced estimator variance for local-type non-Gaussianity for $\ell_{\mathrm{max}} \agt 3500$~\cite{Babich:2004yc} (and hence will be a concern for experiments like Simons Observatory and CMB-S4). The total effect on the estimator covariance will depend on the shape of the primordial bispectra that are estimated: local, or squeezed, shapes will likely be affected the most. 

We focus primarily on bispectra with a single $B$-mode component; in the previous sections we argued that such bispectra are less susceptible to secondary biases. However, this argument does not hold for the variance of the estimator:  when lensing is introduced, it is expected that the estimator covariance is affected in a way that is rather similar to the temperature-only case mentioned above. For example, consider the $ \langle TTB \rangle$ bispectrum; the variance of its estimate will be approximately proportional to the  $\langle TTTT BB \rangle$ $6$-point function. In the noiseless Gaussian case, this $6$-point function reduces to terms proportional to $C_{\ell}^{TT} C_{\ell'}^{TT} C_{\ell''}^{BB}$. When lensing is introduced, the power spectra are replaced by their lensed versions (which has a large effect on $C_{\ell}^{BB}$). However, there should also be a contribution proportional to the connected $\langle TTTT \rangle$ $4$-point function from lensing. One would expect this contribution to saturate the constraining power for $\ell_{\mathrm{max}} \agt 3500$, just like it does for the temperature-only case mentioned above. For the variance on estimates using the $\langle EEB \rangle$ or $\langle TEB \rangle$ bispectra a similar argument applies~\cite{lewis2006weak}. In other words, we expect that an estimate of the $\zeta \zeta h$ $3$-point function using high-resolution data will have large non-Gaussian contributions to its (co)variance, at least for squeezed bispectrum shapes with a $B$-mode contribution on large angular scales.\footnote{Because the effect should only become dominant for $\ell \agt 3500$ there should be negligible effect on primordial bispectra with more than one $B$-mode component and/or shapes that are more equilateral. In these cases, the signal drops sharply for $\ell_B \agt 200$.} Note that this non-Gaussian contribution to the variance is not included in the Fisher forecasts presented in Sec.~\ref{sec:Fisher}. 

In a future study we hope to identify all these contributions to the covariance and estimate their effects on our ability extract the primordial signal. We would like to note that, in principle, secondary biases and non-Gaussian contributions to the covariance from lensing can likely be reduced significantly by delensing~\cite{Green:2016cjr}. As some of the contributions to the covariance might be hard to compute analytically, applying the developed estimator on a suite of realistically lensed simulations would be an important aspect of such a study.

\section{\label{sec:conclusion} Conclusions}

The CMB bispectrum sourced by primordial scalar-tensor interactions is a well-defined observable that can be probed effectively with upcoming CMB polarization data. 
Inference on these types of primordial interactions probes non-standard early-universe models that are essentially unconstrained by current studies. In addition, inference on the squeezed $\zeta \zeta h$ $3$-point function provides a powerful consistency test of the standard inflationary paradigm.

In this work we derived a numerically efficient and optimal estimator for the amplitude of CMB bispectra sourced by primordial $\zeta \zeta h$ $3$-point correlation functions. We demonstrated that despite the intrinsic geometrical complexity of the bispectrum, an efficient estimator can be formulated, see Eq.~\eqref{estimator_cubic_2}. There is a limited computational overhead compared to standard $\zeta \zeta \zeta$ bispectrum estimation, see Eq.~\eqref{eq:ksw_scalar}, but the same asymptotic scaling with data resolution is reached. The derived estimator provides complementarity to the more general modal and binned bispectrum estimators~\cite{FergussonShellard2009,ShellardBispectrum2006,ShellardModeExpansion2009, bucher_2016, Shiraishi:2014roa} and should, due to its numerical advantage, be the preferred method for high-resolution data. 

We studied the bispectrum sourced by a squeezed $\zeta \zeta h$ $3$-point function in more detail. We presented a set of Fisher forecasts that form a baseline to which more realistic forecasts will be compared in future work. The presented forecasts demonstrate a relatively strong dependence on the size of the largest angular scale accessible in the data.  
We also demonstrated how constraints from the combination of temperature, $E$- and $B$-mode data are significantly better than those only from temperature and $B$-mode data or only from $E$- and $B$-mode data. Finally, we found that the lensing contribution to the $B$-mode data starts to significantly impact the constraints from experiments like the Simons Observatory and \emph{LiteBIRD}. For a more futuristic experiment like CMB-S4, delensing of the large-scale $B$-mode data will be crucial. 

Although the Fisher forecasts provide us with a good indication of the ultimate constraining power of future CMB experiments, future forecasts will need to include more realism. This requires applying the estimator directly to simulated sky maps. Besides allowing the characterization of standard complications like non-trivial noise properties and sky cuts, this approach is the appropriate way to study effects that are more specific to e.g.\ the $\zeta \zeta h$ bispectrum. 
Examples of such effects include the incomplete removal of Galactic $B$-mode signal or non-Gaussian polarized secondary sources. Lensed sky simulations will also allow one to quantify the expected extra estimator variance due to non-Gaussian $4$- and $6$-point correlation functions in the lensed CMB fields, as well as the effects of delensing these fields. Although current data are inconclusive, it seems likely that the eventual limit on future constraints will be from foreground uncertainty on large angular scales and the non-Gaussian lensing contribution on small-scales. Before this point is reached however, the data will contain a large amount of unexplored cosmological information. With an efficient estimator in hand, we should now turn towards map-based simulations to predict the exact amount of information.

In the next decade we will significantly improve our measurements of the CMB polarization field. With this in mind, we should consider interesting science targets beyond the tensor-to-scalar ratio that can provide insight into the early Universe. One of these targets is probing the primordial interactions between scalars and tensors as well as tensor self-interactions. Currently, the most sensitive probe of these interactions comes from including the $B$-mode field into CMB bispectrum inference. The work presented here is a contribution towards the development of a complete framework to constrain these interactions with upcoming CMB data.

\section*{Acknowledgments}
The authors like to thank Joel Meyers, Alex van Engelen, William Coulton, Eiichiro Komatsu and Sebastian Baum for useful discussions. P.D.M. acknowledges support from Senior Kavli
Institute Fellowships at the University of Cambridge and the Netherlands organization for scientific
research (NWO) VIDI grant (dossier 639.042.730). 
A.J.D. and K.F. acknowledge support by the Vetenskapsr\aa det (Swedish Research Council) through contract No.\ 638-2013-8993 and the Oskar Klein Centre for Cosmoparticle Physics. KF acknowledges support from the Jeff and Gail Kodosky Endowed Chair in Physics, DoE grant DE- SC007859 and the LCTP at the University of Michigan. Some computations have been performed at the Owl Cluster funded by the University of Oslo and the Research Council of Norway through grant 250672.

\begin{widetext}

\appendix

\section{Estimator for other angular terms}\label{appendix:angular_templates}

In this appendix we show how the $\hat{f}_{\mathrm{NL}}$ estimator for $3$-point functions with other angular terms. Besides providing a few useful examples, it can be seen how each estimator still asymptotically scales as $\mathcal{O}(\ell_{\mathrm{max}}^3)$. 
For each template we show the expression for the bispectrum and the cubic part of the estimator. As demonstrated in Sec.~\ref{sec:estimator_cubic}, it is straightforward to derive the linear term of the estimator given the cubic term. 

\subsection{Scalar-scalar-scalar}
\subsubsection{Standard scalar-only template}
For comparison and completeness, we first treat the standard $\zeta \zeta \zeta$ template, i.e.\ a template with no contracted angular term. Assuming a shape template like Eq.~\eqref{eq:fact_cond}, the expression for the bispectrum in Eq.~\eqref{eq:theoretical_bispectrum} simplifies to:
\begin{align}\label{eq:factored_sss_cmb}
\begin{split}
B_{m_1 m_2 m_3 X_1 X_2 X_3}^{ \ell_1 \ell_2 \ell_3 (\zeta \zeta \zeta)} =  \frac{1}{6} 
 \int_{S^2} \mathrm{d}\Omega (\bm{\hat{n}})  \sum_{i=1}^{N_{\mathrm{prim}}} \int_{0}^{\infty} \! r^2 \mathrm{d}r    \sum_{\ell_1 , m_1} \left[   \big( \mathcal{K}^{(\zeta)}_{(X_1)}[ f^{(i)} ]\big)_{\ell_1, \ell_1} (r) \right] Y_{\ell_1 m_1}(\bm{\hat{n}})&  \\  \quad \quad \quad \quad \times \sum_{\ell_2 , m_2} \left[   \big( \mathcal{K}^{(\zeta)}_{(X_2)}[ g^{(i)} ]\big)_{\ell_2, \ell_2} (r) \right] Y_{\ell_2 m_2}(\bm{\hat{n}})  \sum_{\ell_3 , m_3} \left[   \big( \mathcal{K}^{(\zeta)}_{(X_3)}[ h^{(i)} ]\big)_{\ell_3, \ell_3} (r) \right] Y_{\ell_3 m_3}(\bm{\hat{n}})&   + (5 \ \mathrm{perm.}) \, .
\end{split}
\end{align}
Note that the $5$ extra terms are permutations of the input functions $f$, $g$ and $h$. With this bispectrum, the cubic term of the estimator becomes:
\begin{align}
\begin{split}
\hat{f}_{\mathrm{NL, cubic}}^{\zeta \zeta \zeta} 
= \ \frac{1}{6\, \mathcal{I}_0}   
\int_{S^2} \mathrm{d}\Omega(\bm{\hat{n}}) \sum_{i=1}^{N_{\mathrm{prim}}} \!
 \int_{0}^{\infty} \! r^2 \mathrm{d}r \, \Big(&
 \mathcal{A}^{(\zeta)}_{( 0, 0)}[ f^{(i)}] \,
 \mathcal{A}^{(\zeta)}_{( 0, 0 )}[ g^{(i)}] \,
 \mathcal{A}^{(\zeta)}_{( 0, 0 )}[ h^{(i)}] 
 \Big) (r, \bm{\hat{n}})  \, ,
\label{estimator_cubic_sss}
\end{split}
\end{align}
which is the standard result~\cite{Yadav:2007rk}, but rephrased in our notation. See Eq.~\eqref{eq:a_map_def} and Eq.~\eqref{eq:a_harm_coeff} for the definition of the $\mathcal{A}_{(S,n)}$ functionals. In the $(S,n) = (0,0)$ case used here, the functionals are much less complicated: the $3$-$j$ symbols reduce to a delta function which simplifies the expression to: 
\begin{align}
\mathcal{A}^{(\zeta)}_{( 0, 0 )}[ f] \,  (r, \bm{\hat{n}})  = \sum_{\ell, m} (-1)^{\ell} \! \!  \sum\limits_{X\in\{T,E\}}   \big( \mathcal{K}^{(\zeta)}[ f ]\big)^X_{\ell, \ell} (r) \, \, (C^{-1}a)^{X}_{\ell m} \, Y_{\ell m} (\bm{\hat{n}})\, .
\end{align}
Note that the $(-1)^{\ell}$ factors are not present in the original expression~\cite{Yadav:2007rk}. They do not change the estimator, as only configurations with $\ell_1 + \ell_2 + \ell_3 = \mathrm{even}$ contribute. The $\mathcal{K}$ functionals are defined in Eq.~\eqref{eq:radial_functional}. 

\subsubsection{Scalar-only template with angular dependence of massive spinning particles}

The second $\zeta \zeta \zeta$ template is inspired by the three-point function template derived in Ref.~\cite{Lee:2016vti}. The template captures the imprint of a massive spin-$s$ field during inflation. Although the template only involves the curvature perturbation, it does include a contracted angular term:
\begin{align}
{}^{(000)}F(\bm{k}_1, \bm{k}_2, \bm{k}_3) &= \frac{1}{6} \,  f^{(\zeta\zeta \zeta)}(k_1, k_2, k_3) P_s(\bm{\hat{k}}_2 \cdot \bm{\hat{k}}_3) + (5 \ \mathrm{perm.}) \, ,
\\
&= \frac{1}{6} \sum_{i=1}^{N_{\mathrm{prim}}}  \, f^{(i)}(k_1) g^{(i)}(k_2) h^{(i)}(k_3) P_s(\bm{\hat{k}}_2 \cdot \bm{\hat{k}}_3)   + (5 \ \mathrm{perm.}) \, .
\end{align}
$P_s$ is a Legendre polynomial of degree $s$. The five additional permutations are permutations of the three wave vectors. 

In order to write down the corresponding bispectrum, we expand the Legendre polynomial in terms of spherical harmonics:
\begin{align}
P_s(\bm{\hat{k}} \cdot \bm{\hat{k}}') = \frac{4 \pi}{2 s + 1} \sum_{m' = -s}^s Y_{s m'} (\bm{\hat{k}}) \, Y^*_{s m'} (\bm{\hat{k}}') \, .
\end{align}
The bispectrum for a spin-$s$ template then becomes:
\begin{align}
\begin{split}
B_{m_1 m_2 m_3 X_1 X_2 X_3}^{ \ell_1 \ell_2 \ell_3 (\zeta \zeta \zeta)} =  \frac{4 \pi}{6(2s + 1)}
\sum_{m'=-s}^s (-1)^{m'}  
 \int_{S^2} \mathrm{d}\Omega (\bm{\hat{n}})  \sum_{i=1}^{N_{\mathrm{prim}}} \int_{0}^{\infty} \! r^2 \, \mathrm{d}r \left[ \big( \mathcal{K}^{(\zeta)}_{(X_1)}[ f^{(i)} ]\big)_{\ell_1, \ell_1} (r) \right]& Y_{\ell_1 m_1}(\bm{\hat{n}}) \\
 \times  \sum_{L_2 , M_2} \left[  i^{\ell_2+L_2}  J_{s L_2 \ell_2}^{000}  
\begin{pmatrix}s & L_2 & \ell_2 \\ -m' & M_2 & m_2 \end{pmatrix} \big( \mathcal{K}^{(\zeta)}_{(X_2)}[ g^{(i)} ]\big)_{\ell_2, L_2} (r) \right]& Y_{L_2 M_2}(\bm{\hat{n}}) \\
 \times  \sum_{L_3 , M_3} \left[ i^{\ell_3+L_3}  
J_{s L_3 \ell_3}^{0 0 0} 
\begin{pmatrix}s & L_3 & \ell_3 \\ m' & M_3 & m_3 \end{pmatrix} \big( \mathcal{K}^{(\zeta)}_{(X_3)}[ h^{(i)} ]\big)_{\ell_3, L_3} (r) \right]&  Y_{L_3 M_3}(\bm{\hat{n}}) \, \\ + (5 \ \mathrm{perm.}) \, .
\end{split}
\end{align}
The five additional terms are obtained by simultaneously permuting $f^{(i)}$, $g^{(i)}$, and $h^{(i)}$ with the $1$, $2$, and $3$ indices. The cubic term of the estimator for this bispectrum is given by:
\begin{align}
\begin{split}
\hat{f}_{\mathrm{NL, cubic}}^{\zeta \zeta \zeta} 
= \ \frac{1}{18\, \mathcal{I}_0}   \frac{4\pi}{2s+1} \sum_{m'=-s}^s (-1)^{m'}
\int_{S^2} \mathrm{d}\Omega(\bm{\hat{n}}) \sum_{i=1}^{N_{\mathrm{prim}}} \!
 \int_{0}^{\infty} \! r^2 \mathrm{d}r \, \Big(&
 \mathcal{A}^{(\zeta)}_{( 0, 0)}[ f^{(i)}] \,
 \mathcal{A}^{(\zeta)}_{( s, -m' )}[ g^{(i)}] \,
 \mathcal{A}^{(\zeta)}_{( s, m' )}[ h^{(i)}] 
 \Big) (r, \bm{\hat{n}})  \\ &+ (2 \ \mathrm{cyclic}) \, .
\label{estimator_cubic_sss_spin}
\end{split}
\end{align}
 The two extra terms are cyclic permutations of $f^{(i)}$, $g^{(i)}$, and $h^{(i)}$.
 
\subsection{Scalar-tensor-tensor}

To illustrate the situation for a scalar-tensor-tensor $3$-point function, we use a template inspired by the SFSR result~\cite{Maldacena:2002vr}:
\begin{align}
{}^{(0\lambda_2 \lambda_3)}F(\bm{k}_1, \bm{k}_2, \bm{k}_3) &=  f^{(\zeta h h)}(k_1, k_2, k_3) e^{\lambda_2}_{ab} (\bm{\hat{k}}_2)\, e_{\lambda_3}^{ab} (\bm{\hat{k}}_3)  \, ,
\\
&=  \sum_{i=1}^{N_{\mathrm{prim}}}  \, f^{(i)}(k_1) g^{(i)}(k_2) h^{(i)}(k_3)  e^{\lambda_2}_{ab} (\bm{\hat{k}}_2)\, e_{\lambda_3}^{ab} (\bm{\hat{k}}_3)\label{eq:stt_template}    \, .
\end{align}
The polarization tensors $e^{\pm2}$ are  defined in Eq.~\eqref{eq:pol_tens_cc} and Eq.~\eqref{eq:pol_tens_ortho}, the $a$ and $b$ indices run over the three spatial dimensions. We use the  Einstein summation convention. Using the notation from  Ref.~\cite{shiraishi2011cmb}, we may expand the polarization tensors as:
\begin{align}\label{eq:pol_tens_exp}
e^{\pm 2}_{ab} = \frac{3}{\sqrt{2 \pi}} \sum_{M, m_a, m_b} \! {}_{\mp 2}Y^*_{2 M} \alpha^{m_a}_{a} \alpha^{m_b}_{b}
\begin{pmatrix} 2 & 1 & 1 \\ M & m_a & m_b \end{pmatrix} \, .
\end{align}
The $\alpha$ coefficients obey the following orthogonality relation:
\begin{align}\label{eq:alpha_ortho}
\alpha_{a}^{m} \alpha^{b}_{m'} = \frac{4\pi}{3} (-1)^m \delta_{m}^{-m'} \, .
\end{align}
Using this relation together with the orthogonality relation of the Wigner $3$-$j$ symbols in Eq.~\eqref{eq:3j_orth_2}, the contraction of two polarization tensors can be expressed as follows: 
\begin{align}
e^{\lambda}_{ab} (\bm{\hat{k}})\, e_{\lambda'}^{ab} (\bm{\hat{k}}') = \frac{8 \pi}{5} \sum_{m'=-2}^2 (-1)^{m'} {}_{-\lambda}Y^*_{2 - m'} (\bm{\hat{k}}) \, {}_{-\lambda'}Y^*_{2 m'} (\bm{\hat{k}}') \, .
\end{align}
The bispectrum corresponding to the template in Eq.~\eqref{eq:stt_template} thus becomes:
\begin{align}
\begin{split}
B_{m_1 m_2 m_3 X_1 X_2 X_3}^{ \ell_1 \ell_2 \ell_3 (\zeta h h)} =  \frac{8 \pi}{5}
\sum_{m'=-2}^2 (-1)^{m'}
 \int_{S^2} \mathrm{d}\Omega (\bm{\hat{n}})  \sum_{i=1}^{N_{\mathrm{prim}}} \int_{0}^{\infty} \! r^2 \, \mathrm{d}r \left[ \big( \mathcal{K}^{(\zeta)}_{(X_1)}[ f^{(i)} ]\big)_{\ell_1, \ell_1} (r) \right]& Y_{\ell_1 m_1}(\bm{\hat{n}}) \\
 \times  \sum_{L_2 , M_2} \left[  i^{\ell_2+L_2}  J_{2 L_2 \ell_2}^{-202} [1 + (-1)^{x_2 + L_2 + \ell_2}]
\begin{pmatrix} 2 & L_2 & \ell_2 \\ -m' & M_2 & m_2 \end{pmatrix} \big( \mathcal{K}^{(h)}_{(X_2)}[ g^{(i)} ]\big)_{\ell_2, L_2} (r) \right]& Y_{L_2 M_2}(\bm{\hat{n}}) \\
 \times  \sum_{L_3 , M_3} \left[ i^{\ell_3+L_3}  
J_{2 L_3 \ell_3}^{-2 0 2} [1 + (-1)^{x_3 + L_3 + \ell_3}]
\begin{pmatrix} 2 & L_3 & \ell_3 \\ m' & M_3 & m_3 \end{pmatrix} \big( \mathcal{K}^{(h)}_{(X_3)}[ h^{(i)} ]\big)_{\ell_3, L_3} (r) \right]&  Y_{L_3 M_3}(\bm{\hat{n}}) \, .
\end{split}
\end{align}
The cubic part of the estimator is given by:
\begin{align}
\begin{split}
\hat{f}_{\mathrm{NL, cubic}}^{\zeta  h h} 
= \ \frac{4 \pi}{15\, \mathcal{I}_0} \sum_{m'=-2}^2 (-1)^{m'}
\int_{S^2} \mathrm{d}\Omega(\bm{\hat{n}}) \sum_{i=1}^{N_{\mathrm{prim}}} \!
 \int_{0}^{\infty} \! r^2 \mathrm{d}r \, \Big(&
 \mathcal{A}^{(\zeta)}_{( 0, 0)}[ f^{(i)}] \,
 \mathcal{A}^{(h)}_{( 2, -m' )}[ g^{(i)}] \,
 \mathcal{A}^{(h)}_{( 2, m' )}[ h^{(i)}] 
 \Big) (r, \bm{\hat{n}}) \, .
\label{estimator_cubic_stt}
\end{split}
\end{align}
The expressions for the $h \zeta h$ and $h h \zeta$ parts are derived in an analogous way.

\subsection{Tensor-tensor-tensor}
Finally, we derive the estimator for a tensor-tensor-tensor $3$-point function. We again take the SFSR prediction~\cite{Maldacena:2002vr} as inspiration for our template:
\begin{align}
{}^{(\lambda_1 \lambda_2 \lambda_3)}F(\bm{k}_1, \bm{k}_2, \bm{k}_3) &=  f^{(h h h)}(k_1, k_2, k_3) \nonumber \\ &\quad \quad \quad \times \left[ \hat{k}^{a}_2 \hat{k}^b_{2} e^{\lambda_1}_{ab} (\bm{\hat{k}}_1)\, e_{\lambda_2}^{cd} (\bm{\hat{k}}_2) e^{\lambda_3}_{cd} (\bm{\hat{k}}_3) - 2 e^{\lambda_1}_{ab}(\bm{\hat{k}}_1) e^{\lambda_2}_{cd}(\bm{\hat{k}}_2)
e_{\lambda_3}^{bc}(\bm{\hat{k}}_3)
\hat{k}^{a}_2  \hat{k}^{d}_3  \right]  + (2 \ \mathrm{cyclic}) \, , \\
&=  \sum_{i=1}^{N_{\mathrm{prim}}}  \, f^{(i)}(k_1) g^{(i)}(k_2) h^{(i)}(k_3) \nonumber \\ &\quad \quad \quad \times \left[ \hat{k}^{a}_2 \hat{k}^b_{2} e^{\lambda_1}_{ab} (\bm{\hat{k}}_1)\, e_{\lambda_2}^{cd} (\bm{\hat{k}}_2) e^{\lambda_3}_{cd} (\bm{\hat{k}}_3) - 2 e^{\lambda_1}_{ab}(\bm{\hat{k}}_1) e^{\lambda_2}_{cd}(\bm{\hat{k}}_2)
e_{\lambda_3}^{bc}(\bm{\hat{k}}_3)
\hat{k}^{a}_2  \hat{k}^{d}_3 \right] + (2 \ \mathrm{cyclic}) \label{eq:ttt_template}\, .
\end{align}
The two extra terms are cyclic permutations of the three wave vectors. 

To derive the bispectrum, we need to expand the unit wave vectors in spherical harmonics~\cite{shiraishi2011cmb}:
\begin{align}
\hat{k}^a = \sum_{m} \alpha^a_{m} Y_{1 m} (\bm{\hat{k}}) \, .
\end{align}
The $\alpha$ coefficients obey the relation in Eq.~\eqref{eq:alpha_ortho}. Together with Eq.~\eqref{eq:pol_tens_exp}, Eq.~\eqref{eq:product_swshs}, and Eq.~\eqref{eq:3j_orth_2} we then expand the first angular term in Eq.~\eqref{eq:ttt_template} as follows:
\begin{align}
\begin{split}
\hat{k}^{a}_2 \hat{k}^b_{2} e^{\lambda_1}_{ab} (\bm{\hat{k}}_1)\, e_{\lambda_2}^{cd} (\bm{\hat{k}}_2) e^{\lambda_3}_{cd} (\bm{\hat{k}}_3) = \frac{64 \pi^2}{75} & \! \sum_{L, M, M', M''} J_{2 2 L}^{0 -\lambda_2 \lambda_2} \begin{pmatrix} 2 & 2 & L \\ M & M'' & M' \end{pmatrix} \\ &\times {}_{-\lambda_1}Y^*_{2 M} (\bm{\hat{k}}_1)
\, {}_{-\lambda_2}Y^*_{LM'} (\bm{\hat{k}}_2)
\, {}_{-\lambda_3}Y^*_{2 M''} (\bm{\hat{k}}_3) \, .
\end{split}
\end{align}
The $J$ symbols are defined in Eq.~\eqref{eq:prefactor_gaunt}. The second angular term in Eq.~\eqref{eq:ttt_template} is expressed in terms of Wigner $6$-$j$ symbols by making use of the relation in Eq.~\eqref{eq:6j_3j}, see also Ref.~\cite{Shiraishi:2011st}. The resulting expression is:
\begin{align}
\begin{split}
e^{\lambda_1}_{ab}(\bm{\hat{k}}_1) e^{\lambda_2}_{cd}(\bm{\hat{k}}_2)
e_{\lambda_3}^{bc}(\bm{\hat{k}}_3)
\hat{k}^{a}_2  \hat{k}^{d}_3 = \frac{(8\pi)^{5/2}}{6} & \! \! 
\sum_{\substack{L, J \\ M, M', M''}}  (-1)^{L+1} J_{21L}^{\lambda_2 0 -\lambda_2} J_{21J}^{ \lambda_3 0 - \lambda_3} \begin{pmatrix} J & L & 2 \\ M'' & M' & M \end{pmatrix}
\begin{Bmatrix} J & L & 2 \\ 1 & 1 & 2 \end{Bmatrix}
\begin{Bmatrix} 1 & 2 & J \\ 1 & 2 & 1 \end{Bmatrix}\\ &\times 
{}_{-\lambda_1}Y^*_{2 M} (\bm{\hat{k}}_1)
\, {}_{-\lambda_2}Y^*_{LM'} (\bm{\hat{k}}_2)
\, {}_{-\lambda_3}Y^*_{JM''} (\bm{\hat{k}}_3) \, .
\end{split}
\end{align}
It is convenient to separate the corresponding bispectrum into a part sourced by the first angular term and a part sourced by the second term. The first part is given by:
\begin{align}
\begin{split}
&B_{m_1 m_2 m_3 X_1 X_2 X_3}^{ \ell_1 \ell_2 \ell_3 (h h h, 1)} =  \frac{64 \pi^2}{75}
\! \sum_{L, M, M', M''}  \begin{pmatrix} 2 & 2 & L \\ M & M'' & M' \end{pmatrix}
 \int_{S^2} \mathrm{d}\Omega (\bm{\hat{n}})  \sum_{i=1}^{N_{\mathrm{prim}}} \int_{0}^{\infty} \! r^2 \, \mathrm{d}r \\ 
 &\times \sum_{L_1 , M_1} \left[  i^{\ell_1+L_1}  J_{2 L_1 \ell_1}^{-202} [1 + (-1)^{x_1 + L_1 + \ell_1}]
\begin{pmatrix} 2 & L_1 & \ell_1 \\ M & M_1 & m_1 \end{pmatrix} \big( \mathcal{K}^{(h)}_{(X_1)}[ f^{(i)} ]\big)_{\ell_1, L_1} (r) \right] Y_{L_1 M_1}(\bm{\hat{n}}) \\
 &\times  \sum_{L_2 , M_2} \left[  i^{\ell_2+L_2}  J_{2 L_2 \ell_2}^{-202} J_{2 2 L}^{0 2 -2} [1 + (-1)^{x_2 + L_2 + \ell_2 + L}]
\begin{pmatrix} L & L_2 & \ell_2 \\ M' & M_2 & m_2 \end{pmatrix} \big( \mathcal{K}^{(h)}_{(X_2)}[ g^{(i)} ]\big)_{\ell_2, L_2} (r) \right] Y_{L_2 M_2}(\bm{\hat{n}}) \\
 &\times  \sum_{L_3 , M_3} \left[ i^{\ell_3+L_3}  
J_{2 L_3 \ell_3}^{-2 0 2} [1 + (-1)^{x_3 + L_3 + \ell_3}]
\begin{pmatrix} 2 & L_3 & \ell_3 \\ M'' & M_3 & m_3 \end{pmatrix} \big( \mathcal{K}^{(h)}_{(X_3)}[ h^{(i)} ]\big)_{\ell_3, L_3} (r) \right]  Y_{L_3 M_3}(\bm{\hat{n}})\\ &+ (2\ \mathrm{cyclic}) \, .
\end{split}\label{eq:b_ttt_1}
\end{align}
The two extra terms are given by cyclic permutations of the $f^{(i)}$, $g^{(i)}$, and $h^{(i)}$ input functions together with the $1$, $2$, and $3$ indices. The second part is given by:
\begin{align}
\begin{split}
&B_{m_1 m_2 m_3 X_1 X_2 X_3}^{ \ell_1 \ell_2 \ell_3 (h h h, 2)} =  \frac{(8\pi)^{5/2}}{3}
\!  \! 
\sum_{\substack{L, J \\ M, M', M''}} (-1)^{L+1}  \begin{pmatrix} J & L & 2 \\ M'' & M' & M \end{pmatrix}
\begin{Bmatrix} J & L & 2 \\ 1 & 1 & 2 \end{Bmatrix}
\begin{Bmatrix} 1 & 2 & J \\ 1 & 2 & 1 \end{Bmatrix}
 \int_{S^2} \mathrm{d}\Omega (\bm{\hat{n}})  \sum_{i=1}^{N_{\mathrm{prim}}} \int_{0}^{\infty} \! r^2 \, \mathrm{d}r \\ 
 &\times \sum_{L_1 , M_1} \left[  i^{\ell_1+L_1}  J_{2 L_1 \ell_1}^{-202} [1 + (-1)^{x_1 + L_1 + \ell_1}]
\begin{pmatrix} 2 & L_1 & \ell_1 \\ M & M_1 & m_1 \end{pmatrix} \big( \mathcal{K}^{(h)}_{(X_1)}[ f^{(i)} ]\big)_{\ell_1, L_1} (r) \right] Y_{L_1 M_1}(\bm{\hat{n}}) \\
 &\times  \sum_{L_2 , M_2} \left[  i^{\ell_2+L_2}  J_{2 L_2 \ell_2}^{-202} J_{2 1 L}^{-2 0 2} [1 + (-1)^{x_2 + L_2 + \ell_2 + L + 1}]
\begin{pmatrix} L & L_2 & \ell_2 \\ M' & M_2 & m_2 \end{pmatrix} \big( \mathcal{K}^{(h)}_{(X_2)}[ g^{(i)} ]\big)_{\ell_2, L_2} (r) \right] Y_{L_2 M_2}(\bm{\hat{n}}) \\
 &\times  \sum_{L_3 , M_3} \left[ i^{\ell_3+L_3}  
J_{2 L_3 \ell_3}^{-2 0 2} J_{2 1 J}^{-2 0 2} [1 + (-1)^{x_3 + L_3 + \ell_3 + J +1}]
\begin{pmatrix} J & L_3 & \ell_3 \\ M'' & M_3 & m_3 \end{pmatrix} \big( \mathcal{K}^{(h)}_{(X_3)}[ h^{(i)} ]\big)_{\ell_3, L_3} (r) \right]  Y_{L_3 M_3}(\bm{\hat{n}}) \\ &+ (2\ \mathrm{cyclic}) \, .
\end{split}\label{eq:b_ttt_2}
\end{align}

The cubic estimator is also most easily expressed in two parts. The part corresponding to the first bispectrum, Eq.~\eqref{eq:b_ttt_1}, is given by:
\begin{align}
\begin{split}
\hat{f}_{\mathrm{NL, cubic}}^{h  h h, 1} 
= \ \frac{32 \pi^2}{225\, \mathcal{I}_0} 
\! \sum_{L, M, M', M''}  &\begin{pmatrix} 2 & 2 & L \\ M & M'' & M' \end{pmatrix} J_{2 2 L}^{0 2 -2}
\int_{S^2} \mathrm{d}\Omega(\bm{\hat{n}}) \sum_{i=1}^{N_{\mathrm{prim}}} \!
 \int_{0}^{\infty} \! r^2 \mathrm{d}r \, \\
 & \quad \quad \quad  \times \Big(
 \mathcal{A}^{(h)}_{( 2, M)}[ f^{(i)}] \,
 \mathcal{B}^{(h)}_{( L, M' )}[ g^{(i)}] \,
 \mathcal{A}^{(h)}_{( 2, M'' )}[ h^{(i)}] 
 \Big) (r, \bm{\hat{n}}) + (2\ \mathrm{cyclic}) \, .
\label{estimator_cubic_ttt_1}
\end{split}
\end{align}
 The two extra terms are cyclic permutations of $f^{(i)}$, $g^{(i)}$, and $h^{(i)}$. The second part, corresponding to Eq.~\eqref{eq:b_ttt_1} is given by:
\begin{align}
\begin{split}
\hat{f}_{\mathrm{NL, cubic}}^{h  h h, 2} 
= \ \frac{(8\pi)^{5/2}}{18\, \mathcal{I}_0} 
\!  \! &
\sum_{\substack{L, J \\ M, M', M''}} \! (-1)^{L+1} \begin{pmatrix} J & L & 2 \\ M'' & M' & M \end{pmatrix} J_{2 1 L}^{-2 0 2} J_{2 1 J}^{-2 0 2}
\begin{Bmatrix} J & L & 2 \\ 1 & 1 & 2 \end{Bmatrix}
\begin{Bmatrix} 1 & 2 & J \\ 1 & 2 & 1 \end{Bmatrix} \\
& \times
\int_{S^2} \mathrm{d}\Omega(\bm{\hat{n}}) \sum_{i=1}^{N_{\mathrm{prim}}} \!
 \int_{0}^{\infty} \! r^2 \mathrm{d}r \, 
  \Big(
 \mathcal{A}^{(h)}_{( 2, M)}[ f^{(i)}] \,
 \mathcal{C}^{(h)}_{( L, M' )}[ g^{(i)}] \,
 \mathcal{C}^{(h)}_{( J, M'' )}[ h^{(i)}] 
 \Big) (r, \bm{\hat{n}}) + (2\ \mathrm{cyclic}) \, .
\label{estimator_cubic_ttt_2}
\end{split}
\end{align}
We have introduced the $\mathcal{B}$ and $\mathcal{C}$ functionals. They are completely analogous to the $\mathcal{A}$ functionals, defined in Eq.~\eqref{eq:a_map_def} and Eq.~\eqref{eq:a_harm_coeff}, but slightly differ in their spherical harmonic coefficients:
\begin{align}
\left(\mathcal{B}^{(h)}_{( S, n )}[ f]   \right)_{LM} \! (r) &\equiv (4\pi)^{1/2} \sum\limits_{\ell, m}  i^{\ell+L}  
J_{S L \ell}^{- 2 0 2}  \begin{pmatrix}S & L & \ell \\ n & M & m \end{pmatrix} \! \!
 \sum\limits_{X}  \left[1 + (-1)^{x + L + \ell + S} \right] \!
\big( \mathcal{K}^{(h)}[ f ]\big)^X_{\ell, L} (r) \, \,  (C^{-1}a)^{X}_{\ell m} \, , \\
\left(\mathcal{C}^{(h)}_{( S, n )}[ f]   \right)_{LM} \! (r) &\equiv (4\pi)^{1/2} \sum\limits_{\ell, m}  i^{\ell+L}  
J_{S L \ell}^{- 2 0 2}  \begin{pmatrix}S & L & \ell \\ n & M & m \end{pmatrix} \! \!
 \sum\limits_{X}  \left[1 + (-1)^{x + L + \ell + S + 1} \right] \!
\big( \mathcal{K}^{(h)}[ f ]\big)^X_{\ell, L} (r) \, \,  (C^{-1}a)^{X}_{\ell m} \, .
\end{align}
Computing the combination of Eq.~\eqref{estimator_cubic_ttt_1} and Eq.~\eqref{estimator_cubic_ttt_1} will still asymptotically scale as $\mathcal{O}(\ell_{\mathrm{max}}^3)$. Although more terms have to be computed compared to the previous templates, this computational overhead is easily outweighed by the fact that the $h$ transfer functions impose $\ell_{\mathrm{max}}\approx 200$.

\section{\label{sec:appendix_fourier} Useful mathematical identities}

\subsection{Spin-weighted spherical harmonics}\label{sec:appendix_swsh}

The spin-weighted spherical harmonics (SWSHs) ${}_s Y_{\ell m }$ are generalizations of the standard spherical harmonics $Y_{\ell m }$.  Both types of spherical harmonics are functions on the sphere $S^2$. 
Indeed, one may relate:
\begin{equation}
{}_0 Y_{\ell m } =  Y_{\ell m } \, .
\end{equation}
The relation between the two sets of functions for nonzero $s$ can be found in the literature~\cite{newman_1966, goldberg_1967}.

The SWSHs are conveniently defined on the standard spherical coordinate system by taking the Wigner $D$-matrices (irreps of the three-dimensional rotation group) parameterized in terms of the Euler angles and fixing the polar axis as follows: 
\begin{align}\label{sylm_as_dlms}
{}_s Y_{\ell m } (\theta, \phi) = \left. (-1)^{m} \sqrt{ \frac{2\ell +1}{4 \pi}} D^{\ell}_{-m s } (\phi, \theta, \psi)  \right|_{\psi = 0} \, .
\end{align}
With a slight abuse of notation, we use $\bm{\hat{n}}$ in the arguments of the spherical harmonics to refer to the $\theta$ and $\phi$ angles that describe the spherical decomposition of the 3D unit vector, i.e.\ $\hat{n} = (\sin \theta \cos \phi, \sin \theta \sin \phi, \cos \theta)$. Similarly, we denote the differential solid angle with $\mathrm{d}\Omega (\bm{\hat{n}})$, i.e.\ $\int_{S^2} \mathrm{d}\Omega (\bm{\hat{n}}) \equiv \int_{0}^{2\pi} \mathrm{d}\phi \int_{0}^{\pi} \mathrm{d}\theta \sin \theta $.

The functions form an orthonormal and complete system for each integer\footnote{Throughout this work we only describe (representation of) 3D rotations so we limit ourselves to (non-negative) integer multipole order ($\ell$) and integer magnetic or `azimuthal' numbers ($m$ and~$s$).} spin weight~$s$:
\begin{align}
\int_{S^2} \mathrm{d}\Omega (\bm{\hat{n}}) \, {}_s Y_{\ell m } (\bm{\hat{n}}) \, {}_s Y^*_{\ell' m' } (\bm{\hat{n}}) &= \delta_{\ell \ell'} \delta_{m m'}  \, , \\
\sum_{\ell, m} {}_s Y_{\ell m } (\bm{\hat{n}}) \, {}_s Y^*_{\ell m } (\bm{\hat{n}'}) &=  \delta (\cos \theta - \cos \theta ') \delta(\phi - \phi ')   \, .
\end{align}
This leads to the following forward and inverse transformations for (square-integrable) spin-weighted functions on the sphere:
\begin{align}
\begin{split}
{}_{s}f_{\ell m} &= \int_{S^2} \mathrm{d}\Omega (\bm{\hat{n}}) \, {}^{(s)}f(\bm{\hat{n}}) \, {}_{s}Y^*_{\ell m} (\bm{\hat{n}}) \quad \forall \ell \in \{|s|, \dots , \ell_{\mathrm{max}}\}, \forall m \in \{-\ell, \ell \} \, , \\
{}^{(s)}f(\bm{\hat{n}}) &= \sum_{\ell = |s|}^{\ell_{\mathrm{max}}} \sum_{m = -\ell}^{\ell} {}_{s}f_{\ell m} \, {}_{s}Y_{\ell m} (\bm{\hat{n}}) \quad \forall \bm{\hat{n}}\in S^2 \, .
\end{split}
\end{align}

We include the Condon-Shortley phase convention in our definition of the SWSHs. Under complex conjugation and parity ($(\theta, \phi) \mapsto (\pi - \theta, \phi + \pi)$) the functions therefore obey:
\begin{align}
{}_s Y_{\ell m}^* (\bm{\hat{n}}) &= (-1)^{s+m}  {}_{-s} Y_{\ell -m} (\bm{\hat{n}}) \, , \\
{}_s Y_{\ell m} (-\bm{\hat{n}}) &= (-1)^{\ell}\, {}_{-s} Y_{\ell m} (\bm{\hat{n}}) \, .
\end{align}
In particular, this implies that ${}_{s} f_{\ell m}^* = {}_{-s} f_{\ell -m} (-1)^{m+s}$ holds for two spin-weighted functions ${}^{(\pm s)} f$ that obey $({}^{(s)}f)^* = {}^{(-s)} f$. For $s=0$ this simply means that $f$ is a real-valued function.

A tensor product of SWSHs may be decomposed into a direct sum by making use of the Wigner $3$-$j$ symbols (see next section):
\begin{align}\label{eq:product_swshs}
{}_{s_1} Y_{\ell_1 m_1} (\bm{\hat{n}}) \, {}_{s_2} Y_{\ell_2 m_2} (\bm{\hat{n}}) =  \sum_{\ell_3 = |\ell_1 - \ell_2|}^{\ell_1 + \ell_2} \sum_{m_3 = -\ell_3}^{\ell_3} \sum_{s_3=-\ell_3}^{\ell_3} J_{\ell_1 \ell_2 \ell_3}^{-s_1 -s_2 -s_3} \begin{pmatrix}\ell_1 & \ell_2 & \ell_3 \\ m_1 & m_2 & m_3 \end{pmatrix} {}_{s_3}Y^*_{\ell_3 m_3} (\bm{\hat{n}}) \, ,
\end{align}
with:
\begin{align}\label{eq:prefactor_gaunt}
J_{\ell_1 \ell_2 \ell_3}^{s_1 s_2 s_3} \equiv \sqrt{  \frac{ (2\ell_1 +1) (2\ell_2 +1) (2\ell_3 +1) }{4 \pi} } \, \begin{pmatrix}\ell_1 & \ell_2 & \ell_3 \\ s_1 & s_2 & s_3 \end{pmatrix}.
\end{align}
Note that the upper limit on the first sum in Eq.~\eqref{eq:product_swshs} implies that the harmonic band-limit of a product of functions on the sphere is given by the sum of their individual band-limits. Eq.~\eqref{eq:product_swshs} also shows that the integral over a product of three SWSHs is given by:
\begin{align}
\begin{split}
\int_{S^2} \mathrm{d} \Omega(\bm{\hat{n}}) \, {}_{s_1} Y_{\ell_1 m_1} (\bm{\hat{n}}) \, {}_{s_2} Y_{\ell_2 m_2} (\bm{\hat{n}}) \, {}_{s_3} Y_{\ell_3 m_3} (\bm{\hat{n}}) = J_{\ell_1 \ell_2 \ell_3}^{-s_1 -s_2 -s_3} \, \begin{pmatrix}\ell_1 & \ell_2 & \ell_3 \\ m_1 & m_2 & m_3 \end{pmatrix} \, .
\end{split}
\label{ext_gaunt_fin}
\end{align}
Although, note that this only holds for the $s_1+s_2+s_3=0$ case. For the $s_1=s_2=s_3=0$ case this integral is referred to as the Gaunt integral.

\subsection{Wigner $3$-$j$, $6$-$j$, and $9$-$j$ symbols}\label{app_wigner}

The Wigner $3$-$j$ symbols are real-valued and serve to describe the decomposition of tensor products of SWSHs into direct sums of SWSHs (see Eq.~\eqref{eq:product_swshs}) (this also holds, in more generality, for irreps of the rotation group like the Wigner-$D$ matrices). The $3$-$j$ symbols are closely related to the Clebsch-Gordan coefficients but are  normalized such that they are the exact coefficients needed to form a rotationally invariant product of three SWSH coefficients (recall the definition of the angle-averaged bispectrum in Eq.~\eqref{eq:isotropy_b}). In the following, we list a limited number of  symbol properties; see Ref.~\cite{edmonds_1957} for an exhaustive description. 

The $3$-$j$ symbols pick up a (real) phase factor when the sign of the three `magnetic' indices is simultaneously changed:
\begin{align} \label{wigner3j:parity}
\begin{pmatrix}\ell_1 & \ell_2 & \ell_3 \\ m_1 & m_2 & m_3 \end{pmatrix} = (-1)^{\ell_1 + \ell_2 + \ell_3}\begin{pmatrix}\ell_1 & \ell_2 & \ell_3 \\ -m_1 & -m_2 & -m_3 \end{pmatrix}.
\end{align}
The symbols are invariant under cyclic permutations of $m_1$, $m_2$, and $m_3$ but pick up a factor $(-1)^{\ell_1+\ell_2+\ell_3}$ for anti-cyclic permutations.
The symbols are only nonzero for $m_1 + m_2 + m_3 = 0$, $|\ell_1 - \ell_2| \leq  \ell_3 \leq \ell_1 + \ell_2$, and  $|m_i| \leq \ell_i$ $\forall i \in \{1, 2, 3 \}$. There are two orthogonality relations:
\begin{align}
\sum_{L,M} (2 L + 1)
\begin{pmatrix}\ell_1 & \ell_2 & L \\ m_1 & m_2 & M \end{pmatrix}
\begin{pmatrix}\ell_1 & \ell_2 & L \\ m'_1 & m'_2 & M \end{pmatrix} &= \delta_{m_1 m'_1} \delta_{m_2 m'_2} \, ,\\
\sum_{m_1,m_2}
\begin{pmatrix}\ell_1 & \ell_2 & L \\ m_1 & m_2 & M \end{pmatrix}
\begin{pmatrix}\ell_1 & \ell_2 & L' \\ m_1 & m_2 & M' \end{pmatrix} &= \frac{\delta_{L L'} \delta_{M M'}}{2 L + 1} \label{eq:3j_orth_2} \, .
\end{align}
In particular, in the case of equal symbols one has: 
\begin{align}\label{eq:squared_3j}
\sum_{m_1=-\ell_1}^{\ell_1} \sum_{m_2=-\ell_2}^{\ell_2} \sum_{m_3=-\ell_3}^{\ell_3} \begin{pmatrix}\ell_1 & \ell_2 & \ell_3 \\ m_1 & m_2 & m_3 \end{pmatrix}^2 = 1  \, .
\end{align}

As mentioned, the Wigner $3$-$j$ symbols are used to couple two SWSHs or, equivalently, find the third angular state that  combines two SWSHs into a rotationally invariant quantity. In general, there is no unique way to couple three SWSHs; there are two distinct sequences of applying Eq.~\eqref{eq:product_swshs} to the product. The Wigner $6$-$j$ symbol is used to transform between these two possible final angular states~\cite{edmonds_1957}:
\begin{align}
\begin{split}
\sum_{L}(2 L + 1) (-1)^{\ell_1 + \ell_3 + m_1 + m_4} 
\begin{Bmatrix} \ell_1 & \ell_2 & \ell_4 \\ \ell_3 & \ell_5 & L \end{Bmatrix}
\begin{pmatrix} \ell_1 & L & \ell_5  \\ m_1 & M & m_5 \end{pmatrix} 
\begin{pmatrix} \ell_3 & L & \ell_2  \\ -m_3 & M & m_2 \end{pmatrix} \\
= \begin{pmatrix} \ell_1 & \ell_4 & \ell_2  \\ m_1 & m_4 & -m_2 \end{pmatrix} 
\begin{pmatrix} \ell_3 & \ell_4 & \ell_5  \\ m_3 & -m_4 & -m_5 \end{pmatrix} \, .
\end{split}
\end{align}
By using one of the orthogonality relations of the $3$-$j$ symbols, the $6$-$j$ symbol may equivalently be expressed as:
\begin{align}\label{eq:6j_3j}
\begin{split}
\begin{Bmatrix} \ell_1 & \ell_2 & \ell_3  \\  \ell_4 & \ell_5 & \ell_6 \end{Bmatrix} 
\begin{pmatrix} \ell_1 & \ell_2 & \ell_3  \\  m_1 & m_2 & m_3 \end{pmatrix}
= \! \! \sum_{m_4, m_5, m_6} \! \!
&(-1)^{\sum_{i=4}^6 \ell_i + m_i}
\begin{pmatrix} \ell_1 & \ell_5 & \ell_6  \\  m_1 & m_5 & -m_6 \end{pmatrix} \!
\begin{pmatrix}  \ell_4 & \ell_2 & \ell_6 \\ -m_4 & m_2 & m_6   \end{pmatrix} \! \\ &\times
\begin{pmatrix} \ell_4 & \ell_5 & \ell_3  \\  m_4 & -m_5 & m_3 \end{pmatrix} \, .
\end{split}
\end{align}
The $6$-$j$ symbols are invariant under all permutations of their columns and under the simultaneous permutation of upper and lower arguments in two columns. The symbols also obey several triangle conditions that can be deduced from the top rows of each of the $3$-$j$ symbols in the above expression. There also exist an orthogonality relation for the $6$-$j$ symbols~\cite{edmonds_1957}.

Finally, the Wigner $9$-$j$ symbols are defined to describe the transformation between different couplings of four SWSHs. The symbols may either be expressed in terms of $6$-$j$ or $3$-$j$ symbols~\cite{edmonds_1957}. The latter expression is given by:
\begin{align}\label{eq:9j_in_3j}
\begin{split}
\begin{Bmatrix} \ell_1 & \ell_2 & \ell_3  \\  \ell_4 & \ell_5 & \ell_6 \\  \ell_7 & \ell_8 & \ell_9 \end{Bmatrix} 
\begin{pmatrix} \ell_1 & \ell_2 & \ell_3  \\  m_1 & m_2 & m_3 \end{pmatrix}
= \! \! \sum_{m_4, \dots, m_9} \! \!
&
\begin{pmatrix} \ell_4 & \ell_5 & \ell_6  \\  m_4 & m_5 & m_6 \end{pmatrix} \!
\begin{pmatrix}  \ell_7 & \ell_8 & \ell_9 \\ m_7 & m_8 & m_9   \end{pmatrix} \!
\begin{pmatrix} \ell_4 & \ell_7 & \ell_1  \\  m_4 & m_7 & m_1 \end{pmatrix} \,  \\ &\times 
\begin{pmatrix}  \ell_5 & \ell_8 & \ell_2 \\ m_5 & m_8 & m_2   \end{pmatrix} \!
\begin{pmatrix} \ell_6 & \ell_9 & \ell_3  \\  m_6 & m_9 & m_3 \end{pmatrix} \,  \, .
\end{split}
\end{align}
The $9$-$j$ symbols are invariant under reflections of their arguments along either diagonal and even permutations of rows or columns; odd permutations result in a factor $(-1)^{\sum_{i =1}^9 \ell_i}$. Elements of each row and column are constrained by the triangle conditions of the $3$-$j$ symbols in the above expression. There also exists an orthogonality relation for the $9$-$j$ symbols; details can be found in Ref.~\cite{edmonds_1957}.

\subsection{Delta function}\label{sec:app_delta}
The delta function is expanded as:
\begin{align}
\delta^{(3)}(\bm{k}_1 + \bm{k}_2 + \bm{k}_3) = \frac{1}{(2\pi)^3} \int \mathrm{d}^3 \bm{x} \,  e^{i (\bm{k}_1 + \bm{k}_2 + \bm{k}_3) \cdot \bm{x}} \, .
\end{align}
By making use of the Rayleigh equation:
\begin{align}
e^{i \bm{k} \cdot \bm{x}} =&  \sum_{\ell} i^{\ell} (2 \ell + 1) j_{\ell} (kr) P_{\ell}(\bm{\hat{k}} \cdot \bm{\hat{n}}) \, , \\
=& \, 4\pi  \sum_{\ell, m} i^{\ell} j_{\ell}(kr) Y^*_{\ell m} (\bm{\hat{k}}) Y_{\ell m} (\bm{\hat{n}}) \, ,
\end{align}
we produce two equivalent expressions for the delta function:
\begin{align}
 \delta^{(3)}(\bm{k}_1 + \bm{k}_2 + \bm{k}_3) =&  \, 8 \int_{0}^{\infty} r^2 \mathrm{d}r \, \sum_{\ell_1 , m_1}  \sum_{\ell_2 , m_2}  \sum_{\ell_3 , m_3} \left[ \prod_{i=1}^3  
 j_{\ell_i} (k_i r) Y^*_{\ell_i m_i}(\bm{\hat{k}}_i)  \right] \int_{S^2} \mathrm{d}\Omega(\bm{\hat{n}}) \, \left[\prod_{i=1}^3 Y_{\ell_i, m_i}(\bm{\hat{n}}) \right] \, , \\
 =& \, 8 \int_{0}^{\infty} r^2 \mathrm{d}r 
 \sum_{\ell_1 , m_1}  \sum_{\ell_2 , m_2}  \sum_{\ell_3 , m_3}
 \left[ \prod_{i=1}^3  i^{\ell_1} 
 j_{\ell_i} (k_i r) Y^*_{\ell_i m_i}(\bm{\hat{k}}_i)  \right] J_{\ell_1 \ell_2 \ell_3 }^{000} \begin{pmatrix} \ell_1 & \ell_2 & \ell_3 \\ m_1 & m_2 & m_3 \end{pmatrix} \, .
\end{align}
Note that $\bm{x}=r\bm{\hat{n}}$ and $\bm{k}=k\bm{\hat{k}}$. We have used Eq.~\eqref{ext_gaunt_fin} to arrive at the second expression. The $J$ symbol is defined in Eq.~\eqref{eq:prefactor_gaunt}. See Ref.~\cite{Mehrem:2009ip} for these and alternative expressions.

\section{\label{sec:templates} Cosmology conventions}

\subsection{Power spectra}\label{sec:prim_temp_intro}

Due to the assumed statistical homogeneity of the super-horizon $2$-point correlation functions of the amplitudes of both the spatial components of $\zeta$ and ${}^{(\lambda)}h$, these correlations are represented as diagonal in the 3D Fourier basis. Statistical isotropy limits the diagonal to only depend on the wavenumber $k$. We use the following conventions for the correlation functions:
\begin{align}
\langle \zeta_{\bm{k}} \, \zeta^*_{\bm{k}'}\rangle &= (2 \pi)^3 \delta^{(3)}(\bm{k} - \bm{k}')  P_{\zeta}(k) \, , \\
\langle {}^{(\lambda)}h_{\bm{k}} \,  {}^{(\lambda') }h^*_{\bm{k}'} \rangle &= (2 \pi)^3 \delta^{(3)}(\bm{k} - \bm{k}')   \delta_{\lambda, \lambda' } \frac{P_{h}(k)}{2} \, .
\end{align}
The power spectra are parameterized as follows:
\begin{align}
P_{\zeta}(k) &= 2 \pi^2 \, \frac{A_s(k_{0})}{k^3} \left(\frac{k}{k_{0}}\right)^{n_s(k_{0}) -1} \, , \\
P_{h}(k) &= 2 \pi^2 \, \frac{r_{k_0} A_s(k_{0})}{k^3} \left(\frac{k}{k_{0}}\right)^{n_t(k_{0})} \, ,
\end{align}
with tensor-to-scalar ratio $r_{k_0}$ (i.e.\ the ratio at the pivot scale), scalar amplitude $A_s$, pivot scale $k_0$ and scalar (tensor) spectral tilt $n_s$ ($n_t$). We have used fixed values for some of these parameters: $\{A_s(k_0) = 2.1056\cdot 10^{-9}, k_0 = 0.05 \ \mathrm{Mpc}^{-1}, n_s(k_0) = 0.9665, n_t(k_0) = 0 \}$. The remaining cosmological parameters that govern the radiation transfer functions  are set to $\{T_{\mathrm{CMB}} = 2.7255\ \mathrm{K}, H_0 = 67.66 \ \mathrm{km}\, \mathrm{s}^{-1}\, \mathrm{Mpc}^{-1}, \Omega_{\mathrm{b}} h^2 = 0.02242, \Omega_{\mathrm{c}} h^2 = 0.11933, \tau = 0.0561 \}$, and the CAMB defaults of December 2018.

We extract the radiation transfer functions from CAMB. We normalize the default output from CAMB such that the CMB power/cross spectra are related to the primordial power spectra defined above as:
\begin{align}\label{eq:cl_pk}
\left \langle a^{(Z)}_{X, \ell m} a^{(Z)*}_{Y, \ell' m'} \right\rangle  &= \delta_{\ell \ell'} \delta_{m m'} C^{(Z)}_{XY, \ell} \, , \\
& =  \delta_{\ell \ell'} \delta_{m m'} \frac{2}{\pi}\int_{0}^{\infty} \! k^2 \mathrm{d}k \, P_{Z}(k) \mathcal{T}^{(Z)}_{X,\ell}(k)  \mathcal{T}^{(Z)}_{Y,\ell}(k) \, ,
\end{align}
with $Z\in \{\zeta, h\}$ and $XY \in \{TT, EE, TE, ET, BB \}$.

\subsection{Local $3$-point correlation function}\label{app:fact_leo_shapes}
The local shape template used in Sec.~\ref{sec:Fisher} is given by~\cite{ade2016planck}:
\begin{align}
f^{\mathrm{local}}(k_1, k_2, k_3) &= 2\left[ \left( \frac{1}{(k_1 k_2)^3}\right) + 2\ \mathrm{perm.}  \right] \label{eq:local_template} \, .
\end{align}
The template is symmetric under permutations of the three wavenumbers and perfectly scale-invariant (i.e. proportional to $k^{-6}$ for $k_1=k_2=k_3$). If desired, including the scalar or tensor spectral tilt simply amounts to the replacement $k \mapsto k (k_0/k)^{(n_s-1)/3}$ or $k \mapsto k (k_0/k)^{n_t /3}$, where $k_0$ is some fiducial pivot scale.

\section{Estimator derivation} \label{app:estimator_der}

We review the derivation of the estimator in Eq.~\eqref{eq:estimator_single_par} and its behavior in the presence of non-Gaussian signal.

\subsection{\label{sec:estimator.estimator.estimation_theory} Estimation Theory}
The statistical estimate of a parameter produced by an unbiased estimator has an expectation value that is equal to the true value of the parameter. If such an unbiased estimator saturates the Cram\'{e}r-Rao bound, it achieves the lowest possible variance (or covariance for multiple parameters) on the  estimate, independent from the true value(s) of the parameter(s). We will briefly introduce the Cram\'{e}r-Rao bound.

Consider a dataset $x = \{x_1, x_2, \cdots, x_n \}$ drawn from the likelihood $\Pr(x|\theta)$: a probability density function (PDF) with unknown fixed parameters $\theta = \{\theta_1, \theta_2, \dots, \theta_d \}$. Under the assumption that the PDF satisfies the following regularity condition:
\begin{align}\label{eq:reg_cond}
\int \mathrm{d}^{n}\!x \frac{\partial \log \Pr(x| \theta)}{\partial \theta_i}  \Pr(x| \theta)  = 0 \, , \ \ \forall i \in \{1, \dots , d \} \, ,
\end{align}
it can be shown that the covariance matrix $C_{\theta}$ of an unbiased estimate of the parameters $\theta$ is bounded by the inverse of the Fisher information matrix:
\begin{align}
C_{\hat{\theta}} \geq \mathcal{I}^{-1}(\theta) \,. 
\end{align}
This bound is the Cram\'{e}r-Rao bound.
In the matrix notation used here, the inequality refers to the positive definiteness of the $C_{\hat{\theta}} - \mathcal{I}^{-1}$ matrix. The elements of the information matrix are directly obtained from the PDF:
\begin{align}
\begin{split}
\mathcal{I}_{ij}(\theta) = \int \mathrm{d}^n\! x &\left( \frac{\partial \log \Pr(x| \theta)}{\partial \theta_i} \right) \left( \frac{\partial \log \Pr(x| \theta)}{\partial \theta_j} \right)   \Pr(x| \theta) \, .
\end{split}
\end{align}
The Fisher information does not depend on the observed data; it only depends on the parameter vector. 

It can be shown that an unbiased estimator $\hat{\theta} = \{\hat{\theta}_1, \hat{\theta}_2, \dots , \hat{\theta}_d \}$ that saturates the bound for all values of the parameters $\theta$ must satisfy:
\begin{align}\label{eq:cramer_rao_condition}
\frac{\partial \log \Pr(x| \theta)}{ \partial\theta_i} = \sum_{j} \mathcal{I}_{ij}(\theta) (\hat{\theta}_j - \theta_j) \, .
\end{align}
Although it is generally non-trivial to construct an estimator that fulfills this relation for all possible values of $\theta$, the relation suggests a simple recipe for the construction of an estimator $\hat{\theta}$ that fulfills the Cram\'{e}r-Rao bound in the case where $\theta$ is known:
\begin{align}
\hat{\theta}_i = \sum_j \mathcal{I}^{-1}_{ij}(\theta)  \frac{\partial \log \Pr(x| \theta)}{ \partial\theta_j}  + \theta_i  \, ,
\end{align}
where $\mathcal{I}^{-1}$ is the inverse of the Fisher matrix. 
In reality, $\theta$ is unknown. However, an estimator constructed in this way may still be useful for estimates of $\theta$ that are close to the assumed value. This is the approach we will take. 

\subsection{\label{sec:estimator.estimator.cmb_estimators}CMB bispectrum estimation}

We now construct the bispectrum estimator and provide a brief discussion of its statistical properties.  We will see that the estimator is  unbiased and becomes statistically optimal (saturates the Cram\'{e}r-Rao bound) in the limit of vanishing non-Gaussianity.

\subsubsection{Probability density function}

It is clear from the previous section that a closed-form expression for the likelihood of the data is required to construct the estimator.  
However, there exists no such expression when the condition of Gaussian initial perturbations is relaxed.  
Without a closed-form expression, we thus construct an approximation to the full non-Gaussian likelihood by perturbing around the Gaussian form. The specifics of this perturbation are determined by the connected moments, or cumulants, predicted by the model. 

Given a characteristic function and its associated probability distribution, one can distinguish between the moments about the origin of the distribution (the $n$-point correlation functions) and the connected moments about the origin (the cumulants). The connected moments are proportional to the MacLaurin coefficients of the natural logarithm of the characteristic function. The connected moments about the origin are proportional to the MacLaurin coefficients of the characteristic function itself (i.e.\ without the logarithm). In more practical terms: the moments about the origin, the $n$-point correlation functions, may be expanded in terms of the connected moments with the help of Wick's theorem~\cite{martinez_2009}. For the mean-zero distributions we are interested in, the first moments of a random field, expressed as a set of spherical harmonic modes $\{a_{\ell m} \}$, are expanded as follows:
\begin{align}
\langle a_{\ell_1 m_1} a_{\ell_2 m_2}  \rangle &= \langle a_{\ell_1 m_1} a_{\ell_2 m_2} \rangle_c \, ,\\
\langle a_{\ell_1 m_1} a_{\ell_2 m_2} a_{\ell_3 m_3}  \rangle &= \langle a_{\ell_1 m_1} a_{\ell_2 m_2} a_{\ell_3 m_3}  \rangle_c  \, , \\
\langle a_{\ell_1 m_1} a_{\ell_2 m_2} a_{\ell_3 m_3} a_{\ell_4 m_4} \rangle &= \langle a_{\ell_1 m_1} a_{\ell_2 m_2} a_{\ell_3 m_3} a_{\ell_4 m_4} \rangle_c + \langle a_{\ell_1 m_1} a_{\ell_2 m_2} \rangle_c \langle a_{\ell_3 m_3} a_{\ell_4 m_4}  \rangle_c \nonumber \\ &\quad + \langle a_{\ell_1 m_1} a_{\ell_3 m_3} \rangle_c \langle a_{\ell_2 m_2} a_{\ell_4 m_4}  \rangle_c + \langle a_{\ell_1 m_1} a_{\ell_4 m_4} \rangle_c \langle a_{\ell_2 m_2} a_{\ell_3 m_3}  \rangle_c \, , \\
\langle a_{\ell_1 m_1} a_{\ell_2 m_2} a_{\ell_3 m_3} a_{\ell_4 m_4} a_{\ell_5 m_5} \rangle &= \langle a_{\ell_1 m_1} a_{\ell_2 m_2} a_{\ell_3 m_3} a_{\ell_4 m_4} a_{\ell_5 m_5}\rangle_c \, , \\
\langle a_{\ell_1 m_1} a_{\ell_2 m_2} a_{\ell_3 m_3} a_{\ell_4 m_4} a_{\ell_5 m_5} a_{\ell_6 m_6} \rangle &= \langle a_{\ell_1 m_1} a_{\ell_2 m_2} a_{\ell_3 m_3} a_{\ell_4 m_4} a_{\ell_5 m_5} a_{\ell_6 m_6}\rangle_c \nonumber \\ 
&\quad + \langle a_{\ell_1 m_1} a_{\ell_2 m_2} a_{\ell_3 m_3} a_{\ell_4 m_4} \rangle_c \langle a_{\ell_5 m_5} a_{\ell_6 m_6} \rangle_c \quad \quad \, + 14 \ \mathrm{perm.} \nonumber \\ 
&\quad + \langle a_{\ell_1 m_1} a_{\ell_2 m_2} a_{\ell_3 m_3} \rangle_c \langle a_{\ell_4 m_4}   a_{\ell_5 m_5} a_{\ell_6 m_6} \rangle_c \quad \quad \, + 9 \ \mathrm{perm.} \nonumber \\ 
&\quad + \langle a_{\ell_1 m_1} a_{\ell_2 m_2}\rangle_c \langle a_{\ell_3 m_3} a_{\ell_4 m_4} \rangle_c \langle a_{\ell_5 m_5} a_{\ell_6 m_6} \rangle_c \quad + 14 \ \mathrm{perm.} \, .
\end{align}
The quantities on the l.h.s.\ represent the moments and the quantities on the r.h.s.\ are the connected moments (denoted by $\langle \dots  \rangle_c$). For a distribution with a vanishing mean, there is no distinction between the moments and connected moments for $n=2$ and $n=3$. For $n=4$ and higher, we see a distinction. A Gaussian distribution is a distribution for which all connected moments with $n>2$ vanish. 

The approximation to the likelihood of the data we will use is known as the Edgeworth series. More specifically: an Edgeworth expansion around a mean zero multivariate Gaussian distribution. We truncate the series such that the only relevant cumulants are the $2$- and $3$-point functions. A detailed derivation of this procedure can be found in Ref.~\cite{taylor2001parameter}. In short: one Taylor expands the non-Gaussian part of a general characteristic function to first order and discards all terms except the third-order  moments. Fourier transforming this truncated series together with the unmodified Gaussian part yields the PDF. 
Although the Edgeworth expansion is an asymptotic series, truncating it to third order does not guarantee a well-defined (i.e.\ positive and normalized) PDF~\cite{2017arXiv170903452S}. However, as long as we are only interested in the weakly non-Gaussian regime, where the third-order moment is subdominant to the second, we assume that these subtleties can be safely ignored.

Representing the likelihood for a measured set of $n$ spherical harmonic modes $a = \{a_{X, \ell m} \}$ as the truncated Edgeworth series yields \cite{babich2005optimal,  Creminelli:2005hu}:
\begin{align}\label{eq:edgeworth}
\begin{split}
\Pr(a | C,B) = \Bigg(1 + \frac{1}{6} \sum_{\substack{\ell_1 , \ell_2 , \ell_3 \\ m_1 , m_2 , m_3}}  B_{m_1 m_2 m_3, X_1 X_2 X_3}^{\ell_1 \ell_2 \ell_3}  \bigg\{ &
\left[(C^{-1}a)^{X_1}_{\ell_1 m_1} (C^{-1}a)^{X_2}_{\ell_2 m_2} (C^{-1}a)^{X_3}_{\ell_3 m_3} \right]   \\
&- \left[(C^{-1})^{X_1 X_2}_{\ell_1 m_1 \ell_2 m_2} (C^{-1}a)^{X_3}_{\ell_3 m_3} + \mathrm{cyclic}\right] \bigg\} \Bigg)  \frac{e^{-\frac{1}{2}  a^{\dagger} C^{-1} a}}{\sqrt{(2\pi)^n \det{C}}} \, ,
\end{split}
\end{align}
where $C$ and $B$ denote the $2$- and $3$-point correlation functions of $a$. The notational shorthand $C^{-1}a$ is defined in Eq.~\eqref{eq:c_inv_a_body}. The above expression is that of a nested model: when $B$ vanishes, we recover the mean zero Gaussian model. The extra terms denoted by `$\mathrm{cyclic}$' are given by the two cyclic permutations of the three $(\ell, m, X)$ triplets.

It is straightforward to incorporate harmonic modes sourced by a combination of primordial scalar and tensor perturbations in the above description. Consider the following decomposition: 
\begin{align}\label{eq:alm_split_simple}
a_{X, \ell m} = a^{(\zeta)}_{X, \ell m} + a^{(h)}_{X, \ell m} + n_{X, \ell m} \, .
\end{align}
Since the noise $n_{X, \ell m}$ is independent from the primordial fields and since all components have zero mean, the most general bispectrum then is expressed as:
\begin{align}\label{eq:bispectra}
B = B^{(\zeta \zeta \zeta)} + 3B^{(\zeta \zeta h)} + 3B^{(\zeta hh)} + B^{(hhh)} \, .
\end{align}
Inserting Eq.~\eqref{eq:bispectra} into Eq.~\eqref{eq:edgeworth} produces a likelihood for $a$ that takes into account the non-Gaussian correlation between the primordial scalar and tensor fields.

\subsubsection{Estimator}

The condition in Eq.~(\ref{eq:cramer_rao_condition}) implies that an unbiased estimator of a vector of parameters $\hat{f}_{\mathrm{NL}} = \{\hat{f}^1_{\mathrm{NL}}, \hat{f}^2_{\mathrm{NL}}, \dots , \hat{f}^d_{\mathrm{NL}}  \}$ constructed as follows saturates the Cram\'{e}r-Rao bound in the limit where the parameter vector goes to the null vector, i.e.\ $f_{\mathrm{NL}}\rightarrow 0$:
\begin{align}\label{eq:opt_est_fnl}
\hat{f}^I_{\mathrm{NL}} = \sum_{J} \mathcal{I}_{IJ}^{-1}(f_{\mathrm{NL}}) \frac{\partial \log \Pr(a | f_{\mathrm{NL}})}{\partial f^J_{\mathrm{NL}}} \, .
\end{align}
The $I$ and $J$ indices run over the dimensions of the parameter vector space. Note that $\hat{f}_{\mathrm{NL}}$ and $f_{\mathrm{NL}}$ can be understood as either scalars or as vectors; in the latter case, the $\mathcal{I}^{-1}$ is the inverse of the $d \times d$ Fisher matrix instead of the scalar Fisher information. To identify $\Pr(a | C, B)$ in Eq.~\eqref{eq:edgeworth} with $\Pr(a | f_{\mathrm{NL}})$, we treat the bispectrum as fixed up to a scaling $f_{\mathrm{NL}} \in \mathbb{R}^d$ and consider the shape of the bispectrum and the covariance as fixed. More specifically, we assume:
\begin{align}\label{eq:b_fnl_vec}
B(f_{\mathrm{NL}}) = f_{\mathrm{NL}} \cdot B_1 \, ,
\end{align}
where the inner product is defined in the parameter vector space. This expression is a generalization of Eq.~\eqref{eq:b_fnl_scal} that allows the bispectrum to consist of a sum of bispectra each with its own $f_{\mathrm{NL}}$ parameter. To construct the estimator we now simply insert Eq.~\eqref{eq:b_fnl_vec} into the expression for the PDF in Eq.~(\ref{eq:edgeworth}) and insert the result into Eq.~\eqref{eq:opt_est_fnl}. We may expand the logarithm in a power series and neglect all terms but the one that is $\mathcal{O}(B)$. This is a valid approach because the second term in the brackets in Eq.~(\ref{eq:edgeworth}) must be $\ll 1$ in the weak non-Gaussian regime. This yields the estimator constructed by Ref.~\cite{Creminelli:2005hu} (which is a refinement to the cubic  expression originally introduced in Ref.~\cite{komatsu2001acoustic}):
\begin{align}
\begin{split}\label{eq:estimator_single_par_app}
\hat{f}^I_{\mathrm{NL}} = \frac{1}{6} \sum_J \mathcal{I}_{0,IJ}^{-1} \sum_{\mathrm{all}\,\ell, m  }   \sum_{\mathrm{all} X} (B_1^J)_{m_1 m_2 m_3, X_1 X_2 X_3}^{\ell_1 \ell_2 \ell_3}  \bigg\{ &
\left[(C^{-1}a)^{X_1}_{\ell_1 m_1} (C^{-1}a)^{X_2}_{\ell_2 m_2} (C^{-1}a)^{X_3}_{\ell_3 m_3} \right] \\
&\ \ \ - \left[(C^{-1})^{X_1 X_2}_{\ell_1 m_1 \ell_2 m_2} (C^{-1}a)^{X_3}_{\ell_3 m_3} + \mathrm{cyclic} \right] \bigg\} \, .
 \end{split}
\end{align}
Note the use of $\mathcal{I}^{-1}_0 \equiv  \mathcal{I}^{-1}(0)$ instead of  $\mathcal{I}^{-1}(f_{\mathrm{NL}})$: strictly speaking, the inverse of the Fisher matrix will depend on the parameter vector. This reflects the fact that a true optimal estimator should vary between datasets based on the value of $f_{\mathrm{NL}}$. Of course, such optimality is not possible with the point estimator we use here: $f_{\mathrm{NL}}$ is unknown. A true optimal weighting would be achieved with a Bayesian approach in which the likelihood of the data is calculated for each value of $f_{\mathrm{NL}}$. In reality, this re-weighting of the estimator is not important for values of $f_{\mathrm{NL}}$ that are of interest~\cite{2010ApJ...724.1262E}. For $f_{\mathrm{NL}} = 0$ the estimator is optimal by construction and the Fisher matrix has a simple analytic solution:
\begin{align}
\begin{split}
\mathcal{I}_{0,IJ} =  \frac{1}{6} \sum_{\mathrm{all}\,\ell, m  }   \sum_{\mathrm{all} X} \left(B^I_1\right)_{m_1 m_2 m_3, X_1 X_2 X_3}^{ \ell_1 \ell_2 \ell_3} \left[ (C^{-1})^{X_1 X_4}_{\ell_1 m_1 \ell_4 m_4} (C^{-1})^{X_2 X_5}_{\ell_2 m_2 \ell_5 m_5} (C^{-1})^{X_3 X_6}_{\ell_3 m_3 \ell_6 m_6} \! \right]   (B_1^{J*} )_{m_4 m_5 m_6, X_4 X_5 X_6}^{\ell_4 \ell_5 \ell_6} \, .
\end{split}
\end{align}

\subsubsection{Statistical properties estimator}

The estimator $\hat{f}_{\mathrm{NL}}$ is a function, or `statistic', of the data $a$, so the statistical properties of the estimator may be derived from the likelihood of the data. Here we present a heuristic overview of the statistical properties, given different models for the data. It should be understood that an analytic approach like the one presented here is mainly useful to gain intuition; characterization of the estimator applied to a real dataset requires the use of simulations.

To derive the bias, covariance and higher-order moments of the estimate, we first define what we mean by the $p$th moment of the estimate: 
\begin{align}
\langle \hat{f}_{\mathrm{NL}}^p\rangle &\equiv \mathrm{E}(\hat{f}^p_{\mathrm{NL}} | f_{\mathrm{NL}}) \\ 
&=\int \mathcal{D}a \, \hat{f}_{\mathrm{NL}}^p \Pr (a | f_{\mathrm{NL}}) \label{eq:moment_int}\, , 
\end{align}
where:
\begin{align}
\int \mathcal{D} a \equiv \prod_{\ell, m} \int \mathrm{d} a_{\ell m} \, ,
\end{align}
and where $\hat{f}_{\mathrm{NL}}^p$ denotes the $p$th power of the estimate. This notation is understood to generalize to the multivariate case as e.g.\ $\langle \hat{f}_{\mathrm{NL}}^2\rangle \rightarrow \langle \hat{f}_{\mathrm{NL}}^I \hat{f}_{\mathrm{NL}}^J \rangle$, $\langle \hat{f}_{\mathrm{NL}}^3\rangle \rightarrow \langle \hat{f}_{\mathrm{NL}}^I \hat{f}_{\mathrm{NL}}^J \hat{f}_{\mathrm{NL}}^K \rangle$, etc. It is then convenient to note that the expression for $\Pr (a | f_{\mathrm{NL}})$ in Eq.~\eqref{eq:edgeworth} consists of two parts: a regular Gaussian PDF and a second part that consists of a Gaussian PDF times terms cubic and linear in $a$. This means that we can divide the integral in Eq.~\eqref{eq:moment_int} into a purely Gaussian integral ($\langle \dots \rangle_G$) and another Gaussian integral ($\langle \dots \rangle_{G'}$) with an integrand that is multiplied with these cubic and linear terms. Since the estimator in Eq.~\eqref{eq:estimator_single_par_app} is an odd function of $a$, the $\langle \dots \rangle_G$ integral will always vanish for $p = \mathrm{odd}$. On the other hand, the $\langle \dots \rangle_{G'}$ part will always vanish for a moment with $p = \mathrm{even}$.

With this knowledge and the likelihood of the data in Eq.~\eqref{eq:edgeworth}, deriving the bias of the estimator comes down to evaluating Eq.~\eqref{eq:moment_int} for $p=1$. This is an odd moment, so only the $\langle \dots \rangle_{G'}$ integral has to be evaluated. The result is that $\langle \hat{f}_{\mathrm{NL}}\rangle =  f_{\mathrm{NL}}$, i.e.\ the estimate is unbiased regardless of the value of $f_{\mathrm{NL}}$. For the (co)variance of the estimate, i.e.\ $\mathrm{Var}(\hat{f}_{\mathrm{NL}}) \equiv \langle \hat{f}^2_{\mathrm{NL}}\rangle - \langle \hat{f}_{\mathrm{NL}}\rangle^2$, we need to additionally evaluate Eq.~\eqref{eq:moment_int} for $p=2$. Doing so, we find $\mathrm{Var}(\hat{f}_{\mathrm{NL}}) = \mathcal{I}^{-1}_{0} - f_{\mathrm{NL}}^2$, which is only equal to the optimal value $\mathcal{I}^{-1}(f_{\mathrm{NL}})$ when $f_{\mathrm{NL}} = 0$. So we establish that in the limit of $f_{\mathrm{NL}} \rightarrow 0$, the estimator is unbiased and optimal. In cases where $f_{\mathrm{NL}} \neq 0$, the estimator is still unbiased\footnote{Of course, any statements about unbiasedness rely on the assumed validity of the  truncated Edgeworth expansion, which, as mentioned, should be reconsidered in cases of large deviations from Gaussianity.} but suffers from non-optimal (co)variance~\cite{Creminelli:2006gc, Liguori:2007sj}.
This is expected, as the situation does not conform to  Eq.~\eqref{eq:cramer_rao_condition} anymore. Finally, note that for $f_{\mathrm{NL}} \neq 0$, the estimate itself becomes (weakly) non-Gaussian. For instance, there will be a nonzero $\langle \hat{f}^3_{\mathrm{NL}}\rangle$ moment with an $\mathcal{O}(f_{\mathrm{NL}} B_1^4  C^{-6} \mathcal{I}_0^{-3})$  amplitude. 

In the above we assumed that the likelihood for the data is described by Eq.~\eqref{eq:edgeworth}. When an additional $3$-point function, not parameterized  by an $f_{\mathrm{NL}}$ parameter, is introduced in the likelihood, the estimator becomes biased. The exact bias depends on the shape of the added $3$-point function, see the discussion in Sec.~\ref{sec:discussion}.

An interesting situation arises when the data are drawn from a distribution with nonzero higher-order connected moments. This situation is not only hypothetical: lensing introduces a significant connected $4$-point function, as well as smaller connected $6$-, $8$-, etc.\ point functions~\cite{lewis2006weak}. To describe the statistical properties of the estimator in the presence of lensing, we thus need to update the likelihood of the data in Eq.~\eqref{eq:edgeworth} with these nonzero higher-order connected moments. Let us focus on the connected $4$-point function, denoted by $T$. The Edgeworth expansion will now include $\mathcal{O}(T a^4/C^4)$, $\mathcal{O}(T a^2/C^3)$, and $\mathcal{O}(T / C^2)$ terms in addition to the $\mathcal{O}(1)$, $\mathcal{O}(B_1 a^3 / C^3)$, and $\mathcal{O}(B_1 a / C^2)$ terms already present in Eq.~\eqref{eq:edgeworth}. With these additions, the bias of the estimator does not change, but the variance of the estimator receives an $\mathcal{O}(B_1^2 T  C^{-5} \mathcal{I}_0^{-2})$ contribution. By extension, the addition of connected $6$-, $8$- or higher-point functions to the likelihood will also contribute to the estimator variance. The estimate itself will also become non-Gaussian with these additions. For instance: there is an $\mathcal{O}(B_1^4 T  C^{-8} \mathcal{I}_0^{-4})$ connected $4$-point function of $\hat{f}_{\mathrm{NL}}$ when a connected $4$-point function $T$ is added to the likelihood. Computing a semi-analytic estimate of the additional estimator variance is highly challenging due to the number of elements that make up the higher-order connected moments. See Ref.~\cite{2013MNRAS.429..344K,Coulton:2017crj} for  details on an semi-analytic approach  in the flat-sky approximation. We briefly discuss the expected additional lensing-induced estimator variance in Sec.~\ref{sec:disc_var}. 

\end{widetext}
\bibliographystyle{apsrev.bst}

\bibliography{references}

\end{document}